\documentclass[a4paper,UKenglish,cleveref, autoref, thm-restate]{lipics-v2021}

\pdfoutput=1 
\hideLIPIcs  

\title{Linear Rank Intersection Types} 

\author{Fábio {Reis}}{DCC-FCUP, University of Porto, Porto, Portugal \and LIACC - Artificial Intelligence and Computer Science Laboratory }{fabio.d.reis@protonmail.ch}{https://orcid.org/0000-0003-1714-8303}{This work was partially financially supported by Base Funding - UIDB/00027/2020 of the Artificial Intelligence and Computer Science Laboratory – LIACC - funded by national funds through the FCT/MCTES (PIDDAC).}

\author{Sandra {Alves}}{DCC-FCUP, University of Porto, Porto, Portugal \and LIACC - Artificial Intelligence and Computer Science Laboratory \and CRACS, INESC-TEC - Centre for Research in Advanced Computing Systems \and \url{http://www.dcc.fc.up.pt/~sandra} }{sandra@fc.up.pt}{https://orcid.org/0000-0001-8840-5587}{Partially supported by National Funds through the Portuguese funding agency, FCT - Fundação para a Ciência e a Tecnologia, within project LA/P/0063/2020.}

\author{Mário {Florido}}{DCC-FCUP, University of Porto, Porto, Portugal \and LIACC - Artificial Intelligence and Computer Science Laboratory }{amflorid@fc.up.pt}{https://orcid.org/0000-0002-0574-7555}{}

\authorrunning{F. Reis, S. Alves and M. Florido} 

\Copyright{Fábio Daniel M. Reis, Sandra Maria M. Alves, and António Mário S.\,M. Florido} 



\category{} 

\relatedversion{} 

\nolinenumbers 

\EventEditors{John Q. Open and Joan R. Access}
\EventNoEds{2}
\EventLongTitle{42nd Conference on Very Important Topics (CVIT 2016)}
\EventShortTitle{CVIT 2016}
\EventAcronym{CVIT}
\EventYear{2016}
\EventDate{December 24--27, 2016}
\EventLocation{Little Whinging, United Kingdom}
\EventLogo{}
\SeriesVolume{42}
\ArticleNo{23}

\usepackage{amsmath}
\usepackage{algorithm}
\usepackage{adjustbox}
\usepackage{mathdots}
\usepackage{textcomp}
\usepackage[noend]{algpseudocode}
\usepackage{amssymb}
\usepackage{mathtools}
\usepackage{multirow}
\usepackage{array,booktabs}
\newcolumntype{L}{@{}>{\kern\tabcolsep}l<{\kern\tabcolsep}}
\newcolumntype{?}{!{\vrule width 1pt}}
\usepackage{colortbl}
\usepackage[table]{xcolor}
\usepackage{siunitx}
\usepackage{syntax}
\usepackage{xparse}
\usepackage{forest}

\let\olddegree\degree  
\let\degree\relax      
\usepackage{gensymb}   
\let\degree\olddegree  

\newenvironment{Dequation}
  {%
  \def\tagform@##1{%
    \maketag@@@{\makebox[0pt][r]{(\ignorespaces##1\unskip\@@italiccorr)}}}%
  \ignorespaces
  }
  {%
  \def\tagform@##1{\maketag@@@{(\ignorespaces##1\unskip\@@italiccorr)}}%
  \ignorespacesafterend
  }

\usepackage{tikz}
\usepackage{tikz-qtree}
\usepackage{bussproofs}

\usetikzlibrary{calc,trees,positioning,arrows,chains,shapes.geometric,%
    decorations.pathreplacing,decorations.pathmorphing,shapes,%
    matrix,shapes.symbols}
\tikzset{
>=stealth',
  punktchain/.style={
    rectangle, 
    rounded corners, 
    draw=black, very thick,
    text width=5em, 
    minimum height=3em, 
    text centered, 
    on chain},
  line/.style={draw, thick, <-},
  element/.style={
    tape,
    top color=white,
    bottom color=blue!50!black!60!,
    minimum width=5em,
    draw=blue!40!black!90, very thick,
    text width=5em, 
    minimum height=3.5em, 
    text centered, 
    on chain},
  every join/.style={->, thick,shorten >=1pt},
  decoration={brace},
  tuborg/.style={decorate},
  tubnode/.style={midway, right=2pt},
}

\newenvironment{mathprooftree}
  {\varwidth{.9\textwidth}\centering\leavevmode}
  {\DisplayProof\endvarwidth}

\newtheorem*{notation}{Notation}
\newtheorem{convention}{Convention}[section]

\newcommand{\tul}[1]{\underline{#1}}
\newcommand{\msf}[1]{\mathsf{#1}}

\newcommand{\inftwo}[3]{#1 \vdash_{2} #2 : #3}

\newcommand{\infertight}[5]{#1 \vdash^{#2} #3 : (#4, #5)}

\newcommand{\tl}[1]{\mathbb{T_L}_{#1}}

\newcommand{\fv}[1]{\mathsf{FV}(#1)}

\def\dom{\mathsf{dom}}

\newcommand{\subst}[1]{[#1]}

\newcommand{\sub}[2]{%
  #1 \subst{#2}%
}

\newcommand{\tsubst}[1]{[#1]}

\newcommand{\tsub}[2]{%
  #1 \tsubst{#2}%
}

\def\s{\mathcal{S}}

\def\ts{\mathbb{S}}

\def\rz{\tau}
\def\ro{\vec{\tau}}
\def\rt{\sigma}

\def\trz{t}
\def\tro{\vec{t}}
\def\trt{s}

\newcommand{\oset}[3][0ex]{%
  \mathrel{\mathop{#3}\limits^{
    \vbox to#1{\kern-2pt
    \hbox{$\scriptstyle#2$}\vss}}}}
\newcommand{\overmultimap}[1]{\oset[.06ex]{\multimap}{#1}}
\def\longmultimap{\hbox{$-$}\kern-2pt\hbox{$-$}\kern-1.5pt\hbox{$\circ$}}
\newcommand{\overlongmultimap}[1]{\oset[.6ex]{\longmultimap}{#1}}

\definecolor{azulciencias}{rgb}{0.0, 0.0, 0.0} 

\begin{document}

\maketitle

\begin{abstract}
Non-idempotent intersection types provide quantitative information about typed programs, and have been used to obtain time and space complexity measures. Intersection type systems characterize termination, so restrictions need to be made in order to make typability decidable. One such restriction consists in using a notion of finite rank for the idempotent intersection types. In this work, we define a new notion of rank for the non-idempotent intersection types. We then define a novel type system and a type inference algorithm for the $\lambda$-calculus, using the new notion of rank~2. In the second part of this work, we extend the type system and the type inference algorithm to use the quantitative properties of the non-idempotent intersection types to infer quantitative information related to resource usage.
\end{abstract}

\section{Introduction}\label{chap:intro}

The ability to determine upper bounds for the number of execution steps of a program in compilation time is a relevant problem, since it allows us to know in advance the computational resources needed to run the program.

Type systems are a powerful and successful tool of static program analysis that are used, for example, to detect errors in programs before running them. Quantitative type systems, besides helping on the detection of errors, can also provide quantitative information related to computational properties.

Intersection types, defined by the grammar $\rt \Coloneqq \alpha \mid \rt_1 \cap \cdots \cap \rt_n \rightarrow \rt$ (where $\alpha$ is a type variable and $n \geq 1$), are used in several type systems for the $\lambda$-calculus \cite{coppo1980extension, DBLP:conf/mfcs/Coppo80, jim1995rank, DBLP:journals/fuin/Bakel96} and allow $\lambda$-terms to have more than one type.
Non-idempotent intersection types \cite{Gardner94, Kfoury2000, Carvalho07, BucciarelliKV17}, also known as \emph{quantitative types}, are a flavour of intersection types in which the type constructor $\cap$ is non-idempotent, and provide more than just qualitative information about programs. They are particularly useful in contexts where we are interested in measuring the use of resources, as they are related to the consumption of time and space in programs.
Type systems based on non-idempotent intersection types, use non-idempotence to count the number of evaluation steps and the size of the result. For instance in \cite{accattoli2018tight}, the authors define several quantitative type systems, corresponding to different evaluation strategies, for which they are able to measure the number of steps taken by that strategy to reduce a term to its normal form, and the size of the term's normal form.
Typability is undecidable for intersection type systems, as it corresponds to termination. One way to get around this is to restrict intersection types to finite ranks, a notion defined by Daniel Leivant in \cite{leivant1983polymorphic} that makes typability decidable - Kfoury and Wells \cite{KfouryW99} define an intersection type system that, when restricted to any finite-rank, has principal typings and decidable type inference. Type systems that use finite-rank intersection types are still very powerful and useful. For instance, rank~2 intersection type systems \cite{jim1995rank, van1993intersection, damiani2007rank} are more powerful, in the sense that they can type strictly more terms, than popular systems like the ML type system \cite{damasmilner}.  Still related to decidability of typability for finite ranks, Dudenhefner and Rehof~\cite{Dudenhefner17b} studied the problem for a notion of bounded-dimensional intersection types. This notion was previously defined in the context of type inhabitation~\cite{Dudenhefner17}, where it was used to prove decidability of type inhabitation for a non-idempotent intersection type system (the problem is known to be undecidable above rank~2, for idempotent intersection types~\cite{Urzyczyn99}). 

In this paper we present a new definition of rank for the quantitative types, which we call \emph{linear rank} and differs from the classical one in the base case -- instead of simple types, linear rank~$0$ intersection types are the linear types. In a non-idempotent intersection type system, every linear term is typable with a simple type (in fact, in many of those systems, only the linear terms are), which is the motivation to use linear types for the base case. The relation between non-idempotent intersection types and linearity has already been studied by Kfoury \cite{Kfoury2000}, de Carvalho \cite{Carvalho07},  Gardner \cite{Gardner94} and Florido and Damas \cite{FloridoD04}.
Our motivation to redefine rank in the first place, has to do with our interest in using non-idempotent intersection types to estimate the number of evaluation steps of a $\lambda$-term to normal form while inferring its type, and the realization that there is a way to define rank that is more suitable for the quantitative types.
We define a new intersection type system for the $\lambda$-calculus, restricted to linear rank~2 non-idempotent intersection types, and a new type inference algorithm that we prove to be sound and complete with respect to the type system.

 Finally we extend our type system and inference algorithm to use the quantitative properties of the linear rank~2 non-idempotent intersection types to infer not only the type of a $\lambda$-term, but also the number of evaluation steps of the term to its normal form.
The new type system is the result of a merge between our Linear Rank~2 Intersection Type System and the system for the leftmost-outermost evaluation strategy presented in \cite{accattoli2018tight}. The type system in~\cite{accattoli2018tight} is a quantitative typing system extended with the notion of tight types (which provide an effective characterisation of minimal typings) that is crucial to extract exact bounds for reduction. We prove that the system gives the correct number of evaluation steps for a kind of derivation.
As for the new type inference algorithm, we show that it is sound and complete with respect to the type system for the inferred types, and conjecture that the inferred measures correspond to the ones given by the type system (i.e., correspond to the number of evaluation steps of the term to its normal form, when using the leftmost-outermost evaluation strategy).

Thus, the main contributions of this paper are the following:

\begin{itemize}
    \item A new definition of rank for non-idempotent intersection types, which we call \emph{linear rank} (\autoref{chap:linearrank});
    
    \item A Linear Rank~2 Intersection Type System for the $\lambda$-calculus (\autoref{chap:linearrank});
    
    \item A type inference algorithm that is sound and complete with respect to the Linear Rank~2 Intersection Type System (\autoref{chap:linearrank});
    
    \item A Linear Rank~2 Quantitative Type System for the $\lambda$-calculus that derives a measure related to the number of evaluation steps for the leftmost-outermost strategy (\autoref{chap:resource});
    
    \item A type inference algorithm that is sound and complete with respect to the Linear Rank~2 Quantitative Type System, for the inferred types, and gives a measure that we conjecture to correspond to the number of evaluation steps of the typed term for the leftmost-outermost strategy (\autoref{chap:resource}).
\end{itemize}

In this paper we assume that the reader is familiar with the $\lambda$-calculus \cite{Barendregt85}.
From now on, in the rest of the paper, terms of the $\lambda$-calculus are considered modulo $\alpha$-equivalence and we use Barendregt's variable convention \cite{barendregt1992lambda}.

\section{Intersection Types}\label{chap:background}

The simply typed $\lambda$-calculus is a typed version of the $\lambda$-calculus, introduced by Alonzo Church in \cite{church1940simple} and by Haskell Curry and Robert Feys in \cite{curry1958combinatory}. One system that uses simple types is the Curry Type System, which was first introduced in \cite{curry1934functionality} for the theory of combinators, and then modified for the $\lambda$-calculus in \cite{curry1958combinatory}. Typability in this system is decidable and there is an algorithm that given a term, returns its principal pair. However, the system presents some disadvantages when comparing to others, one of them being the large number of terms that cannot be typed. For example, in the Curry Type System we cannot assign a type to the $\lambda$-term $\lambda x.x x$. This term, on the other hand, can be typed in systems that use intersection types, which allow terms to have more than one type. Such a system is the Coppo-Dezani Type System \cite{coppo1980extension}, which was one of the first to use intersection types, and a basis for subsequent systems.

\begin{definition}[Intersection types]\label{intertypes}
Intersection types $\rt,\rt_1,\rt_2, \ldots \in \mathbb{T}$ are defined by the following grammar:
    \begin{grammar}\centering
        <$\rt$> $\Coloneqq$ $\alpha \mid \rt_1 \cap \cdots \cap \rt_n \rightarrow \rt$
    \end{grammar}
 where $n \geq 1$ and $\rt_1 \cap \cdots \cap \rt_n$ is called a \emph{sequence} of types.
    
Note that intersections arise in different systems in different scopes. Here we follow several previous presentations where intersections are only allowed directly on the left-hand side of arrow types and sequences are non-empty \cite{coppo1980extension, DBLP:conf/mfcs/Coppo80, jim1995rank, DBLP:journals/fuin/Bakel96}.
\end{definition}

\begin{notation}
The intersection type constructor $\cap$ binds stronger than $\rightarrow$: $\alpha_1 \cap \alpha_2 \rightarrow \alpha_3$ stands for $(\alpha_1 \cap \alpha_2) \rightarrow \alpha_3$.
\end{notation}

\begin{example}
Some examples of intersection types are:
\begin{center}
    $\alpha$;\\
    $\alpha_1 \rightarrow \alpha_2$;\\
    $\alpha_1 \cap \alpha_2 \rightarrow \alpha_3$;\\
    $(\alpha_1 \cap \alpha_2 \rightarrow \alpha_3) \rightarrow \alpha_4$;\\
    $\alpha_1 \cap (\alpha_1 \rightarrow \alpha_2) \rightarrow \alpha_3$.
\end{center}
\end{example}

\begin{definition}[Coppo-Dezani Type System]
In the Coppo-Dezani Type System, we say that $M$ has type $\rt$ given the environment $\Gamma$ (where the predicates of declarations are sequences), and write $\Gamma \vdash_\mathcal{CD} M:\rt$, if $\Gamma \vdash_\mathcal{CD} M:\rt$ can be obtained from the \emph{derivation rules} in \autoref{fig:cd}, where $1 \leq i \leq n$:

\begin{figure}
\begin{equation}\tag{Axiom}
\begin{mathprooftree}
    \AxiomC{$\Gamma \cup \{x:\rt_1 \cap \cdots \cap \rt_n\} \vdash_\mathcal{CD} x:\rt_i$}
\end{mathprooftree}
\end{equation}
\begin{equation}\tag{$\rightarrow$ Intro}
\begin{mathprooftree}
    \AxiomC{$\Gamma \cup \{x:\rt_1 \cap \cdots \cap \rt_n\} \vdash_\mathcal{CD} M:\rt$}
    \UnaryInfC{$\Gamma \vdash_\mathcal{CD} \lambda x.M:\rt_1 \cap \cdots \cap \rt_n \rightarrow \rt$}
\end{mathprooftree}
\end{equation}
\begin{equation}\tag{$\rightarrow$ Elim}
\begin{mathprooftree}
    \AxiomC{$\Gamma \vdash_\mathcal{CD} M_1:\rt_1 \cap \cdots \cap \rt_n \rightarrow \rt$}
    \AxiomC{$\Gamma \vdash_\mathcal{CD} M_2:\rt_1 \; \cdots \; \Gamma \vdash_\mathcal{CD} M_2:\rt_n$}
    \BinaryInfC{$\Gamma \vdash_\mathcal{CD} M_1 M_2:\rt$}
\end{mathprooftree}
\end{equation}
\caption{Coppo-Dezani Type System}\label{fig:cd}
\end{figure}

\end{definition}

\begin{example} For the $\lambda$-term $\lambda x.x x$ the following derivation is obtained:
    \begin{equation*}
    \begin{mathprooftree}
        \AxiomC{$\{x:\rt_1 \cap (\rt_1 \rightarrow \rt_2)\} \vdash_\mathcal{CD} x:\rt_1 \rightarrow \rt_2$}
        \AxiomC{$\{x:\rt_1 \cap (\rt_1 \rightarrow \rt_2)\} \vdash_\mathcal{CD} x:\rt_1$}
        \BinaryInfC{$\{x:\rt_1 \cap (\rt_1 \rightarrow \rt_2)\} \vdash_\mathcal{CD} x x:\rt_2$}
        \UnaryInfC{$\vdash_\mathcal{CD} \lambda x.x x:\rt_1 \cap (\rt_1 \rightarrow \rt_2) \rightarrow \rt_2$}
    \end{mathprooftree}
    \end{equation*}
\end{example}

This system is a true extension of the Curry Type System, allowing term variables to have more than one type in the ($\rightarrow$ Intro) derivation rule and the right-hand term to also have more than one type in the ($\rightarrow$ Elim) derivation rule.

\subsection{Finite Rank}\label{subsec:rank}

Intersection type systems, like the Coppo-Dezani Type System, characterize termination, in the sense that a $\lambda$-term is strongly normalizable if and only if it is typable in an intersection type system. Thus, typability is undecidable for these systems.

To get around this, some current intersection type systems are restricted to types of finite rank \cite{jim1995rank, van1993intersection, KfouryW99, damiani2007rank} using a notion of rank first defined by Daniel Leivant in \cite{leivant1983polymorphic}. This restriction makes typability decidable \cite{KfouryW99}. Despite using finite-rank intersection types, these systems are still very powerful and useful. For instance, rank~2 intersection type systems \cite{jim1995rank, van1993intersection, damiani2007rank} are more powerful, in the sense that they can type strictly more terms, than popular systems like the ML type system \cite{damasmilner}.

The \emph{rank} of an intersection type is related to the depth of the nested intersections and it can be easily determined by examining the type in tree form: a type is of rank~$k$ if no path from the root of the type to an intersection type constructor $\cap$ passes to the left of $k$ arrows.

\begin{example}
The intersection type $\alpha_1 \cap (\alpha_1 \rightarrow \alpha_2) \rightarrow \alpha_2$ (tree on the left) is a rank~2 type and $(\alpha_1 \cap \alpha_2 \rightarrow \alpha_3) \rightarrow \alpha_4$ (tree on the right) is a rank~3 type:

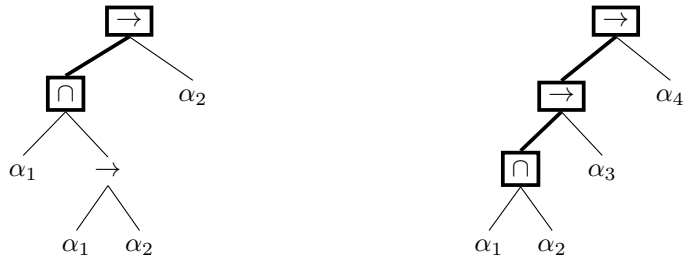
\forestset{
    sn edges/.style={for tree={parent anchor=south, child anchor=north, s sep=2mm}},
    azulciencias subtree/.style={for tree={text=azulciencias},for descendants={edge=azulciencias}}
}

\begin{center}
\begin{forest}
sn edges
[$\rightarrow$, s sep=11mm, draw={azulciencias, line width=1.4pt}
              [$\cap$, s sep=5mm, edge={azulciencias, line width=1.4pt}, draw={azulciencias, line width=1.4pt}
                      [$\alpha_1$]
                      [$\rightarrow$
                                      [$\alpha_1$]
                                      [$\alpha_2$]
                      ]
              ]
              [$\alpha_2$]
]
\end{forest}
\qquad\qquad\qquad\qquad\quad
\begin{forest}
sn edges
[$\rightarrow$, s sep=8mm, draw={azulciencias, line width=1.4pt}
              [$\rightarrow$, s sep=5mm, edge={azulciencias, line width=1.4pt}, draw={azulciencias, line width=1.4pt}
                      [$\cap$, edge={azulciencias, line width=1.4pt}, draw={azulciencias, line width=1.4pt}
                                      [$\alpha_1$]
                                      [$\alpha_2$]
                      ]
                      [$\alpha_3$]
              ]
              [$\alpha_4$]
]
\end{forest}
\end{center}
\end{example}

\begin{definition}[Rank of intersection types]\label{rank}
Let $\mathbb{T}_0$ be the set of simple types and $\mathbb{T}_1 = \{ \rz_1 \cap \dots \cap \rz_m \mid \rz_1, \dots, \rz_m \in \mathbb{T}_0, m \geq 1 \}$ the set of sequences of simple types (written as $\ro$). 
The set $\mathbb{T}_k$, of rank $k$ intersection types (for $k \geq 2$), can be defined recursively in the following way ($n \geq 3$, $m \geq 1$):
\begin{alignat*}{3}
    &\mathbb{T}_2 &&= &&\; \mathbb{T}_{0}
        \cup \{ \ro \rightarrow \rt \mid \ro \in \mathbb{T}_{1}, \rt \in \mathbb{T}_2 \}\\
    &\mathbb{T}_n &&= &&\; \mathbb{T}_{n-1}
        \cup \{ \rho_1 \cap \dots \cap \rho_m \rightarrow \rt \mid \rho_1, \dots, \rho_m \in \mathbb{T}_{n-1}, \rt \in \mathbb{T}_n \}
\end{alignat*}
\end{definition}

\begin{notation}
We consider the intersection type constructor $\cap$ to be associative, commutative and non-idempotent (meaning that $\alpha \cap \alpha$ is not equivalent to $\alpha$).
\end{notation}

We are particularly interested in non-idempotent intersection types, also known as quantitative types, because they provide more quantitative information than the idempotent ones.

\section{Linear Rank Intersection Types}\label{chap:linearrank}

In the previous chapter, we mentioned several intersection type systems in which intersection is idempotent and types are rank-restricted. There are also many quantitative type systems \cite{Gardner94, Kfoury2000, Carvalho07, BucciarelliKV17} that, on the other hand, make use of non-idempotent intersection types, for which there is no specific definition of rank.

The generalization of ranking for non-idempotent intersection types is not trivial and raises interesting questions that we will address in this chapter, along with a definition of a new non-idempotent intersection type system and a type inference algorithm.

This and the following sections cover original work that we presented at the TYPES~2022 conference \cite{typesabstract}.

\subsection{Linear Rank}\label{sec:linearrank}

The set of terms typed using idempotent rank~2 intersection types and non-idempotent rank~2 intersection types is not the same. For instance, the term  $(\lambda x.x x) (\lambda f x. f (f x))$ is typable with a simple type when using idempotent intersection types, but not when using non-idempotent intersection types. This comes from the two different occurrences of $f$ in $\lambda f x. f (f x)$, which even if typed with the same type, are not contractible because intersection is non-idempotent. Note that this is strongly related to the linearity features of terms. A $\lambda$-term $M$ is called a \emph{linear term} if and only if, for each subterm of the form $\lambda x.N$ in $M$, $x$ occurs free in $N$ exactly once, and if each free variable of $M$ has just one occurrence free in $M$. So the term $(\lambda x.x x) (\lambda f x. f (f x))$ is not typable with a non-idempotent rank~2 intersection type precisely because the term $\lambda f x. f (f x)$ is not linear.

Note that in a non-idempotent intersection type system, every linear term is typable with a simple type (in fact, in many of those systems, only the linear terms are). This motivated us to come up with a new notion of rank for non-idempotent intersection types, based on linear types (the ones derived in a linear type system -- a substructural type system in which each assumption must be used exactly once, corresponding to the implicational fragment of linear logic \cite{Girard87}).
The relation between non-idempotent intersection types and linearity was first introduced by Kfoury \cite{Kfoury2000} and further explored by de Carvalho \cite{Carvalho07}, who established its relation with linear logic.

Here we propose a new definition of rank for intersection types, which we call \emph{linear rank} and differs from the classical one in the base case -- instead of simple types, linear rank~$0$ intersection types are the linear types -- and in the introduction of the functional type constructor `linear arrow' $\multimap$.

\begin{definition}[Linear rank of intersection types]\label{linearrank}
Let $\tl{0} = \mathbb{V} \cup \{ \rz_1 \multimap \rz_2 \mid \rz_1, \rz_2 \in \tl{0} \}$ be the set of \textbf{linear types} and $\tl{1} = \{ \rz_1 \cap \dots \cap \rz_m \mid \rz_1, \dots, \rz_m \in \tl{0}, m \geq 1 \}$ the set of sequences of linear types. The set $\tl{k}$, of \emph{linear rank}~$k$ intersection types (for $k \geq 2$), can be defined recursively in the following way ($n \geq 3$, $m \geq 2$):
\begin{alignat*}{3}
    &\tl{2} &&= &&\; \tl{0}
        \cup \{ \rz \multimap \rt \mid \rz \in \tl{0}, \rt \in \tl{2} \}\\
        & && &&\; \cup \{ \rz_1 \cap \dots \cap \rz_m \rightarrow \rt \mid \rz_1, \dots, \rz_m \in \tl{0}, \rt \in \tl{2} \}\\
    &\tl{n} &&= &&\; \tl{n-1}
        \cup \{ \rho \multimap \rt \mid \rho \in \tl{n-1}, \rt \in \tl{n} \}\\
        & && &&\; \cup \{ \rho_1 \cap \dots \cap \rho_m \rightarrow \rt \mid \rho_1, \dots, \rho_m \in \tl{n-1}, \rt \in \tl{n} \}
\end{alignat*}
\end{definition}

Initially, the idea for the change arose from our interest in using rank-restricted intersection types to estimate the number of evaluation steps of a $\lambda$-term while inferring its type. While defining the intersection type system to obtain quantitative information, we realized that the ranks could be potentially more useful for that purpose if the base case was changed to types that give more quantitative information in comparison to simple types, which is the case for linear types -- for instance, if a term is typed with a linear rank~2 intersection type, one knows that each occurrence of its arguments is linear, meaning that they will be used exactly once.

The relation between the standard definition of rank and our definition of linear rank is not clear, and most likely non-trivial. Note that the set of terms typed using standard rank~2 intersection types \cite{jim1995rank, van1993intersection} and linear rank~2 intersection types is not the same. For instance, again, the term $(\lambda x.x x) (\lambda f x. f (f x))$, typable with a simple type in the standard Rank~2 Intersection Type System, is not typable in the Linear Rank~2 Intersection Type System, because, as the term $(\lambda f x. f (f x))$ is not linear and intersection is not idempotent, by \autoref{linearrank}, the type of $(\lambda x.x x) (\lambda f x. f (f x))$ is now (linear) rank~3. This relation between rank and linear rank is an interesting question that will not be covered here, but one that we would like to explore in the future.

\subsection{Type System}\label{sec:typesystem}

We now define a new type system for the $\lambda$-calculus with linear rank~2 non-idempotent intersection types.

\begin{definition}[Substitution] Let $S=\subst{N/x}$ denote a \emph{substitution}. Then the result of  substituting the term $N$ for each free occurrence of $x$ in the term $M$, denoted by $\sub{M}{N/x}$ (or $\s(M)$), is inductively defined as follows:
\begin{align*}
    \sub{x}{N/x} &= N;\\
    \sub{x_1}{N/x_2} &= x_1, \text{ if } x_1 \neq x_2;\\
    \sub{(M_1 M_2)}{N/x} &= (\sub{M_1}{N/x}) (\sub{M_2}{N/x});\\
    \sub{(\lambda x.M)}{N/x} &= \lambda x.M;\\
    \sub{(\lambda x_1.M)}{N/x_2} &= \lambda x_1.(\sub{M}{N/x_2}), \text{ if } x_1 \neq x_2.
\end{align*}
\end{definition}

\begin{notation}
We write $\sub{M}{M_1/x_1, M_2/x_2, \dots, M_n/x_n}$ for $\sub{(\dots (\sub{(\sub{M}{M_1/x_1})}{M_2/x_2}) \dots)}{M_n/x_n}$.
\end{notation}

Composing two substitutions $\s_1$ and $\s_2$ results in a substitution $\s_2 \circ \s_1$ that when applied, has the same effect as applying $\s_1$ followed by $\s_2$.

\begin{definition}[Substitution composition]
The \emph{composition} of two substitutions $\s_1 = \subst{N_1/x_1}$ and $\s_2 = \subst{N_2/x_2}$, denoted by $\s_2 \circ \s_1$, is defined as:
\[
    \s_2 \circ \s_1 (M) = \sub{M}{N_1/x_1, N_2/x_2}.
\]

Also, we consider that the operation is right-associative:
\[
\s_1 \circ \s_2 \circ \cdots \circ \s_{n-1} \circ \s_n = \s_1 \circ (\s_2 \circ \cdots \circ (\s_{n-1} \circ \s_n) \dots).
\]
\end{definition}

\begin{notation}
From now on, we will use $\alpha$ to range over a countable infinite set $\mathbb{V}$ of type variables, $\rz$ to range over the set $\tl{0}$ of linear types, $\ro$ to range over the set $\tl{1}$ of linear type sequences and $\rt$ to range over the set $\tl{2}$ of linear rank~2 intersection types. In all cases, we may use or not single quotes and/or number subscripts.
\end{notation}


\begin{definition}
    ~\begin{itemize}
        \item A \emph{statement} is an expression of the form $M:\ro$, where $\ro$ is called the \emph{predicate}, and the term $M$ is called the \emph{subject} of the statement.
        
        \item A \emph{declaration} is a statement where the subject is a term variable.
        
        \item The comma operator (,) appends a declaration to the end of a list (of declarations). The list $(\Gamma_1, \Gamma_2)$ is the list that results from appending the  list $\Gamma_2$ to the end of the list $\Gamma_1$.
        
        \item A finite list of declarations is \emph{consistent} if and only if the term variables are all distinct.
        
        \item An \emph{environment} is a consistent finite list of declarations which predicates are sequences of linear types (i.e., elements of $\tl{1}$) and we use $\Gamma$ (possibly with single quotes and/or number subscripts) to range over environments.
        
        \item An environment $\Gamma = [x_1:\ro_1, \dots, x_n:\ro_n]$ induces a partial function $\Gamma$ with domain $\dom(\Gamma) = \{x_1, \dots, x_n\}$, and $\Gamma(x_i) = \ro_i$.
        
        \item We write $\Gamma_x$ for the resulting environment of eliminating the declaration of $x$ from $\Gamma$ (if there is no declaration of $x$ in $\Gamma$, then $\Gamma_x = \Gamma$).

        \item We extend the notion of substitution to environments in the following way:
        \[
            \s(\Gamma) = [\s(x_1):\ro_1,\dots,\s(x_n):\ro_n] \qquad \text{if } \Gamma = [x_1:\ro_1,\dots,x_n:\ro_n]
        \]
        
        \item We write $\Gamma_1 \equiv \Gamma_2$ if the environments $\Gamma_1$ and $\Gamma_2$ are equal up to the order of the declarations.
        
        \item If $\Gamma_1$ and $\Gamma_2$ are environments, the environment $\Gamma_1 + \Gamma_2$ is defined as follows:
        
        for each $x \in \dom(\Gamma_1) \cup \dom(\Gamma_2)$,
            \[
                (\Gamma_1 + \Gamma_2)(x) = \left\{
                \begin{array}{ll}
                    \Gamma_1(x) &\text{if } x \notin \dom(\Gamma_2)\\
                    \Gamma_2(x) &\text{if } x \notin \dom(\Gamma_1)\\
                    \Gamma_1(x) \cap \Gamma_2(x) &\text{otherwise}
                \end{array}
                \right.
            \]
        with the declarations of the variables in $\dom(\Gamma_1)$ in the beginning of the list, by the same order they appear in $\Gamma_1$, followed by the declarations of the variables in $\dom(\Gamma_2) \setminus \dom(\Gamma_1)$, by the order they appear in $\Gamma_2$.
    \end{itemize}
\end{definition}

\begin{definition}[Linear Rank~2 Intersection Type System]\label{lineartypesystem}
In the Linear Rank~2 Intersection Type System, we say that $M$ has type $\rt$ given the environment $\Gamma$, and write $\inftwo{\Gamma}{M}{\rt}$, if it can be obtained from the \emph{derivation rules} in \autoref{fig:lr2its}.
\begin{figure}[t]
\begin{equation}\tag{Axiom}
\begin{mathprooftree}
    \AxiomC{$\inftwo{ [x:\rz] }{ x }{ \rz }$}
\end{mathprooftree}
\end{equation}

\begin{equation}\tag{Exchange}
\begin{mathprooftree}
    \AxiomC{$\inftwo{ \Gamma_1, x:\ro_1, y:\ro_2, \Gamma_2 }{ M }{ \rt }$}
    \UnaryInfC{$\inftwo{ \Gamma_1, y:\ro_2, x:\ro_1, \Gamma_2 }{ M }{ \rt }$}
\end{mathprooftree}
\end{equation}

\begin{equation}\tag{Contraction}
\begin{mathprooftree}
    \AxiomC{$\inftwo{ \Gamma_1, x_1:\ro_1, x_2:\ro_2, \Gamma_2 }{ M }{ \rt }$}
    \UnaryInfC{$\inftwo{ \Gamma_1, x:\ro_1 \cap \ro_2, \Gamma_2 }{ \sub{M}{x/x_1, x/x_2} }{ \rt }$}
\end{mathprooftree}
\end{equation}

\begin{equation}\tag{$\rightarrow$ Intro}
\begin{mathprooftree}
    \AxiomC{$\inftwo{ \Gamma, x:\rz_1 \cap \cdots \cap \rz_n }{ M }{ \rt }$}
    \AxiomC{$n \geq 2$}
    \BinaryInfC{$\inftwo{ \Gamma }{ \lambda x.M }{ \rz_1 \cap \cdots \cap \rz_n \rightarrow \rt}$}
\end{mathprooftree}
\end{equation}

\begin{Dequation}
\begin{equation}\tag{$\rightarrow$ Elim}
\begin{mathprooftree}
    \AxiomC{$\inftwo{ \Gamma }{ M_1 }{ \rz_1 \cap \cdots \cap \rz_n \rightarrow \rt }$}
    \AxiomC{$\inftwo{ \Gamma_1 }{ M_2 }{ \rz_1 } \; \cdots \; \inftwo{ \Gamma_n }{ M_2 }{ \rz_n }$}
    \AxiomC{$n \geq 2$}
    \TrinaryInfC{$\inftwo{ \Gamma, \sum_{i=1}^n \Gamma_i }{ M_1 M_2 }{ \rt }$}
\end{mathprooftree}
\end{equation}
\end{Dequation}

\begin{equation}\tag{$\multimap$ Intro}
\begin{mathprooftree}
    \AxiomC{$\inftwo{ \Gamma, x:\rz }{ M }{ \rt }$}
    \UnaryInfC{$\inftwo{ \Gamma }{ \lambda x.M }{ \rz \multimap \rt }$}
\end{mathprooftree}
\end{equation}

\begin{equation}\tag{$\multimap$ Elim}
\begin{mathprooftree}
    \AxiomC{$\inftwo{ \Gamma_1 }{ M_1 }{ \rz \multimap \rt }$}
    \AxiomC{$\inftwo{ \Gamma_2 }{ M_2 }{ \rz }$}
    \BinaryInfC{$\inftwo{ \Gamma_1, \Gamma_2 }{ M_1 M_2 }{ \rt }$}
\end{mathprooftree}
\end{equation}
    \caption{Linear Rank~2 Intersection Type System}
    \label{fig:lr2its}
\end{figure}
\end{definition}

\begin{example}
Let us write $\overmultimap{\alpha}$ for the type $(\alpha \multimap \alpha)$. For the $\lambda$-term $(\lambda x.x x)(\lambda y.y)$, the following derivation is obtained:

\scalebox{0.92}{
\begin{mathprooftree}
    \AxiomC{$\inftwo{ [x_1: \; \overmultimap{\alpha} \multimap \overmultimap{\alpha}] }{ x_1 }{ \; \overmultimap{\alpha} \multimap \overmultimap{\alpha} }$}
    
    \AxiomC{$\inftwo{ [x_2:\; \overmultimap{\alpha}] }{ x_2 }{ \; \overmultimap{\alpha} }$}
    
    \BinaryInfC{$\inftwo{ [x_1:\; \overmultimap{\alpha} \multimap \overmultimap{\alpha}, x_2:\; \overmultimap{\alpha}] }{ x_1 x_2 }{ \; \overmultimap{\alpha} }$}
    \UnaryInfC{$\inftwo{ [x:(\overmultimap{\alpha} \multimap \overmultimap{\alpha}) \; \cap \overmultimap{\alpha}] }{ x x }{ \; \overmultimap{\alpha} }$}
    \UnaryInfC{$\inftwo{ [\;] }{ \lambda x.x x }{ (\overmultimap{\alpha} \multimap \overmultimap{\alpha}) \; \cap \overmultimap{\alpha} \rightarrow \overmultimap{\alpha} }$}
    
    \AxiomC{$\inftwo{ [y:\; \overmultimap{\alpha}] }{ y }{ \; \overmultimap{\alpha} }$}
    \UnaryInfC{$\inftwo{ [\;] }{ \lambda y.y }{ \; \overmultimap{\alpha} \multimap \overmultimap{\alpha} }$}
    
    \AxiomC{$\inftwo{ [y:\alpha] }{ y }{ \alpha }$}
    \UnaryInfC{$\inftwo{ [\;] }{ \lambda y.y }{ \; \overmultimap{\alpha} }$}
    
    \TrinaryInfC{$\inftwo{ [\;] }{ (\lambda x.x x)(\lambda y.y) }{ \; \overmultimap{\alpha} }$}
\end{mathprooftree}
}
\end{example}

\subsection{Type Inference Algorithm}\label{sec:inference}

In this section we define a new type inference algorithm for the $\lambda$-calculus (\autoref{ti2}), which is sound (\autoref{soundness}) and complete (\autoref{completeness}) with respect to the Linear Rank~2 Intersection Type System.

Our algorithm is based on Trevor Jim's type inference algorithm \cite{jim1995rank} for a Rank~2 Intersection Type System that was introduced by Daniel Leivant in \cite{leivant1983polymorphic}, where the algorithm was briefly covered. Different versions of the algorithm were later defined by Steffen van Bakel in \cite{van1993intersection} and by Trevor Jim in \cite{jim1995rank}.

Part of the definitions, properties and proofs here presented are also adapted from \cite{jim1995rank}.

\begin{definition}[Type substitution]
Let $\ts = \tsubst{\rz_1/\alpha_1, \dots, \rz_n/\alpha_n}$ denote a \emph{type substitution}, where $\alpha_1,\dots,\alpha_n$ are distinct type variables in $\mathbb{V}$ and $\rz_1,\dots,\rz_n$ are types in $\tl{0}$.

For any $\rz$ in $\tl{0}$, $\ts(\rz) = \tsub{\rz}{\rz_1/\alpha_1, \dots, \rz_n/\alpha_n}$ is the type obtained by simultaneously substituting $\alpha_i$ by $\rz_i$ in $\rz$, with $1 \leq i \leq n$.

The type $\ts(\rz)$ is called an \emph{instance} of the type $\rz$.

The notion of type substitution can be extended to environments in the following way:
\[
    \ts(\Gamma) = [x_1:\ts(\ro_1),\dots,x_n:\ts(\ro_n)] \qquad \text{if } \Gamma = [x_1:\ro_1,\dots,x_n:\ro_n]
\]

The environment $\ts(\Gamma)$ is called an \emph{instance} of the environment $\Gamma$.

If $\ts_1 = \tsubst{\rz_1/\alpha_1, \dots, \rz_n/\alpha_n}$ and $\ts_2 = \tsubst{\rz_1'/\alpha_1', \dots, \rz_n'/\alpha_n'}$ are type substitutions such that the variables $\alpha_1, \dots, \alpha_n, \alpha_1', \dots, \alpha_n'$ are all distinct, then the type substitution $\ts_1 \cup \ts_2$ is defined as $\ts_1 \cup \ts_2 = \tsubst{\rz_1/\alpha_1, \dots, \rz_n/\alpha_n, \rz_1'/\alpha_1', \dots, \rz_n'/\alpha_n'}$.
\end{definition}

Composing two type substitutions $\ts_1$ and $\ts_2$ results in a type substitution $\ts_2 \circ \ts_1$ that when applied, has the same effect as applying $\ts_1$ followed by $\ts_2$.

\begin{definition}[Type substitution composition]
The \emph{composition} of two type substitutions $\ts_1 = \tsubst{\rz_1/\alpha_1, \dots, \rz_n/\alpha_n}$ and $\ts_2 = \tsubst{\rz_1'/\alpha_1', \dots, \rz_m'/\alpha_m'}$, denoted by $\ts_2 \circ \ts_1$, is defined as:
\[
    \ts_2 \circ \ts_1 = \tsubst{\rz_{i_1}'/\alpha_{i_1}', \dots, \rz_{i_k}'/\alpha_{i_k}', \ts_2(\rz_1)/\alpha_1, \dots, \ts_2(\rz_n)/\alpha_n},
\]
where $\{\alpha_{i_1}', \dots, \alpha_{i_k}'\} = \{\alpha_1', \dots, \alpha_m'\} \setminus \{\alpha_1, \dots, \alpha_n\}$.

Also, we consider that the operation is right-associative:
\[
\ts_1 \circ \ts_2 \circ \cdots \circ \ts_{n-1} \circ \ts_n = \ts_1 \circ (\ts_2 \circ \cdots \circ (\ts_{n-1} \circ \ts_n) \dots).
\]
\end{definition}

\subsubsection{Unification}\label{subsec:unification}
We now recall Robinson's unification~\cite{robinson1965machine}, for the special case of equations involving simple types.
For the unification algorithm we follow a latter (more efficient) presentation by Martelli and Montanari~\cite{Martelli1982AnEU}.
\begin{definition} [Unification problem]
A \emph{unification problem} is a finite set of equations $P = \{\rz_1 = \rz_1', \dots, \rz_n=\rz_n'\}$.
A \emph{unifier} (or \emph{solution}) is a substitution $\ts$, such that $\ts(\rz_i) = \ts(\rz_i')$, for $1 \leq i \leq n$.
We call $\ts(\rz_i)$ (or $\ts(\rz_i')$) a \emph{common instance} of $\rz_i$ and $\rz_i'$.
$P$ is \emph{unifiable} if it has at least one unifier. $\mathcal{U}(P)$ is the set of unifiers of $P$.
\end{definition}

\begin{example}
The types $\alpha_1 \multimap \alpha_2 \multimap \alpha_1$ and $(\alpha_3 \multimap \alpha_3) \multimap \alpha_4$ are \emph{unifiable}. For the type substitution $\ts = \tsubst{(\alpha_3 \multimap \alpha_3)/\alpha_1, (\alpha_2 \multimap (\alpha_3 \multimap \alpha_3))/\alpha_4}$, the \emph{common instance} is $(\alpha_3 \multimap \alpha_3) \multimap \alpha_2 \multimap (\alpha_3 \multimap \alpha_3)$.
\end{example}

\begin{definition} [Most general unifier]\label{mgu}
A substitution $\ts$ is a \emph{most general unifier} (MGU) of $P$ if $\ts$ is the least element of $\mathcal{U}(P)$. That is,
\[
    \ts \in \mathcal{U}(P) \text{ and } \forall \ts_1 \in \mathcal{U}(P) \ldotp \exists \ts_2 \ldotp \ts_1 = \ts_2 \circ \ts.
\]
\end{definition}

\begin{example}\label{eg:mgu}
Consider the types $\rz_1 = (\alpha_1 \multimap \alpha_1)$ and $\rz_2 = (\alpha_2 \multimap \alpha_3)$.

The type substitution $\ts' = \tsubst{(\alpha_4 \multimap \alpha_5)/\alpha_1, (\alpha_4 \multimap \alpha_5)/\alpha_2, (\alpha_4 \multimap \alpha_5)/\alpha_3}$ is a unifier of $\rz_1$ and $\rz_2$, but it is not the MGU.

The MGU of $\rz_1$ and $\rz_2$ is $\ts = \tsubst{\alpha_3/\alpha_1, \alpha_3/\alpha_2}$. The common instance of $\rz_1$ and $\rz_2$ by $\ts'$, $(\alpha_4 \multimap \alpha_5) \multimap (\alpha_4 \multimap \alpha_5)$, is an instance of $(\alpha_3 \multimap \alpha_3)$, the common instance by $\ts$.
\end{example}

\begin{definition}[Solved form]
A unification problem $P = \{\alpha_1=\rz_1,\dots,\alpha_n=\rz_n\}$ is in \emph{solved form} if $\alpha_1,\dots,\alpha_n$ are all pairwise distinct variables that do not occur in any of the $\rz_i$. In this case, we define $\ts_P = \tsubst{\rz_1/\alpha_1,\dots,\rz_n/\alpha_n}$.
\end{definition}

\begin{definition}[Type unification]\label{transformuni}
We define the following relation $\Rightarrow$ on type unification problems (for types in $\tl{0}$):
\begin{alignat*}{3}
    &\{\rz = \rz\} \cup P \quad && \Rightarrow \quad P &&\\
    &\{\rz_1 \multimap \rz_2 = \rz_3 \multimap \rz_4\} \cup P \quad && \Rightarrow \quad \{\rz_1 = \rz_3, \rz_2 = \rz_4\} \cup P &&\\
    &\{\rz_1 \multimap \rz_2  = \alpha\} \cup P \quad && \Rightarrow \quad \{\alpha = \rz_1 \multimap \rz_2\} \cup P &&\\
    &\{\alpha = \rz\} \cup P \quad && \Rightarrow \quad \{\alpha = \rz\} \cup \tsub{P}{\rz/\alpha} &&\quad \text{if } \alpha \in  \msf{fv}(P) \setminus  \msf{fv}(\rz)\\
    &\{\alpha = \rz\} \cup P \quad && \Rightarrow \quad \msf{FAIL} &&\quad \text{if } \alpha \in  \msf{fv}(\rz) \text{ and } \alpha \neq \rz
\end{alignat*}
where $\tsub{P}{\rz/\alpha}$ corresponds to the notion of type substitution extended to type unification problems. If $P = \{\rz_1 = \rz_1', \dots, \rz_n=\rz_n'\}$, then $\tsub{P}{\rz/\alpha} = \{\tsub{\rz_1}{\rz/\alpha} = \tsub{\rz_1'}{\rz/\alpha}, \dots, \tsub{\rz_n}{\rz/\alpha} = \tsub{\rz_n'}{\rz/\alpha}\}$.
And $ \msf{fv}(P)$ and $ \msf{fv}(\rz)$ are the sets of free type variables in $P$ and $\rz$, respectively. Since in our system all occurrences of type variables are free, $ \msf{fv}(P)$ and $ \msf{fv}(\rz)$ are the sets of type variables in $P$ and $\rz$, respectively.
\end{definition}

\begin{definition}[Unification algorithm]\label{unify}
Let $P$ be a unification problem (with types in $\tl{0}$). The unification function $\msf{UNIFY}(P)$ that decides whether $P$ has a solution and, if so, returns the MGU of $P$ (see \cite{robinson1965machine}), is defined as:
\begin{algorithmic}
    \Function{$\msf{UNIFY}$}{$P$}
        \While{$P \Rightarrow P'$}
                \State $P \coloneqq P'$;
        \EndWhile
        \If{$P$ is in solved form}
            \State \Return $\ts_P$;
        \Else
            \State $\msf{FAIL}$;
        \EndIf
    \EndFunction
\end{algorithmic}
\end{definition}


\begin{example}\label{eg:unify}
Consider again the types $\alpha_1 \multimap \alpha_1$ and $\alpha_2 \multimap \alpha_3$ in \autoref{eg:mgu}. For the unification problem $P = \{ \alpha_1 \multimap \alpha_1 = \alpha_2 \multimap \alpha_3 \}$, $\msf{UNIFY}(P)$ performs the following transformations over $P$:
\begin{align*}
    \{ \alpha_1 \multimap \alpha_1 = \alpha_2 \multimap \alpha_3 \}
    &\Rightarrow \{ \alpha_1 = \alpha_2, \alpha_1 = \alpha_3 \} \cup \{\;\} &=&& \{ \alpha_1 = \alpha_2, \alpha_1 = \alpha_3 \}\\
    &\Rightarrow \{ \alpha_1 = \alpha_2\} \cup \tsub{\{\alpha_1 = \alpha_3\}}{\alpha_2/\alpha_1} &=&& \{ \alpha_1 = \alpha_2, \alpha_2 = \alpha_3 \}\\
    &\Rightarrow \{ \alpha_2 = \alpha_3\} \cup \tsub{\{\alpha_1 = \alpha_2\}}{\alpha_3/\alpha_2} &=&& \{ \alpha_1 = \alpha_3, \alpha_2 = \alpha_3 \}
\end{align*}

and, since $\{ \alpha_1 = \alpha_3, \alpha_2 = \alpha_3 \}$ is in solved form, it  returns the type substitution $\tsubst{\alpha_3/\alpha_1, \alpha_3/\alpha_2}$.
\end{example}

\subsubsection{Type Inference}\label{subsec:inference}

\begin{definition}[Type inference algorithm]\label{ti2}
Let $\Gamma$ be an environment, $M$ a $\lambda$-term, $\rt$ a linear rank~2 intersection type and $\msf{UNIFY}$ the function in \autoref{unify}. The function $\msf{T}(M) = (\Gamma, \rt)$ defines a type inference algorithm for the $\lambda$-calculus in the Linear Rank~2 Intersection Type System, in the following way:
\begin{enumerate}
    \item \tul{If} $M=x$, \tul{then} $\Gamma = [x:\alpha]$ and $\rt = \alpha$, where $\alpha$ is a new variable;
    
    \item \tul{If} $M=\lambda x.M_1$ and $\msf{T}(M_1) = (\Gamma_1, \rt_1)$ \tul{then}:
    \begin{enumerate}
        \item \tul{if} $x \notin \dom(\Gamma_1)$, \tul{then} $\msf{FAIL}$;
        
        \item \tul{if} $(x:\rz) \in \Gamma_1$, \tul{then} $\msf{T}(M) = ({\Gamma_1}_x, \rz \multimap \rt_1)$;
            
        \item \tul{if} $(x:\rz_1 \cap \cdots \cap \rz_n) \in \Gamma_1$ (with $n \geq 2$), \tul{then} $\msf{T}(M) = ({\Gamma_1}_x, \rz_1 \cap \cdots \cap \rz_n \rightarrow \rt_1)$.
        
    \end{enumerate}
    \renewcommand*\thefootnote{\Alph{footnote}}
    \item \tul{If} $M=M_1 M_2$, \tul{then}:
    \begin{enumerate}
        \item \tul{if} $\msf{T}(M_1) = (\Gamma_1, \alpha_1)$ and $\msf{T}(M_2) = (\Gamma_2, \rz_2)$,\\
        \tul{then} $\msf{T}(M) = (\ts(\Gamma_1 + \Gamma_2), \ts(\alpha_3))$,\\
        where $\ts = \msf{UNIFY}(\{\alpha_1 = \alpha_2 \multimap \alpha_3, \rz_2 = \alpha_2\})$ and $\alpha_2,\alpha_3$ are new variables;
        
        \item \tul{if} $\msf{T}(M_1) = (\Gamma_1', \rz_1' \cap \cdots \cap \rz_n' \rightarrow \rt_1')$ (with $n \geq 2$) and, for each $1 \leq i \leq n$,\\
        $\msf{T}(M_2) = (\Gamma_i, \rz_i)$\footnote{Note that $\Gamma_i, \rz_i$ can all be different up to renaming of variables.}, \\
        \tul{then} $\msf{T}(M) = (\ts(\Gamma_1' + \sum_{i=1}^n \Gamma_i), \ts(\rt_1'))$,\\
        where $\ts = \msf{UNIFY}(\{\rz_i = \rz_i' \mid 1 \leq i \leq n\})$;
        
        \item \tul{if} $\msf{T}(M_1) = (\Gamma_1, \rz \multimap \rt_1)$ and $\msf{T}(M_2) = (\Gamma_2, \rz_2)$,\\
        \tul{then} $\msf{T}(M) = (\ts(\Gamma_1 + \Gamma_2), \ts(\rt_1))$,\\
        where $\ts = \msf{UNIFY}(\{\rz_2 = \rz\})$;
        
        \item \tul{otherwise} $\msf{FAIL}$.\\
    \end{enumerate}
\end{enumerate}
\end{definition}

\begin{example}
Let us show the type inference process for the $\lambda$-term $\lambda x.x x$.

\begin{itemize}
    \item By rule 1., $\msf{T}(x) = ([x:\alpha_1], \alpha_1)$.
    
    \item By rule 1., again, $\msf{T}(x) = ([x:\alpha_2], \alpha_2)$.
    
    \item Then by rule 3.(a), $\msf{T}(x x) = (\ts([x:\alpha_1] + [x:\alpha_2]), \ts(\alpha_4)) = (\ts([x:\alpha_1 \cap \alpha_2]), \ts(\alpha_4))$,
    
    where $\ts = \msf{UNIFY}(\{\alpha_1 = \alpha_3 \multimap \alpha_4, \alpha_2 = \alpha_3\}) = \tsubst{\alpha_3 \multimap \alpha_4/\alpha_1, \alpha_3/\alpha_2}$.
    
    So $\msf{T}(x x) = ([x:(\alpha_3 \multimap \alpha_4) \cap \alpha_3], \alpha_4)$.
    
    \item Finally, by rule 2.(c), $\msf{T}(\lambda x.x x) = ([\;], (\alpha_3 \multimap \alpha_4) \cap \alpha_3 \rightarrow \alpha_4)$.
\end{itemize}
\end{example}

\begin{example}
Let us now show the type inference process for the $\lambda$-term $(\lambda x.x x) (\lambda y.y)$.

\begin{itemize}
    \item From the previous example, we have $\msf{T}(\lambda x.x x) = ([\;], (\alpha_3 \multimap \alpha_4) \cap \alpha_3 \rightarrow \alpha_4)$.
    
    \item By rules 1. and 2.(b), for the identity, the algorithm gives $\msf{T}(\lambda y.y) = ([\;], \alpha_1 \multimap \alpha_1)$.
    
    \item By rules 1. and 2.(b), again, for the identity, $\msf{T}(\lambda y.y) = ([\;], \alpha_2 \multimap \alpha_2)$.
    
    \item Then by rule 3.(b), $\msf{T}((\lambda x.x x) (\lambda y.y)) = (\ts([\;] + [\;] + [\;]), \ts(\alpha_4)) = ([\;], \ts(\alpha_4))$,
    
    where $\ts = \msf{UNIFY}(\{\alpha_1 \multimap \alpha_1 = \alpha_3 \multimap \alpha_4, \alpha_2 \multimap \alpha_2 = \alpha_3\})$, calculated by performing the following transformations:
    \begin{align*}
    \{ \alpha_1 \multimap \alpha_1 = \alpha_3 \multimap \alpha_4, \alpha_2 \multimap \alpha_2 = \alpha_3 \}
    &\Rightarrow \{ \alpha_1 = \alpha_3, \alpha_1 = \alpha_4, \alpha_2 \multimap \alpha_2 = \alpha_3 \}\\
    &\Rightarrow \{ \alpha_1 = \alpha_3, \alpha_3 = \alpha_4, \alpha_2 \multimap \alpha_2 = \alpha_3 \}\\
    &\Rightarrow \{ \alpha_1 = \alpha_4, \alpha_3 = \alpha_4, \alpha_2 \multimap \alpha_2 = \alpha_4 \}\\
    &\Rightarrow \{ \alpha_1 = \alpha_4, \alpha_3 = \alpha_4, \alpha_4 = \alpha_2 \multimap \alpha_2 \}\\
    &\Rightarrow \{ \alpha_1 = \alpha_2 \multimap \alpha_2, \alpha_3 = \alpha_2 \multimap \alpha_2, \alpha_4 = \alpha_2 \multimap \alpha_2 \}
\end{align*}
    
    So $\ts = \tsubst{(\alpha_2 \multimap \alpha_2) / \alpha_1, (\alpha_2 \multimap \alpha_2) / \alpha_3, (\alpha_2 \multimap \alpha_2) / \alpha_4}$
    
    and $\msf{T}((\lambda x.x x) (\lambda y.y)) = ([\;], \alpha_2 \multimap \alpha_2)$.
\end{itemize}
\end{example}

Now we show several properties of our type system and type inference algorithm, in order to prove the soundness and completeness of the algorithm with respect to the system.

\begin{notation}
We write $\Phi \rhd \inftwo{ \Gamma }{ M }{ \rt }$ to denote that $\Phi$ is a derivation tree ending with $\inftwo{ \Gamma }{ M }{ \rt }$. In this case, $|\Phi|$ is the depth of the derivation tree $\Phi$.
\end{notation}

\begin{lemma}[Substitution]\label{lemSubstitutivity} 
If $\Phi \rhd \inftwo{ \Gamma }{ M }{ \rt }$, then $\inftwo{ \ts(\Gamma) }{ M }{ \ts(\rt) }$ for any substitution~$\ts$.

\begin{proof}
By induction on $|\Phi|$.

\begin{enumerate}
    \item \tul{(Axiom)}: Then $\Gamma = [x:\rz]$, $M = x$ and $\rt = \rz$.\\
    
    So $\ts(\Gamma) = [x:\ts(\rz)]$ and $\ts(\rt) = \ts(\rz)$,
    
    and by rule (Axiom) we have $\inftwo{ \ts(\Gamma) }{ x }{ \ts(\rt) }$.\\

    \item \tul{(Exchange)}: Then $\Gamma = (\Gamma_1, y:\ro_2, x:\ro_1, \Gamma_2)$, $M = M_1$, $\rt = \rt_1$, and assuming that the premise $\inftwo{ \Gamma_1, x:\ro_1, y:\ro_2, \Gamma_2 }{ M_1 }{ \rt_1 }$ holds.\\
    
    By the induction hypothesis, for any substitution $\ts$, $\inftwo{ \ts(\Gamma_1, x:\ro_1, y:\ro_2, \Gamma_2) }{ M_1 }{ \ts(\rt_1) }$,
    
    which is the same as $\inftwo{ \ts(\Gamma_1), x:\ts(\ro_1), y:\ts(\ro_2), \ts(\Gamma_2) }{ M_1 }{ \ts(\rt_1) }$.\\
    
    By rule (Exchange) we get $\inftwo{ \ts(\Gamma_1), y:\ts(\ro_2), x:\ts(\ro_1), \ts(\Gamma_2) }{ M_1 }{ \ts(\rt_1) }$,
    
    which is the same as $\inftwo{ \ts(\Gamma_1, y:\ro_2, x:\ro_1, \Gamma_2) }{ M_1 }{ \ts(\rt_1) }$, i.e., $\inftwo{ \ts(\Gamma) }{ M }{ \ts(\rt) }$.\\

    \item \tul{(Contraction)}: Then $\Gamma = (\Gamma_1, x:\ro_1 \cap \ro_2, \Gamma_2)$, $M = \sub{M_1}{x/x_1, x/x_2}$, $\rt = \rt_1$, and assuming that the premise $\inftwo{ \Gamma_1, x_1:\ro_1, x_2:\ro_2, \Gamma_2 }{ M_1 }{ \rt_1 }$ holds.\\
    
    By the induction hypothesis, for any substitution $\ts$, $\inftwo{ \ts(\Gamma_1, x_1:\ro_1, x_2:\ro_2, \Gamma_2) }{ M_1 }{ \ts(\rt_1) }$,
    
    which is the same as $\inftwo{ \ts(\Gamma_1), x_1:\ts(\ro_1), x_2:\ts(\ro_2), \ts(\Gamma_2) }{ M_1 }{ \ts(\rt_1) }$.\\
    
    By rule (Contraction) we get $\inftwo{ \ts(\Gamma_1), x:\ts(\ro_1) \cap \ts(\ro_2), \ts(\Gamma_2) }{ \sub{M_1}{x/x_1, x/x_2} }{ \ts(\rt_1) }$,
    
    which is the same as $\inftwo{ \ts(\Gamma_1, x:\ro_1 \cap \ro_2, \Gamma_2) }{ \sub{M_1}{x/x_1, x/x_2} }{ \ts(\rt_1) }$, i.e., $\inftwo{ \ts(\Gamma) }{ M }{ \ts(\rt) }$.\\

    \item \tul{($\rightarrow$ Intro)}: Then $\Gamma = \Gamma_1$, $M = \lambda x.M_1$, $\rt = \rz_1 \cap \cdots \cap \rz_n \rightarrow \rt_1 $, and assuming that the premise $\inftwo{ \Gamma_1, x:\rz_1 \cap \cdots \cap \rz_n }{ M_1 }{ \rt_1 }$ (with $n \geq 2$) holds.\\
    
    By the induction hypothesis, for any substitution $\ts$, $\inftwo{ \ts(\Gamma_1, x:\rz_1 \cap \cdots \cap \rz_n) }{ M_1 }{ \ts(\rt_1) }$,
    
    which is the same as $\inftwo{ \ts(\Gamma_1), x:\ts(\rz_1) \cap \cdots \cap \ts(\rz_n) }{ M_1 }{ \ts(\rt_1) }$.\\
    
    By rule ($\rightarrow$ Intro) we get $\inftwo{ \ts(\Gamma_1) }{ \lambda x.M_1 }{ \ts(\rz_1) \cap \cdots \cap \ts(\rz_n) \rightarrow \ts(\rt_1) }$,
    
    which is the same as $\inftwo{ \ts(\Gamma_1) }{ \lambda x.M_1 }{ \ts(\rz_1 \cap \cdots \cap \rz_n \rightarrow \rt_1) }$, i.e., $\inftwo{ \ts(\Gamma) }{ M }{ \ts(\rt) }$.\\

    \item \tul{($\rightarrow$ Elim)}: Then $\Gamma = (\Gamma_0, \sum_{i=1}^n \Gamma_i)$, $M = M_1 M_2$, $\rt = \rt_1$, and assuming that the premises $\inftwo{ \Gamma_0 }{ M_1 }{ \rz_1 \cap \cdots \cap \rz_n \rightarrow \rt_1 }$ and $\inftwo{ \Gamma_i }{ M_2 }{ \rz_i }$, for $1 \leq i \leq n$ (with $n \geq 2$), hold.\\
    
    By the induction hypothesis, for any substitution $\ts$:
    \begin{itemize}
        \item $\inftwo{ \ts(\Gamma_0) }{ M_1 }{ \ts(\rz_1 \cap \cdots \cap \rz_n \rightarrow \rt_1) }$, 
        
        which is the same as $\inftwo{ \ts(\Gamma_0) }{ M_1 }{ \ts(\rz_1) \cap \cdots \cap \ts(\rz_n) \rightarrow \ts(\rt_1) }$;
    
        \item $\inftwo{ \ts(\Gamma_i) }{ M_2 }{ \ts(\rz_i) }$, for $1 \leq i \leq n$.
    \end{itemize}
    
    By rule ($\rightarrow$ Elim) we get $\inftwo{ \ts(\Gamma_0), \sum_{i=1}^n \ts(\Gamma_i) }{ M_1 M_2 }{ \ts(\rt_1) }$,
    
    which is the same as $\inftwo{ \ts(\Gamma_0, \sum_{i=1}^n \Gamma_i) }{ M_1 M_2 }{ \ts(\rt_1) }$, i.e., $\inftwo{ \ts(\Gamma) }{ M }{ \ts(\rt) }$.\\

    \item \tul{($\multimap$ Intro)}: Then $\Gamma = \Gamma_1$, $M = \lambda x.M_1$, $\rt = \rz \multimap \rt_1 $, and assuming that the premise $\inftwo{ \Gamma_1, x:\rz }{ M_1 }{ \rt_1 }$ holds.\\
    
    By the induction hypothesis, for any substitution $\ts$, $\inftwo{ \ts(\Gamma_1, x:\rz) }{ M_1 }{ \ts(\rt_1) }$,
    
    which is the same as $\inftwo{ \ts(\Gamma_1), x:\ts(\rz) }{ M_1 }{ \ts(\rt_1) }$.\\
    
    By rule ($\multimap$ Intro) we get $\inftwo{ \ts(\Gamma_1) }{ \lambda x.M_1 }{ \ts(\rz) \multimap \ts(\rt_1) }$,
    
    which is the same as $\inftwo{ \ts(\Gamma_1) }{ \lambda x.M_1 }{ \ts(\rz \multimap \rt_1) }$, i.e., $\inftwo{ \ts(\Gamma) }{ M }{ \ts(\rt) }$.\\

    \item \tul{($\multimap$ Elim)}: Then $\Gamma = (\Gamma_1, \Gamma_2)$, $M = M_1 M_2$, $\rt = \rt_1$, and assuming that the premises $\inftwo{ \Gamma_1 }{ M_1 }{ \rz \multimap \rt_1 }$ and $\inftwo{ \Gamma_2 }{ M_2 }{ \rz }$ hold.\\
    
    By the induction hypothesis, for any substitution $\ts$:
    \begin{itemize}
        \item $\inftwo{ \ts(\Gamma_1) }{ M_1 }{ \ts(\rz \multimap \rt_1) }$, which is the same as $\inftwo{ \ts(\Gamma_1) }{ M_1 }{ \ts(\rz) \multimap \ts(\rt_1) }$;
    
        \item $\inftwo{ \ts(\Gamma_2) }{ M_2 }{ \ts(\rz) }$.
    \end{itemize}
    
    By rule ($\multimap$ Elim) we get $\inftwo{ \ts(\Gamma_1), \ts(\Gamma_2) }{ M_1 M_2 }{ \ts(\rt_1) }$,
    
    which is the same as $\inftwo{ \ts(\Gamma_1, \Gamma_2) }{ M_1 M_2 }{ \ts(\rt_1) }$, i.e., $\inftwo{ \ts(\Gamma) }{ M }{ \ts(\rt) }$.
\end{enumerate}
\end{proof}
\end{lemma}

\begin{lemma}[Relevance]\label{xfreesys}
If $\Phi \rhd \inftwo{ \Gamma }{ M }{ \rt }$, then $x \in \dom(\Gamma)$ if and only if $x \in \fv{M}$.

\begin{proof}
Easy induction on $|\Phi|$.
\end{proof}
\end{lemma}

\begin{lemma}\label{xfreealg} 
If $\msf{T}(M) = (\Gamma, \rt)$, then $x \in \dom(\Gamma)$ if and only if $x \in \fv{M}$.

\begin{proof}
Easy induction on the definition of $\msf{T}(M)$.
\end{proof}
\end{lemma}

\begin{corollary}\label{equaldom}
From \autoref{xfreesys} and \autoref{xfreealg}, it follows that if $\msf{T}(M) = (\Gamma, \rt)$ and $\inftwo{ \Gamma' }{ M }{ \rt' }$, then $\dom(\Gamma) = \dom(\Gamma')$.
\end{corollary}

\begin{lemma}\label{eqsubstsystem}
If $\Phi_1 \rhd \inftwo{ \Gamma }{ M }{ \rt }$, $x \in \fv{M}$ and $y$ does not occur in $M$, then there exists $\Phi_2 \rhd \inftwo{ \sub{\Gamma}{y/x} }{ \sub{M}{y/x} }{ \rt }$, with $|\Phi_1| = |\Phi_2|$.

\begin{proof}
By induction on $|\Phi_1|$.

(We will only prove the first part of the lemma, since the second ($|\Phi_1| = |\Phi_2|$) can be shown with a trivial induction proof.)\\

Let $x$ be a variable that occurs free in $M$ and $y$ a new variable not occurring in $M$.

\begin{enumerate}
    \item \tul{(Axiom)}: Then $\Gamma = [x_1:\rz]$, $M = x_1$, $\rt = \rz$ and $x = x_1$.\\
    
    By rule (Axiom) we have $\inftwo{ [y:\rz] }{ y }{ \rz }$,
    
    which is the same as $\inftwo{ \sub{\Gamma}{y/x} }{ \sub{M}{y/x} }{ \rt }$.\\

    \item \tul{(Exchange)}: Then $\Gamma = (\Gamma_1, y_1:\ro_2, x_1:\ro_1, \Gamma_2)$, $M = M_1$, $\rt = \rt_1$, and assuming that the premise $\inftwo{ \Gamma_1, x_1:\ro_1, y_1:\ro_2, \Gamma_2 }{ M_1 }{ \rt_1 }$ holds.\\
    
    Since $x \in \fv{M_1}$ and $y$ does not occur in $M_1$,
    
    by induction, $\inftwo{ \sub{(\Gamma_1, x_1:\ro_1, y_1:\ro_2, \Gamma_2)}{y/x} }{ \sub{M_1}{y/x} }{ \rt_1 }$,
    
    which is the same as $\inftwo{ (\sub{\Gamma_1}{y/x}), \sub{x_1}{y/x}:\ro_1, \sub{y_1}{y/x}:\ro_2, (\sub{\Gamma_2}{y/x}) }{ \sub{M_1}{y/x} }{ \rt_1 }$.\\
    
    Then by rule (Exchange), $\inftwo{ (\sub{\Gamma_1}{y/x}), \sub{y_1}{y/x}:\ro_2, \sub{x_1}{y/x}:\ro_1, (\sub{\Gamma_2}{y/x}) }{ \sub{M_1}{y/x} }{ \rt_1 }$,
    
    which is the same as $\inftwo{ \sub{\Gamma}{y/x} }{ \sub{M}{y/x} }{ \rt }$.\\

    \item \tul{(Contraction)}: Then $\Gamma = (\Gamma_1, x':\ro_1 \cap \ro_2, \Gamma_2)$, $M = \sub{M_1}{x'/x_1, x'/x_2}$, $\rt = \rt_1$, and assuming that the premise $\inftwo{ \Gamma_1, x_1:\ro_1, x_2:\ro_2, \Gamma_2 }{ M_1 }{ \rt_1 }$ holds.\\
    
    There are two possible cases regarding $x$:
    
    \begin{enumerate}
        \item $x = x'$:\\
        
        Since $y$ does not occur in $M$, $y \notin \fv{M}$, so by $\autoref{xfreesys}$, $y \notin \dom(\Gamma)$. So $y \notin \dom(\Gamma_1)$ and $y \notin \dom(\Gamma_2)$.\\
        
        Then we can apply the rule (Contraction) to $\inftwo{ \Gamma_1, x_1:\ro_1, x_2:\ro_2, \Gamma_2 }{ M_1 }{ \rt_1 }$
        
        and get $\inftwo{ \Gamma_1, y:\ro_1 \cap \ro_2, \Gamma_2 }{ \sub{M_1}{y/x_1, y/x_2} }{ \rt_1 }$,
            
        which is equivalent to $\inftwo{ \sub{(\Gamma_1, x':\ro_1 \cap \ro_2, \Gamma_2)}{y/x'} }{ \sub{(\sub{M_1}{x'/x_1, x'/x_2})}{y/x'} }{ \rt_1 }$
            
        and the same as $\inftwo{ \sub{(\Gamma_1, x':\ro_1 \cap \ro_2, \Gamma_2)}{y/x} }{ \sub{(\sub{M_1}{x'/x_1, x'/x_2})}{y/x} }{ \rt_1 }$.\\
    
        So $\inftwo{ \sub{\Gamma}{y/x} }{ \sub{M}{y/x} }{ \rt }$.\\

        \item $x \neq x'$ (and so $x \in \fv{M_1}$):\\
        
        There are three possible cases regarding $y$:
        
        \begin{enumerate}
            \item $y \neq x_1$ and $y \neq x_2$:\\
            
            Since $x \in \fv{M_1}$ and $y$ does not occur in $M_1$,
            
            by induction, $\inftwo{ \sub{(\Gamma_1, x_1:\ro_1, x_2:\ro_2, \Gamma_2)}{y/x} }{ \sub{M_1}{y/x} }{ \rt_1 }$,
            
            which is the same as $\inftwo{ (\sub{\Gamma_1}{y/x}), \sub{x_1}{y/x}:\ro_1, \sub{x_2}{y/x}:\ro_2, (\sub{\Gamma_2}{y/x}) }{ \sub{M_1}{y/x} }{ \rt_1 }$.\\
    
            Then by rule (Contraction),
            
            $\inftwo{ (\sub{\Gamma_1}{y/x}), x':\ro_1 \cap \ro_2, (\sub{\Gamma_2}{y/x}) }{ \sub{(\sub{M_1}{y/x})}{x'/(\sub{x_1}{y/x}), x'/(\sub{x_2}{y/x})} }{ \rt_1 }$.\\
    
            Since $x \neq x'$, $x' = \sub{x'}{y/x}$.
            
            So $(\sub{\Gamma_1}{y/x}), x':\ro_1 \cap \ro_2, (\sub{\Gamma_2}{y/x}) = \sub{(\Gamma_1, x':\ro_1 \cap \ro_2, \Gamma_2)}{y/x} = \sub{\Gamma}{y/x}$.\\
    
            And $\sub{x_1}{y/x} = x_1$, $\sub{x_2}{y/x} = x_2$ because $x \neq x_1, x \neq x_2$ (otherwise it would contradict the assumption that $x \in \fv{M}$),
            
            so $\sub{(\sub{M_1}{y/x})}{x'/(\sub{x_1}{y/x}), x'/(\sub{x_2}{y/x})} = \sub{(\sub{M_1}{y/x})}{x'/x_1, x'/x_2}$.\\
            
            And since $x \neq x_1$, $x \neq x_2$, $y \neq x_1$, $y \neq x_2$ and $x \neq x'$,
            
            then $\sub{(\sub{M_1}{y/x})}{x'/x_1, x'/x_2} = \sub{(\sub{M_1}{x'/x_1, x'/x_2})}{y/x} = \sub{M}{y/x}$.\\
    
            So $\inftwo{ \sub{\Gamma}{y/x} }{ \sub{M}{y/x} }{ \rt }$.\\
            
            \item $y = x_1$:\\
            
            So the premise can be written as $\inftwo{ \Gamma_1, y:\ro_1, x_2:\ro_2, \Gamma_2 }{ M_1 }{ \rt_1 }$.
            
            Let $y'$ be a fresh variable not occurring in any of the terms and environments mentioned.\\
            
            Then by induction, we have $\inftwo{ \sub{(\Gamma_1, y:\ro_1, x_2:\ro_2, \Gamma_2)}{y'/y} }{ \sub{M_1}{y'/y} }{ \rt_1 }$,
            
            which is the same as $\inftwo{ \Gamma_1, y':\ro_1, x_2:\ro_2, \Gamma_2 }{ \sub{M_1}{y'/y} }{ \rt_1 }$.\\
            
            As $x \in \fv{M}$, by \autoref{xfreesys}, $x \in \dom(\Gamma)$.
            
            And since $x \neq x'$, then either $x \in \dom(\Gamma_1)$ or $x \in \dom(\Gamma_2)$.
            
            This means that $x \in \dom(\Gamma_1, y':\ro_1, x_2:\ro_2, \Gamma_2)$, and so by $\autoref{xfreesys}$, $x \in \fv{\sub{M_1}{y'/y}}$.\\
            
            And $y$ does not occur in $\sub{M_1}{y'/y}$.\\
            
            So we can then apply the induction hypothesis to the derivation ending with $\inftwo{ \Gamma_1, y':\ro_1, x_2:\ro_2, \Gamma_2 }{ M_1 [y'/y] }{ \rt_1 }$
            
            and get $\inftwo{ (\Gamma_1, y':\ro_1, x_2:\ro_2, \Gamma_2) [y/x] }{ (M_1 [y'/y]) [y/x] }{ \rt_1 }$,
            
            which is equivalent to $\inftwo{ \Gamma_1[y/x], y':\ro_1, x_2:\ro_2, \Gamma_2[y/x] }{ (M_1 [y'/y]) [y/x] }{ \rt_1 }$.\\
            
            Then by rule (Contraction),
            
            $\inftwo{ \Gamma_1[y/x], x':\ro_1 \cap \ro_2, \Gamma_2[y/x] }{ ((M_1 [y'/y]) [y/x]) [x'/y', x'/x_2] }{ \rt_1 }$,
            
            which is the same as
            
            $\inftwo{ (\Gamma_1, x':\ro_1 \cap \ro_2, \Gamma_2)[y/x] }{ ((M_1 [y'/x_1]) [y/x]) [x'/y', x'/x_2] }{ \rt_1 }$.\\
            
            This is equivalent to
            
            $\inftwo{ (\Gamma_1, x':\ro_1 \cap \ro_2, \Gamma_2)[y/x] }{ ((M_1 [y'/x_1]) [x'/y', x'/x_2]) [y/x] }{ \rt_1 }$,
            
            which is the equivalent to
            
            $\inftwo{ (\Gamma_1, x':\ro_1 \cap \ro_2, \Gamma_2)[y/x] }{ ((M_1 [x'/x_1, x'/x_2]) [y/x] }{ \rt_1 }$.\\
            
            So $\inftwo{ \Gamma[y/x] }{ M[y/x] }{ \rt }$.\\
            
            \item $y = x_2$:\\
            
            Analogous to the case where $y = x_1$.\\
        \end{enumerate}
    \end{enumerate}

    \item \tul{($\rightarrow$ Intro)}: Then $\Gamma = \Gamma_1$, $M = \lambda x_1.M_1$, $\rt = \rz_1 \cap \cdots \cap \rz_n \rightarrow \rt_1 $, and assuming that the premise $\inftwo{ \Gamma_1, x_1:\rz_1 \cap \cdots \cap \rz_n }{ M_1 }{ \rt_1 }$ (with $n \geq 2$) holds.\\
    
    Since $x \in \fv{M_1}$ and $y$ does not occur in $M_1$,
    
    by induction, $\inftwo{ (\Gamma_1, x_1:\rz_1 \cap \cdots \cap \rz_n)[y/x] }{ M_1[y/x] }{ \rt_1 }$,
    
    which is the same as $\inftwo{ (\Gamma_1[y/x]), x_1[y/x]:\rz_1 \cap \cdots \cap \rz_n }{ M_1[y/x] }{ \rt_1 }$.\\
    
    Then by rule ($\rightarrow$ Intro), $\inftwo{ \Gamma_1[y/x] }{ \lambda (x_1[y/x]).M_1[y/x] }{ \rz_1 \cap \cdots \cap \rz_n \rightarrow \rt_1 }$.\\
    
    Since $x \neq x_1$ (otherwise it would contradict the assumption that $x \in \fv{M}$), $x_1[y/x] = x_1$ and $\lambda x_1.M_1[y/x] = (\lambda x_1.M_1)[y/x]$.\\
    
    So we have $\inftwo{ \Gamma_1[y/x] }{ (\lambda x_1.M_1)[y/x] }{ \rz_1 \cap \cdots \cap \rz_n \rightarrow \rt_1 }$,
    
    which is the same as $\inftwo{ \Gamma[y/x] }{ M[y/x] }{ \rt }$.\\

    \item \tul{($\rightarrow$ Elim)}: Then $\Gamma = (\Gamma', \sum_{i=1}^n \Gamma_i)$, $M = M_1 M_2$, $\rt = \rt_1$, and assuming that the premises $\inftwo{ \Gamma' }{ M_1 }{ \rz_1 \cap \cdots \cap \rz_n \rightarrow \rt_1 }$ and $\inftwo{ \Gamma_i }{ M_2 }{ \rz_i }$, for $1 \leq i \leq n$ (with $n \geq 2$), hold.\\
    
    Since $x \in \fv{M}$ and $y$ does not occur in $M$, then $y$ does not occur in $M_1$ nor in $M_2$ and there are three possible cases regarding $x$:
        
        \begin{enumerate}
            \item $x \in \fv{M_1}$ and $x \in \fv{M_2}$:\\
            
            Then by induction, $\inftwo{ \Gamma'[y/x] }{ M_1[y/x] }{ \rz_1 \cap \cdots \cap \rz_n \rightarrow \rt_1 }$ and, for $1 \leq i \leq n$, $\inftwo{ \Gamma_i[y/x] }{ M_2[y/x] }{ \rz_i }$.\\
            
            So by rule ($\rightarrow$ Elim), $\inftwo{ \Gamma'[y/x], \sum_{i=1}^n \Gamma_i[y/x] }{ (M_1[y/x]) (M_2[y/x]) }{ \rt_1 }$,
            
            which is equivalent to $\inftwo{ \Gamma[y/x] }{ M[y/x] }{ \rt }$.\\

            \item $x \in \fv{M_1}$ and $x \notin \fv{M_2}$:\\
            
            Then $M_2[y/x] = M_2$ and $\Gamma_i[y/x] = \Gamma_i$, for $1 \leq i \leq n$.
            
            So $\inftwo{ \Gamma_i[y/x] }{ M_2[y/x] }{ \rz_i }$ is equivalent to $\inftwo{ \Gamma_i }{ M_2 }{ \rz_i }$, for $1 \leq i \leq n$.\\
            
            By induction, $\inftwo{ \Gamma'[y/x] }{ M_1[y/x] }{ \rz_1 \cap \cdots \cap \rz_n \rightarrow \rt_1 }$.\\
            
            So by rule ($\rightarrow$ Elim), $\inftwo{ \Gamma'[y/x], \sum_{i=1}^n \Gamma_i[y/x] }{ (M_1[y/x]) (M_2[y/x]) }{ \rt_1 }$,
            
            which is equivalent to $\inftwo{ \Gamma[y/x] }{ M[y/x] }{ \rt }$.\\

            \item $x \notin \fv{M_1}$ and $x \in \fv{M_2}$:\\
            
            Then $M_1[y/x] = M_1$ and $\Gamma'[y/x] = \Gamma'$.
            
            So $\inftwo{ \Gamma'[y/x] }{ M_1[y/x] }{ \rz_1 \cap \cdots \cap \rz_n \rightarrow \rt_1 }$ is equivalent to $\inftwo{ \Gamma' }{ M_1 }{ \rz_1 \cap \cdots \cap \rz_n \rightarrow \rt_1 }$.\\
            
            By induction, $\inftwo{ \Gamma_i[y/x] }{ M_2[y/x] }{ \rz_i }$, for $1 \leq i \leq n$.\\
            
            So by rule ($\rightarrow$ Elim), $\inftwo{ \Gamma'[y/x], \sum_{i=1}^n \Gamma_i[y/x] }{ (M_1[y/x]) (M_2[y/x]) }{ \rt_1 }$,
            
            which is equivalent to $\inftwo{ \Gamma[y/x] }{ M[y/x] }{ \rt }$.\\
        \end{enumerate}

    \item \tul{($\multimap$ Intro)}: Then $\Gamma = \Gamma_1$, $M = \lambda x_1.M_1$, $\rt = \rz \multimap \rt_1 $, and assuming that the premise $\inftwo{ \Gamma_1, x_1:\rz }{ M_1 }{ \rt_1 }$ holds.\\
    
    Since $x \in \fv{M_1}$ and $y$ does not occur in $M_1$,
    
    by induction, $\inftwo{ (\Gamma_1, x_1:\rz)[y/x] }{ M_1[y/x] }{ \rt_1 }$,
    
    which is the same as $\inftwo{ (\Gamma_1[y/x]), x_1[y/x]:\rz }{ M_1[y/x] }{ \rt_1 }$.\\
    
    Then by rule ($\multimap$ Intro), $\inftwo{ \Gamma_1[y/x] }{ \lambda (x_1[y/x]).M_1[y/x] }{ \rz \multimap \rt_1 }$.\\
    
    Since $x \neq x_1$ (otherwise it would contradict the assumption that $x \in \fv{M}$), $x_1[y/x] = x_1$ and $\lambda x_1.M_1[y/x] = (\lambda x_1.M_1)[y/x]$.\\
    
    So we have $\inftwo{ \Gamma_1[y/x] }{ (\lambda x_1.M_1)[y/x] }{ \rz \multimap \rt_1 }$, which is the same as $\inftwo{ \Gamma[y/x] }{ M[y/x] }{ \rt }$.\\

    \item \tul{($\multimap$ Elim)}: Then $\Gamma = (\Gamma_1, \Gamma_2)$, $M = M_1 M_2$, $\rt = \rt_1$, and assuming that the premises $\inftwo{ \Gamma_1 }{ M_1 }{ \rz \multimap \rt_1 }$ and $\inftwo{ \Gamma_2 }{ M_2 }{ \rz }$ hold.\\
    
    Since $x \in \fv{M}$ and $y$ does not occur in $M$, then $y$ does not occur in $M_1$ nor in $M_2$ and there are three possible cases regarding $x$:
        
        \begin{enumerate}
            \item $x \in \fv{M_1}$ and $x \in \fv{M_2}$:\\
            
            Then by induction, $\inftwo{ \Gamma_1[y/x] }{ M_1[y/x] }{ \rz \multimap \rt_1 }$ and $\inftwo{ \Gamma_2[y/x] }{ M_2[y/x] }{ \rz }$.\\
            
            So by rule ($\multimap$ Elim), $\inftwo{ \Gamma_1[y/x], \Gamma_2[y/x] }{ (M_1[y/x]) (M_2[y/x]) }{ \rt_1 }$,
            
            which is equivalent to $\inftwo{ \Gamma[y/x] }{ M[y/x] }{ \rt }$.\\

            \item $x \in \fv{M_1}$ and $x \notin \fv{M_2}$:\\
            
            Then $M_2[y/x] = M_2$ and $\Gamma_2[y/x] = \Gamma_2$.
            
            So $\inftwo{ \Gamma_2[y/x] }{ M_2[y/x] }{ \rz }$ is equivalent to $\inftwo{ \Gamma_2 }{ M_2 }{ \rz }$.\\
            
            By induction, $\inftwo{ \Gamma_1[y/x] }{ M_1[y/x] }{ \rz \multimap \rt_1 }$.\\
            
            So by rule ($\multimap$ Elim), $\inftwo{ \Gamma_1[y/x], \Gamma_2[y/x] }{ (M_1[y/x]) (M_2[y/x]) }{ \rt_1 }$,
            
            which is equivalent to $\inftwo{ \Gamma[y/x] }{ M[y/x] }{ \rt }$.\\

            \item $x \notin \fv{M_1}$ and $x \in \fv{M_2}$:\\
            
            Then $M_1[y/x] = M_1$ and $\Gamma_1[y/x] = \Gamma_1$.
            
            So $\inftwo{ \Gamma_1[y/x] }{ M_1[y/x] }{ \rz \multimap \rt_1 }$ is equivalent to $\inftwo{ \Gamma_1 }{ M_1 }{ \rz \multimap \rt_1 }$.\\
            
            By induction, $\inftwo{ \Gamma_2[y/x] }{ M_2[y/x] }{ \rz }$.\\
            
            So by rule ($\multimap$ Elim), $\inftwo{ \Gamma_1[y/x], \Gamma_2[y/x] }{ (M_1[y/x]) (M_2[y/x]) }{ \rt_1 }$,
            
            which is equivalent to $\inftwo{ \Gamma[y/x] }{ M[y/x] }{ \rt }$.
        \end{enumerate}
\end{enumerate}
\end{proof}
\end{lemma}

\begin{corollary}\label{eqsubstsystemcol}
From \autoref{eqsubstsystem}, it follows that if $\inftwo{ \Gamma }{ M }{ \rt }$, $\{x_1, \dots, x_n\} \subseteq \fv{M}$ and $y_1, \dots, y_n$ are all different variables not occurring in $M$, then $\inftwo{ \Gamma[y_1/x_1, \dots, y_n/x_n] }{ M[y_1/x_1, \dots, y_n/x_n] }{ \rt }$.
\end{corollary}

\begin{theorem}[Soundness]\label{soundness} 
If $\msf{T}(M) = (\Gamma, \rt)$, then $\inftwo{ \Gamma }{ M }{ \rt }$.

\begin{proof}
By induction on the definition of $\msf{T}(M)$.

\begin{enumerate}
    \item If $M=x$, then $(\Gamma, \rt) = ([x:\alpha], \alpha)$, and we have $\inftwo{ \Gamma }{ x }{ \rt }$ by rule (Axiom).\\
    
    \item If $M=\lambda x.M_1$, we have the following cases:
    
    \begin{enumerate}
        \item $x \in \fv{M_1}$ and $(\Gamma, \rt) = ({\Gamma_1}_x, {\Gamma_1} (x) \multimap \rt_1)$, where $\msf{T}(M_1) = (\Gamma_1, \rt_1)$ and $\Gamma_1 (x) = \rz \in \tl{0}$.\\
        
        By induction, $\inftwo{ \Gamma_1 }{ M_1 }{ \rt_1 }$, and by \autoref{xfreealg}, $x \in \dom(\Gamma_1)$.\\
        
        So by applying the rule (Exchange) zero or more times successively, we obtain $\inftwo{ {\Gamma_1}_x, x:\rz }{ M_1 }{ \rt_1 }$.\\
        
        So $\inftwo{ \Gamma }{ \lambda x.M_1 }{ \rt }$ by rule ($\multimap$ Intro).\\

        \item $x \in \fv{M_1}$ and $(\Gamma, \rt) = ({\Gamma_1}_x, {\Gamma_1} (x) \rightarrow \rt_1)$, where $\msf{T}(M_1) = (\Gamma_1, \rt_1)$ and $\Gamma_1 (x) = \rz_1 \cap \cdots \cap \rz_n$, with $n \geq 2$.\\
        
        By induction, $\inftwo{ \Gamma_1 }{ M_1 }{ \rt_1 }$, and by \autoref{xfreealg}, $x \in \dom(\Gamma_1)$.\\
        
        So by applying the rule (Exchange) zero or more times successively, we obtain $\inftwo{ {\Gamma_1}_x, x:\rz_1 \cap \cdots \cap \rz_n }{ M_1 }{ \rt_1 }$.\\
        
        So $\inftwo{ \Gamma }{ \lambda x.M_1 }{ \rt }$ by rule ($\rightarrow$ Intro).\\
    \end{enumerate}
    
    \item If $M = M_1 M_2$, we have the following cases:
    
    \begin{enumerate}
        \item $(\Gamma, \rt) = (\ts(\Gamma_1 + \Gamma_2), \ts(\alpha_3))$, where $\msf{T}(M_1) = (\Gamma_1, \alpha_1)$, $\msf{T}(M_2) = (\Gamma_2, \rz_2)$, $\ts = \msf{UNIFY}(\{\alpha_1 = \alpha_2 \multimap \alpha_3, \rz_2 = \alpha_2\})$ and $\alpha_2, \alpha_3$ do not occur in $\Gamma_1, \Gamma_2, \alpha_1, \rz_2$.\\
        
        By induction, $\inftwo{ \Gamma_1 }{ M_1 }{ \alpha_1 }$ and $\inftwo{ \Gamma_2 }{ M_2 }{ \rz_2 }$.\\
        
        Let $\s_1 = [y_1/x_1, \dots, y_n/x_n]$ and $\s_2 = [z_1/x_1, \dots, z_n/x_n]$, where $\dom(\Gamma_1) \cap \dom(\Gamma_2) = \{x_1, \dots, x_n\}$ (which by \autoref{xfreesys}, occur free in $M_1$ and $M_2$) and $y_1, \dots, y_n, z_1, \dots, z_n$ are all distinct fresh term variables, not occurring in $M_1$ nor in $M_2$ (and consequently, by \autoref{xfreesys}, not occurring in $\Gamma_1$ nor in $\Gamma_2$).\\
        
        By \autoref{eqsubstsystemcol}, $\inftwo{ \s_1(\Gamma_1) }{ \s_1(M_1) }{ \alpha_1 }$ and $\inftwo{ \s_2(\Gamma_2) }{ \s_2(M_2) }{ \rz_2 }$.\\
        
        By \autoref{lemSubstitutivity}, $\inftwo{ \ts (\s_1(\Gamma_1)) }{ \s_1(M_1) }{ \ts (\alpha_1) }$ and $\inftwo{ \ts (\s_2(\Gamma_2)) }{ \s_2(M_2) }{ \ts (\rz_2) }$.\\
        
        Since $\ts(\rz_2) = \ts(\alpha_2)$, $\ts(\alpha_1) = \ts(\alpha_2) \multimap \ts(\alpha_3)$ and $(\s_1(\Gamma_1), \s_2(\Gamma_2))$ is consistent,
        
        by rule ($\multimap$ Elim) we have $\inftwo{ (\ts (\s_1(\Gamma_1)), \ts (\s_2(\Gamma_2))) }{ (\s_1(M_1)) (\s_2(M_2)) }{ \ts(\alpha_3) }$,
        
        which is the same as $\inftwo{ \ts(\s_1(\Gamma_1), \s_2(\Gamma_2)) }{ (\s_1(M_1)) (\s_2(M_2)) }{ \ts(\alpha_3) }$.\\
        
        For each pair $(y_i:\ro_i, z_i:\ro_i')$ (for $1 \leq i \leq n$) in the environment $\ts(\s_1(\Gamma_1), \s_2(\Gamma_2))$ in the previous derivation, let us apply the rule (Contraction) to obtain the environment with $x_i:\ro_i \cap \ro_i'$ instead (and applying the rule (Exchange) as necessary).\\
        
        After these applications of the rules (Contraction) and (Exchange) (and consequent applications of (Exchange), if necessary), and by looking at the definition of $(+)$, we end up with $\inftwo{ \ts(\Gamma_1 + \Gamma_2) }{ M_1 M_2 }{ \ts(\alpha_3) }$.\\

        \item $(\Gamma, \rt) = (\ts(\Gamma' + \sum_{i=1}^n \Gamma_i), \ts(\rt_1'))$, where $\msf{T}(M_1) = (\Gamma', \rz_1' \cap \cdots \cap \rz_n' \rightarrow \rt_1')$, with $n \geq 2$, $\msf{T}(M_2) = (\Gamma_i, \rz_i)$ for $1 \leq i \leq n$, and $\ts = \msf{UNIFY}(\{\rz_i = \rz_i' \mid 1 \leq i \leq n\})$.\\
        
        By induction, $\inftwo{ \Gamma' }{ M_1 }{ \rz_1' \cap \cdots \cap \rz_n' \rightarrow \rt_1' }$
        
        and $\inftwo{ \Gamma_i }{ M_2 }{ \rz_i }$ (for $1 \leq i \leq n$).
        
        Note that $\dom(\Gamma_1) = \dom(\Gamma_2) = \dots = \dom(\Gamma_{n-1}) = \dom(\Gamma_n)$.\\
        
        Let $\s_1 = [y_1/x_1, \dots, y_n/x_n]$ and $\s_2 = [z_1/x_1, \dots, z_n/x_n]$, where $\dom(\Gamma') \cap \dom(\Gamma_1) = \{x_1, \dots, x_n\}$ (which by \autoref{xfreesys}, occur free in $M_1$ and in $M_2$) and $y_1, \dots, y_n, z_1, \dots, z_n$ are all distinct fresh term variables, not occurring in $M_1$ nor in $M_2$ (and consequently, by \autoref{xfreesys}, not occurring in $\Gamma'$ nor in $\Gamma_i$, for all $1 \leq i \leq n$).\\
        
        By \autoref{eqsubstsystemcol}, $\inftwo{ \s_1(\Gamma') }{ \s_1(M_1) }{ \rz_1' \cap \cdots \cap \rz_n' \rightarrow \rt_1' }$
        
        and $\inftwo{ \s_2(\Gamma_i) }{ \s_2(M_2) }{ \rz_i }$ (for $1 \leq i \leq n$).\\
        
        By \autoref{lemSubstitutivity}, $\inftwo{ \ts(\s_1(\Gamma')) }{ \s_1(M_1) }{ \ts(\rz_1' \cap \cdots \cap \rz_n' \rightarrow \rt_1') }$
        
        and $\inftwo{ \ts(\s_2(\Gamma_i)) }{ \s_2(M_2) }{ \ts(\rz_i) }$ (for $1 \leq i \leq n$).\\
        
        Since $\ts(\rz_i) = \ts(\rz_i')$ for all $1 \leq i \leq n$ and $(\s_1(\Gamma'), (\s_2(\Gamma_1) + \dots + \s_2(\Gamma_n)))$ is consistent,
        
        by rule ($\rightarrow$ Elim) we have $\inftwo{ (\ts(\s_1(\Gamma')), (\ts(\s_2(\Gamma_1)) + \dots + \ts(\s_2(\Gamma_n)))) }{ (\s_1(M_1)) (\s_2(M_2)) }{ \ts(\rt_1') }$,
        
        which is the same as $\inftwo{ \ts(\s_1(\Gamma'), (\s_2(\Gamma_1) + \dots + \s_2(\Gamma_n))) }{ (\s_1(M_1)) (\s_2(M_2)) }{ \ts(\rt_1') }$.\\
        
        For each pair $(y_i:\ro_i, z_i:\ro_i')$ (for $1 \leq i \leq n$) in the environment $\ts(\s_1(\Gamma'), (\s_2(\Gamma_1) + \dots + \s_2(\Gamma_n)))$ in the previous derivation, let us apply the rule (Contraction) to obtain the environment with $x_i:\ro_i \cap \ro_i'$ instead (and applying the rule (Exchange) as necessary).\\
        
        After these applications of the rules (Contraction) and (Exchange) (and consequent applications of (Exchange), if necessary), and by looking at the definition of $(+)$, we end up with $\inftwo{ \ts(\Gamma' + \sum_{i=1}^n \Gamma_i) }{ M_1 M_2 }{ \ts(\rt_1') }$.\\

        \item $(\Gamma, \rt) = (\ts(\Gamma_1 + \Gamma_2), \ts(\rt_1))$, where $\msf{T}(M_1) = (\Gamma_1, \rz \multimap \rt_1)$, $\msf{T}(M_2) = (\Gamma_2, \rz_2)$ and $\ts = \msf{UNIFY}(\{\rz_2 = \rz\})$.\\
        
        By induction, $\inftwo{ \Gamma_1 }{ M_1 }{ \rz \multimap \rt_1 }$ and $\inftwo{ \Gamma_2 }{ M_2 }{ \rz_2 }$.\\
        
        Let $\s_1 = [y_1/x_1, \dots, y_n/x_n]$ and $\s_2 = [z_1/x_1, \dots, z_n/x_n]$, where $\dom(\Gamma_1) \cap \dom(\Gamma_2) = \{x_1, \dots, x_n\}$ (which by \autoref{xfreesys}, occur free in $M_1$ and $M_2$) and $y_1, \dots, y_n, z_1, \dots, z_n$ are all distinct fresh term variables, not occurring in $M_1$ nor in $M_2$ (and consequently, by \autoref{xfreesys}, not occurring in $\Gamma_1$ nor in $\Gamma_2$).\\
        
        By \autoref{eqsubstsystemcol}, $\inftwo{ \s_1(\Gamma_1) }{ \s_1(M_1) }{ \rz \multimap \rt_1 }$ and $\inftwo{ \s_2(\Gamma_2) }{ \s_2(M_2) }{ \rz_2 }$.\\
        
        By \autoref{lemSubstitutivity}, $\inftwo{ \ts (\s_1(\Gamma_1)) }{ \s_1(M_1) }{ \ts (\rz \multimap \rt_1) }$ and $\inftwo{ \ts (\s_2(\Gamma_2)) }{ \s_2(M_2) }{ \ts (\rz_2) }$.\\
        
        Since $\ts(\rz_2) = \ts(\rz)$ and $(\s_1(\Gamma_1), \s_2(\Gamma_2))$ is consistent,
        
        by rule ($\multimap$ Elim) we have $\inftwo{ (\ts (\s_1(\Gamma_1)), \ts (\s_2(\Gamma_2))) }{ (\s_1(M_1)) (\s_2(M_2)) }{ \ts(\rt_1) }$,
        
        which is the same as $\inftwo{ \ts(\s_1(\Gamma_1), \s_2(\Gamma_2)) }{ (\s_1(M_1)) (\s_2(M_2)) }{ \ts(\rt_1) }$.\\
        
        For each pair $(y_i:\ro_i, z_i:\ro_i')$ (for $1 \leq i \leq n$) in the environment $\ts(\s_1(\Gamma_1), \s_2(\Gamma_2))$ in the previous derivation, let us apply the rule (Contraction) to obtain the environment with $x_i:\ro_i \cap \ro_i'$ instead (and applying the rule (Exchange) as necessary).\\
        
        After these applications of the rules (Contraction) and (Exchange) (and consequent applications of (Exchange), if necessary), and by looking at the definition of $(+)$, we end up with $\inftwo{ \ts(\Gamma_1 + \Gamma_2) }{ M_1 M_2 }{ \ts(\rt_1) }$.
    \end{enumerate}
\end{enumerate}

For any other possible case, the algorithm fails (by rules 2.(a) and 3.(d)), thus making the left side of the implication ($\msf{T}(M) = (\Gamma, \rt)$) false, which makes the statement true.
\end{proof}
\end{theorem}

\begin{lemma}\label{eqsubst1}
If $\msf{T}(M) = (\Gamma, \rt)$, $x \in \fv{M}$ and $y$ does not occur in $M$, then $\msf{T}(M[y/x]) = (\Gamma[y/x], \rt)$.

\begin{proof}
By induction on the definition of $\msf{T}(M)$.

\begin{enumerate}
    \item If $M = x_1$ and let $x = x_1$ and $y \neq x_1$, then $(\Gamma, \rt) = ([x_1:\alpha], \alpha)$
    
    and $\msf{T}(M[y/x]) = \msf{T}(M[y/x_1]) = \msf{T}(y) = ([y:\alpha], \alpha) = (\Gamma[y/x], \rt)$.\\
    
    (Note that we can choose the same type variable $\alpha$ from $\msf{T}(M)$ in $\msf{T}(y)$ as these are independent, so $\alpha$ is fresh in $\msf{T}(y)$.)\\

    \item If $M = \lambda x_1.M_1$ and let $x$ be a variable that occurs free in $M$ and $y$ a new variable not occurring in $M$, we have the following cases:
    
    \begin{enumerate}
        \item $(\Gamma, \rt) = ({\Gamma_1}_{x_1}, \Gamma_1 (x_1) \multimap \rt_1)$, where $\msf{T}(M_1) = (\Gamma_1, \rt_1)$ and $\Gamma_1 (x_1) = \rz \in \tl{0}$.\\
        
        Since $x \in \fv{M_1}$ and $y$ does not occur in $M_1$ (otherwise it would contradict the assumption that $x \in \fv{M}$ and $y$ does not occur in $M$),
        
        by induction, $\msf{T}(M_1[y/x]) = (\Gamma_1[y/x], \rt_1)$.\\
        
        And $(\Gamma_1[y/x]) (x_1) = \Gamma_1 (x_1) = \rz \in \tl{0}$.\\
        
        So by rule 2.(b) of the inference algorithm, $\msf{T}(\lambda x_1.(M_1[y/x])) = ({(\Gamma_1[y/x])}_{x_1}, \rz \multimap \rt_1)$.\\
        
        And $M[y/x] = (\lambda x_1.M_1)[y/x] = \lambda x_1.(M_1[y/x])$, so
        \begin{align*}
            \msf{T}(M[y/x])
            &= \msf{T}(\lambda x_1.(M_1[y/x]))\\
            &= ({(\Gamma_1[y/x])}_{x_1}, \rz \multimap \rt_1)\\
            &= ({\Gamma_1}_{x_1}[y/x], \Gamma_1 (x_1) \multimap \rt_1)\\
            &= (\Gamma[y/x], \rt).
        \end{align*}

        \item $(\Gamma, \rt) = ({\Gamma_1}_{x_1}, {\Gamma_1} (x_1) \rightarrow \rt_1)$, where $\msf{T}(M_1) = (\Gamma_1, \rt_1)$ and $\Gamma_1 (x_1) = \rz_1 \cap \cdots \cap \rz_n$, with $n \geq 2$.\\
        
        Since $x \in \fv{M_1}$ and $y$ does not occur in $M_1$, by induction, $\msf{T}(M_1[y/x]) = (\Gamma_1[y/x], \rt_1)$.\\
        
        And $(\Gamma_1[y/x]) (x_1) = \Gamma_1 (x_1) = \rz_1 \cap \cdots \cap \rz_n$.\\
        
        So by rule 2.(c) of the inference algorithm, $\msf{T}(\lambda x_1.(M_1[y/x])) = ({(\Gamma_1[y/x])}_{x_1}, \rz_1 \cap \cdots \cap \rz_n \rightarrow \rt_1)$.\\
        
        And $M[y/x] = (\lambda x_1.M_1)[y/x] = \lambda x_1.(M_1[y/x])$, so
        \begin{align*}
            \msf{T}(M[y/x])
            &= \msf{T}(\lambda x_1.(M_1[y/x]))\\
            &= ({(\Gamma_1[y/x])}_{x_1}, \rz_1 \cap \cdots \cap \rz_n \rightarrow \rt_1)\\
            &= ({\Gamma_1}_{x_1}[y/x], \Gamma_1 (x_1) \rightarrow \rt_1)\\
            &= (\Gamma[y/x], \rt).
        \end{align*}
    \end{enumerate}
    
    \item If $M = M_1 M_2$ and let $x$ be a variable that occurs free in $M$ and $y$ a new variable not occurring in $M$, we have the following cases:
    
    \begin{enumerate}
        \item $(\Gamma, \rt) = (\ts(\Gamma_1 + \Gamma_2), \ts(\alpha_3))$, where $\msf{T}(M_1) = (\Gamma_1, \alpha_1)$, $\msf{T}(M_2) = (\Gamma_2, \rz_2)$, $\ts = \msf{UNIFY}(\{\alpha_1 = \alpha_2 \multimap \alpha_3, \rz_2 = \alpha_2\})$ and $\alpha_2, \alpha_3$ do not occur in $\Gamma_1, \Gamma_2, \alpha_1, \rz_2$.\\
        
        Since $x \in \fv{M}$ and $y$ does not occur in $M$, then $y$ does not occur in $M_1$ nor in $M_2$ and there are three possible cases regarding $x$:
        
        \begin{enumerate}
            \item $x \in \fv{M_1}$ and $x \in \fv{M_2}$:\\
            
            Then by induction, $\msf{T}(M_1[y/x]) = (\Gamma_1[y/x], \alpha_1)$ and $\msf{T}(M_2[y/x]) = (\Gamma_2[y/x], \rz_2)$.\\
        
            So by rule 3.(a) of the inference algorithm,
            
            $\msf{T}((M_1[y/x]) (M_2[y/x])) = (\ts((\Gamma_1[y/x]) + (\Gamma_2[y/x])), \ts(\alpha_3))$.\\
        
             (As before, as well as in the following cases, note that we can choose the same type variables $\alpha_2, \alpha_3$ (and, consequently, the same $\ts$) in $\msf{T}((M_1[y/x]) (M_2[y/x]))$ because they are fresh in this inference and, since they do not occur in $\Gamma_1$ and $\Gamma_2$, they also do not occur in $\Gamma_1[y/x]$ and $\Gamma_2[y/x]$ (nor in $\alpha_1, \rz_2$).)\\
        
            And $M[y/x] = (M_1[y/x]) (M_2[y/x])$, so
            \begin{align*}
                \msf{T}(M[y/x])
                &= \msf{T}((M_1[y/x]) (M_2[y/x]))\\
                &= (\ts((\Gamma_1[y/x]) + (\Gamma_2[y/x])), \ts(\alpha_3))\\
                &= (\ts((\Gamma_1 + \Gamma_2)[y/x]), \ts(\alpha_3))\\
                &= ((\ts(\Gamma_1 + \Gamma_2))[y/x], \ts(\alpha_3))\\
                &= (\Gamma[y/x], \rt).
            \end{align*}

            \item $x \in \fv{M_1}$ and $x \notin \fv{M_2}$:\\
            
            Then $M_2[y/x] = M_2$ and $\Gamma_2[y/x] = \Gamma_2$.
            
            So $\msf{T}(M_2[y/x]) = \msf{T}(M_2) = (\Gamma_2, \rz_2) = (\Gamma_2[y/x], \rz_2)$.\\
            
            By induction, $\msf{T}(M_1[y/x]) = (\Gamma_1[y/x], \alpha_1)$.\\
        
            So by rule 3.(a) of the inference algorithm,
            
            $\msf{T}((M_1[y/x]) (M_2[y/x])) = (\ts((\Gamma_1[y/x]) + (\Gamma_2[y/x])), \ts(\alpha_3))$.\\
        
            And $M[y/x] = (M_1[y/x]) (M_2[y/x])$, so
            \begin{align*}
                \msf{T}(M[y/x])
                &= \msf{T}((M_1[y/x]) (M_2[y/x]))\\
                &= (\ts((\Gamma_1[y/x]) + (\Gamma_2[y/x])), \ts(\alpha_3))\\
                &= (\ts((\Gamma_1 + \Gamma_2)[y/x]), \ts(\alpha_3))\\
                &= ((\ts(\Gamma_1 + \Gamma_2))[y/x], \ts(\alpha_3))\\
                &= (\Gamma[y/x], \rt).
            \end{align*}

            \item $x \notin \fv{M_1}$ and $x \in \fv{M_2}$:\\
            
            Then $M_1[y/x] = M_1$ and $\Gamma_1[y/x] = \Gamma_1$.
            
            So $\msf{T}(M_1[y/x]) = \msf{T}(M_1) = (\Gamma_1, \alpha_1) = (\Gamma_1[y/x], \alpha_1)$.\\
            
            By induction, $\msf{T}(M_2[y/x]) = (\Gamma_2[y/x], \rz_2)$.\\
        
            So by rule 3.(a) of the inference algorithm,
            
            $\msf{T}((M_1[y/x]) (M_2[y/x])) = (\ts((\Gamma_1[y/x]) + (\Gamma_2[y/x])), \ts(\alpha_3))$.\\
        
            And $M[y/x] = (M_1[y/x]) (M_2[y/x])$, so
            \begin{align*}
                \msf{T}(M[y/x])
                &= \msf{T}((M_1[y/x]) (M_2[y/x]))\\
                &= (\ts((\Gamma_1[y/x]) + (\Gamma_2[y/x])), \ts(\alpha_3))\\
                &= (\ts((\Gamma_1 + \Gamma_2)[y/x]), \ts(\alpha_3))\\
                &= ((\ts(\Gamma_1 + \Gamma_2))[y/x], \ts(\alpha_3))\\
                &= (\Gamma[y/x], \rt).
            \end{align*}
        \end{enumerate}

        \item $(\Gamma, \rt) = (\ts(\Gamma' + \sum_{i=1}^n \Gamma_i), \ts(\rt_1'))$, where $\msf{T}(M_1) = (\Gamma', \rz_1' \cap \cdots \cap \rz_n' \rightarrow \rt_1')$, with $n \geq 2$, $\msf{T}(M_2) = (\Gamma_i, \rz_i)$ for $1 \leq i \leq n$, and $\ts = \msf{UNIFY}(\{\rz_i = \rz_i' \mid 1 \leq i \leq n\})$.\\
        
        Since $x \in \fv{M}$ and $y$ does not occur in $M$, then $y$ does not occur in $M_1$ nor in $M_2$ and there are three possible cases regarding $x$:
        
        \begin{enumerate}
            \item $x \in \fv{M_1}$ and $x \in \fv{M_2}$:\\
            
            Then by induction, $\msf{T}(M_1[y/x]) = (\Gamma'[y/x], \rz_1' \cap \cdots \cap \rz_n' \rightarrow \rt_1')$
            
            and $\msf{T}(M_2[y/x]) = (\Gamma_i[y/x], \rz_i)$, for all $1 \leq i \leq n$.\\
        
            So by rule 3.(b) of the inference algorithm,
            
            $\msf{T}((M_1[y/x]) (M_2[y/x])) = (\ts((\Gamma'[y/x]) + \sum_{i=1}^n (\Gamma_i[y/x])), \ts(\rt_1'))$.\\
        
            And $M[y/x] = (M_1[y/x]) (M_2[y/x])$, so
            \begin{align*}
                \msf{T}(M[y/x])
                &= \msf{T}((M_1[y/x]) (M_2[y/x]))\\
                &= (\ts((\Gamma'[y/x]) + \sum_{i=1}^n (\Gamma_i[y/x])), \ts(\rt_1'))\\
                &= (\ts((\Gamma' + \sum_{i=1}^n \Gamma_i)[y/x]), \ts(\rt_1'))\\
                &= ((\ts(\Gamma' + \sum_{i=1}^n \Gamma_i))[y/x], \ts(\rt_1'))\\
                &= (\Gamma[y/x], \rt).
            \end{align*}

            \item $x \in \fv{M_1}$ and $x \notin \fv{M_2}$:\\
            
            Then $M_2[y/x] = M_2$ and $\Gamma_i[y/x] = \Gamma_i$, for all $1 \leq i \leq n$.
            
            So $\msf{T}(M_2[y/x]) = \msf{T}(M_2) = (\Gamma_i, \rz_i) = (\Gamma_i[y/x], \rz_i)$, for all $1 \leq i \leq n$.\\
            
            By induction, $\msf{T}(M_1[y/x]) = (\Gamma'[y/x], \rz_1' \cap \cdots \cap \rz_n' \rightarrow \rt_1')$.\\
        
            So by rule 3.(b) of the inference algorithm,
            
            $\msf{T}((M_1[y/x]) (M_2[y/x])) = (\ts((\Gamma'[y/x]) + \sum_{i=1}^n (\Gamma_i[y/x])), \ts(\rt_1'))$.\\
        
            And $M[y/x] = (M_1[y/x]) (M_2[y/x])$, so
            \begin{align*}
                \msf{T}(M[y/x])
                &= \msf{T}((M_1[y/x]) (M_2[y/x]))\\
                &= (\ts((\Gamma'[y/x]) + \sum_{i=1}^n (\Gamma_i[y/x])), \ts(\rt_1'))\\
                &= (\ts((\Gamma' + \sum_{i=1}^n \Gamma_i)[y/x]), \ts(\rt_1'))\\
                &= ((\ts(\Gamma' + \sum_{i=1}^n \Gamma_i))[y/x], \ts(\rt_1'))\\
                &= (\Gamma[y/x], \rt).
            \end{align*}

            \item $x \notin \fv{M_1}$ and $x \in \fv{M_2}$:\\
            
            Then $M_1[y/x] = M_1$ and $\Gamma'[y/x] = \Gamma'$.
            
            So $\msf{T}(M_1[y/x]) = \msf{T}(M_1) = (\Gamma', \rz_1' \cap \cdots \cap \rz_n' \rightarrow \rt_1') = (\Gamma'[y/x], \rz_1' \cap \cdots \cap \rz_n' \rightarrow \rt_1')$.\\
            
            By induction, $\msf{T}(M_2[y/x]) = (\Gamma_i[y/x], \rz_i)$, for all $1 \leq i \leq n$.\\
        
            So by rule 3.(b) of the inference algorithm,
            
            $\msf{T}((M_1[y/x]) (M_2[y/x])) = (\ts((\Gamma'[y/x]) + \sum_{i=1}^n (\Gamma_i[y/x])), \ts(\rt_1'))$.\\
        
            And $M[y/x] = (M_1[y/x]) (M_2[y/x])$, so
            \begin{align*}
                \msf{T}(M[y/x])
                &= \msf{T}((M_1[y/x]) (M_2[y/x]))\\
                &= (\ts((\Gamma'[y/x]) + \sum_{i=1}^n (\Gamma_i[y/x])), \ts(\rt_1'))\\
                &= (\ts((\Gamma' + \sum_{i=1}^n \Gamma_i)[y/x]), \ts(\rt_1'))\\
                &= ((\ts(\Gamma' + \sum_{i=1}^n \Gamma_i))[y/x], \ts(\rt_1'))\\
                &= (\Gamma[y/x], \rt).
            \end{align*}
        \end{enumerate}

        \item $(\Gamma, \rt) = (\ts(\Gamma_1 + \Gamma_2), \ts(\rt_1))$, where $\msf{T}(M_1) = (\Gamma_1, \rz \multimap \rt_1)$, $\msf{T}(M_2) = (\Gamma_2, \rz_2)$ and $\ts = \msf{UNIFY}(\{\rz_2 = \rz\})$.\\
        
        Since $x \in \fv{M}$ and $y$ does not occur in $M$, then $y$ does not occur in $M_1$ nor in $M_2$ and there are three possible cases regarding $x$:
        
        \begin{enumerate}
            \item $x \in \fv{M_1}$ and $x \in \fv{M_2}$:\\
            
            Then by induction, $\msf{T}(M_1[y/x]) = (\Gamma_1[y/x], \rz \multimap \rt_1)$ and $\msf{T}(M_2[y/x]) = (\Gamma_2[y/x], \rz_2)$.\\
        
            So by rule 3.(c) of the inference algorithm,
            
            $\msf{T}((M_1[y/x]) (M_2[y/x])) = (\ts((\Gamma_1[y/x]) + (\Gamma_2[y/x])), \ts(\rt_1))$.\\
        
            And $M[y/x] = (M_1[y/x]) (M_2[y/x])$, so
            \begin{align*}
                \msf{T}(M[y/x])
                &= \msf{T}((M_1[y/x]) (M_2[y/x]))\\
                &= (\ts((\Gamma_1[y/x]) + (\Gamma_2[y/x])), \ts(\rt_1))\\
                &= (\ts((\Gamma_1 + \Gamma_2)[y/x]), \ts(\rt_1))\\
                &= ((\ts(\Gamma_1 + \Gamma_2))[y/x], \ts(\rt_1))\\
                &= (\Gamma[y/x], \rt).
            \end{align*}

            \item $x \in \fv{M_1}$ and $x \notin \fv{M_2}$:\\
            
            Then $M_2[y/x] = M_2$ and $\Gamma_2[y/x] = \Gamma_2$.
            
            So $\msf{T}(M_2[y/x]) = \msf{T}(M_2) = (\Gamma_2, \rz_2) = (\Gamma_2[y/x], \rz_2)$.\\
            
            By induction, $\msf{T}(M_1[y/x]) = (\Gamma_1[y/x], \rz \multimap \rt_1)$.\\
        
            So by rule 3.(c) of the inference algorithm,
            
            $\msf{T}((M_1[y/x]) (M_2[y/x])) = (\ts((\Gamma_1[y/x]) + (\Gamma_2[y/x])), \ts(\rt_1))$.\\
        
            And $M[y/x] = (M_1[y/x]) (M_2[y/x])$, so
            \begin{align*}
                \msf{T}(M[y/x])
                &= \msf{T}((M_1[y/x]) (M_2[y/x]))\\
                &= (\ts((\Gamma_1[y/x]) + (\Gamma_2[y/x])), \ts(\rt_1))\\
                &= (\ts((\Gamma_1 + \Gamma_2)[y/x]), \ts(\rt_1))\\
                &= ((\ts(\Gamma_1 + \Gamma_2))[y/x], \ts(\rt_1))\\
                &= (\Gamma[y/x], \rt).
            \end{align*}

            \item $x \notin \fv{M_1}$ and $x \in \fv{M_2}$:\\
            
            Then $M_1[y/x] = M_1$ and $\Gamma_1[y/x] = \Gamma_1$.
            
            So $\msf{T}(M_1[y/x]) = \msf{T}(M_1) = (\Gamma_1, \rz \multimap \rt_1) = (\Gamma_1[y/x], \rz \multimap \rt_1)$.\\
            
            By induction, $\msf{T}(M_2[y/x]) = (\Gamma_2[y/x], \rz_2)$.\\
        
            So by rule 3.(c) of the inference algorithm,
            
            $\msf{T}((M_1[y/x]) (M_2[y/x])) = (\ts((\Gamma_1[y/x]) + (\Gamma_2[y/x])), \ts(\rt_1))$.\\
        
            And $M[y/x] = (M_1[y/x]) (M_2[y/x])$, so
            \begin{align*}
                \msf{T}(M[y/x])
                &= \msf{T}((M_1[y/x]) (M_2[y/x]))\\
                &= (\ts((\Gamma_1[y/x]) + (\Gamma_2[y/x])), \ts(\rt_1))\\
                &= (\ts((\Gamma_1 + \Gamma_2)[y/x]), \ts(\rt_1))\\
                &= ((\ts(\Gamma_1 + \Gamma_2))[y/x], \ts(\rt_1))\\
                &= (\Gamma[y/x], \rt).
            \end{align*}
        \end{enumerate}
    \end{enumerate}
\end{enumerate}

Any other possible case makes the left side of the implication ($\msf{T}(M) = (\Gamma, \rt)$, $x \in \fv{M}$ and $y$ does not occur in $M$) false, which makes the statement true.
\end{proof}
\end{lemma}

\begin{lemma}\label{eqsubst2}
If $\msf{T}(M) = (\Gamma, \rt)$, with $\Gamma \equiv (\Gamma', y_1:\ro_1, y_2:\ro_2)$, and $y$ does not occur in  $M$, then $\msf{T}(M[y/y_1, y/y_2]) = (\Gamma'', \rt)$, with $\Gamma'' \equiv (\Gamma', y:\ro_1 \cap \ro_2)$.

\begin{proof}
By induction on the definition of $\msf{T}(M)$.

\begin{enumerate}
    \item If $M = \lambda x_1.M_1$ and let $y$ be a new variable not occurring in $M$, we have the following cases:
    
    \begin{enumerate}
        \item $(\Gamma, \rt) = ({\Gamma_1}_{x_1}, \Gamma_1 (x_1) \multimap \rt_1)$, with $\Gamma \equiv (\Gamma', y_1:\ro_1, y_2:\ro_2)$, where $\msf{T}(M_1) = (\Gamma_1, \rt_1)$ and $\Gamma_1 (x_1) = \rz \in \tl{0}$.\\
        
        Since ${\Gamma_1}_{x_1} \equiv (\Gamma', y_1:\ro_1, y_2:\ro_2)$, then $\Gamma_1 \equiv (\Gamma', x_1:\rz, y_1:\ro_1, y_2:\ro_2)$.\\
        
        And since $y$ does not occur in $M_1$ (otherwise it would contradict the assumption that $y$ does not occur in $M$),
        
        by induction, $\msf{T}(M_1[y/y_1, y/y_2]) = (\Gamma_1', \rt_1)$,
        
        with $\Gamma_1' \equiv (\Gamma', x_1:\rz, y:\ro_1 \cap \ro_2)$.\\
        
        So by rule 2.(b) of the inference algorithm,
        
        $\msf{T}(\lambda x_1.(M_1[y/y_1, y/y_2])) = ({\Gamma_1'}_{x_1}, \rz \multimap \rt_1)$.
        
        And ${\Gamma_1'}_{x_1} \equiv (\Gamma', y:\ro_1 \cap \ro_2)$.\\
        
        Let $\Gamma'' = {\Gamma_1'}_{x_1}$.\\
        
        Also, $M[y/y_1, y/y_2] = \lambda x_1.(M_1[y/y_1, y/y_2])$, so
        \begin{align*}
            \msf{T}(M[y/y_1, y/y_2])
            &= \msf{T}(\lambda x_1.(M_1[y/y_1, y/y_2]))\\
            &= (\Gamma'', \rz \multimap \rt_1)\\
            &= (\Gamma'', \rt),
        \end{align*}
        
        with $\Gamma'' \equiv (\Gamma', y:\ro_1 \cap \ro_2)$.\\

        \item $(\Gamma, \rt) = ({\Gamma_1}_{x_1}, \Gamma_1 (x_1) \rightarrow \rt_1)$, with $\Gamma \equiv (\Gamma', y_1:\ro_1, y_2:\ro_2)$, where $\msf{T}(M_1) = (\Gamma_1, \rt_1)$ and $\Gamma_1 (x_1) = \rz_1 \cap \cdots \cap \rz_n$, with $n \geq 2$.\\
        
        Since ${\Gamma_1}_{x_1} \equiv (\Gamma', y_1:\ro_1, y_2:\ro_2)$, then $\Gamma_1 \equiv (\Gamma', x_1:\rz_1 \cap \cdots \cap \rz_n, y_1:\ro_1, y_2:\ro_2)$.\\
        
        And since $y$ does not occur in $M_1$ (otherwise it would contradict the assumption that $y$ does not occur in $M$),
        
        by induction, $\msf{T}(M_1[y/y_1, y/y_2]) = (\Gamma_1', \rt_1)$,
        
        with $\Gamma_1' \equiv (\Gamma', x_1:\rz_1 \cap \cdots \cap \rz_n, y:\ro_1 \cap \ro_2)$.\\
        
        So by rule 2.(c) of the inference algorithm,
        
        $\msf{T}(\lambda x_1.(M_1[y/y_1, y/y_2])) = ({\Gamma_1'}_{x_1}, \rz_1 \cap \cdots \cap \rz_n \rightarrow \rt_1)$.
        
        And ${\Gamma_1'}_{x_1} \equiv (\Gamma', y:\ro_1 \cap \ro_2)$.\\
        
        Let $\Gamma'' = {\Gamma_1'}_{x_1}$.\\
        
        Also, $M[y/y_1, y/y_2] = \lambda x_1.(M_1[y/y_1, y/y_2])$, so
        \begin{align*}
            \msf{T}(M[y/y_1, y/y_2])
            &= \msf{T}(\lambda x_1.(M_1[y/y_1, y/y_2]))\\
            &= (\Gamma'', \rz_1 \cap \cdots \cap \rz_n \rightarrow \rt_1)\\
            &= (\Gamma'', \rt),
        \end{align*}
        
        with $\Gamma'' \equiv (\Gamma', y:\ro_1 \cap \ro_2)$.\\
    \end{enumerate}
    
    \item If $M = M_1 M_2$ and let $y$ be a new variable not occurring in $M$, we have the following cases:
        
    \begin{enumerate}
        \item $(\Gamma, \rt) = (\ts(\Gamma_1 + \Gamma_2), \ts(\alpha_3))$, with $\Gamma \equiv (\Gamma', y_1:\ro_1, y_2:\ro_2)$, where $\msf{T}(M_1) = (\Gamma_1, \alpha_1)$, $\msf{T}(M_2) = (\Gamma_2, \rz_2)$, $\ts = \msf{UNIFY}(\{\alpha_1 = \alpha_2 \multimap \alpha_3, \rz_2 = \alpha_2\})$ and $\alpha_2, \alpha_3$ do not occur in $\Gamma_1, \Gamma_2, \alpha_1, \rz_2$.\\
        
        Because $y$ does not occur in $M$, then $y$ does not occur in $M_1$ nor in $M_2$.\\
        
        Since $\ts(\Gamma_1 + \Gamma_2) \equiv (\Gamma', y_1:\ro_1, y_2:\ro_2)$ and $M_1$ is a term variable (otherwise its type given by the algorithm would not be a type variable), then there are five possible cases regarding the presence of $y_1$ and $y_2$ in $\dom(\Gamma_1)$ and $\dom(\Gamma_2)$:
        
        \begin{enumerate}
            \item $y_1, y_2 \notin \dom(\Gamma_1)$ and $y_1, y_2 \in \dom(\Gamma_2)$:\\
            
            So $\Gamma_2 \equiv (\Gamma_2', y_1:\ro_3, y_2:\ro_4)$ (for some $\ro_3, \ro_4$ such that $\ts(\ro_3) = \ro_1$ and $\ts(\ro_4) = \ro_2$).\\
            
            By induction,
            \begin{equation}
                \msf{T}(M_2[y/y_1, y/y_2]) = (\Gamma_2'', \rz \multimap \rz_2), \tag{1}\label{eq:2.a.1.1}
            \end{equation}
            with $\Gamma_2'' \equiv (\Gamma_2', y:\ro_3 \cap \ro_4)$.\\
            
            And since $y_1, y_2 \notin \dom(\Gamma_1)$, by \autoref{xfreealg}, $y_1, y_2 \notin \fv{M_1}$,
            
            so $M_1[y/y_1, y/y_2] = M_1$
            
            and then
            \begin{equation}
                \msf{T}(M_1[y/y_1, y/y_2]) = \msf{T}(M_1) = (\Gamma_1, \alpha_1). \tag{2}\label{eq:2.a.1.2}
            \end{equation}
            
            \hfill
            
            So by rule 3.(a) of the inference algorithm (and \eqref{eq:2.a.1.1}, \eqref{eq:2.a.1.2}),
        
            $\msf{T}(M_1[y/y_1, y/y_2] M_2[y/y_1, y/y_2]) = (\ts(\Gamma_1 + \Gamma_2''), \ts(\alpha_3))$.\\
            
            (Note that we can choose the same type variables $\alpha_2, \alpha_3$ (and, consequently, the same $\ts$) in $\msf{T}(M_1[y/y_1, y/y_2] M_2[y/y_1, y/y_2])$ because they are fresh in this inference and, since they do not occur in $\Gamma_2$, they also do not occur in $\Gamma_2''$ (nor in $\Gamma_1$, $\alpha_1, \rz_2$). For analogous reasons, the same can and will be done in the following cases.)\\
            
            Since $\Gamma_2'' \equiv (\Gamma_2', y:\ro_3 \cap \ro_4)$, $\Gamma_2 \equiv (\Gamma_2', y_1:\ro_3, y_2:\ro_4)$ and $\ts(\Gamma_1 + \Gamma_2) \equiv (\Gamma', y_1:\ro_1, y_2:\ro_2)$,
            
            we have $\ts(\Gamma_1 + \Gamma_2'') \equiv (\Gamma', y:\ro_1 \cap \ro_2)$.\\
            
            Let $\Gamma'' = \ts(\Gamma_1 + \Gamma_2'')$.\\
            
            Also, $M[y/y_1, y/y_2] = M_1[y/y_1, y/y_2] M_2[y/y_1, y/y_2]$, so
            \begin{align*}
                \msf{T}(M[y/y_1, y/y_2])
                &= \msf{T}(M_1[y/y_1, y/y_2] M_2[y/y_1, y/y_2])\\
                &= (\Gamma'', \ts(\alpha_3))\\
                &= (\Gamma'', \rt),
            \end{align*}
        
            with $\Gamma'' \equiv (\Gamma', y:\ro_1 \cap \ro_2)$.\\

            \item $y_1, y_2 \in \dom(\Gamma_2)$, $y_1 \in \dom(\Gamma_1)$ and $y_2 \notin \dom(\Gamma_1)$:\\
            
            So $\Gamma_2 \equiv (\Gamma_2', y_1:\ro_3, y_2:\ro_4)$ and $\Gamma_1 \equiv (\Gamma_1', y_1:\ro_3')$ (for some $\ro_3, \ro_4, \ro_3'$ such that $\ts(\ro_3' \cap \ro_3) = \ro_1$ and $\ts(\ro_4) = \ro_2$).\\
            
            By induction,
            \begin{equation}
                \msf{T}(M_2[y/y_1, y/y_2]) = (\Gamma_2'', \rz_2), \tag{1}\label{eq:2.a.2.1}
            \end{equation}
            with $\Gamma_2'' \equiv (\Gamma_2', y:\ro_3 \cap \ro_4)$.\\
            
            Since $y_1 \in \dom(\Gamma_1)$, by \autoref{xfreealg}, $y_1 \in \fv{M_1}$.
            
            So by \autoref{eqsubst1}, we have $\msf{T}(M_1[y/y_1]) = (\Gamma_1[y/y_1], \alpha_1)$.\\
            
            And since $y_2 \notin \dom(\Gamma_1)$, by \autoref{xfreealg}, $y_2 \notin \fv{M_1}$,
            
            so $M_1[y/y_1, y/y_2] = M_1[y/y_1]$
            
            and then
            \begin{equation}
                \msf{T}(M_1[y/y_1, y/y_2]) = \msf{T}(M_1[y/y_1]) = (\Gamma_1[y/y_1], \alpha_1). \tag{2}\label{eq:2.a.2.2}
            \end{equation}
            
            \hfill
            
            So by rule 3.(a) of the inference algorithm (and \eqref{eq:2.a.2.1}, \eqref{eq:2.a.2.2}),
        
            $\msf{T}(M_1[y/y_1, y/y_2] M_2[y/y_1, y/y_2]) = (\ts(\Gamma_1[y/y_1] + \Gamma_2''), \ts(\alpha_3))$.\\
            
            Since $\Gamma_2'' \equiv (\Gamma_2', y:\ro_3 \cap \ro_4)$, $\Gamma_2 \equiv (\Gamma_2', y_1:\ro_3, y_2:\ro_4)$, $\Gamma_1[y/y_1] \equiv (\Gamma_1', y:\ro_3')$, $\Gamma_1 \equiv (\Gamma_1', y_1:\ro_3')$, $\ts(\Gamma_1 + \Gamma_2) \equiv (\Gamma', y_1:\ro_1, y_2:\ro_2)$ and $\ro_3' \cap (\ro_3 \cap \ro_4) = (\ro_3' \cap \ro_3) \cap \ro_4$,
            
            we have $\ts(\Gamma_1[y/y_1] + \Gamma_2'') \equiv (\Gamma', y:\ro_1 \cap \ro_2)$.\\
            
            Let $\Gamma'' = \ts(\Gamma_1[y/y_1] + \Gamma_2'')$.\\
            
            Also, $M[y/y_1, y/y_2] = M_1[y/y_1, y/y_2] M_2[y/y_1, y/y_2]$, so
            \begin{align*}
                \msf{T}(M[y/y_1, y/y_2])
                &= \msf{T}(M_1[y/y_1, y/y_2] M_2[y/y_1, y/y_2])\\
                &= (\Gamma'', \ts(\alpha_3))\\
                &= (\Gamma'', \rt),
            \end{align*}
        
            with $\Gamma'' \equiv (\Gamma', y:\ro_1 \cap \ro_2)$.\\

            \item $y_1, y_2 \in \dom(\Gamma_2)$, $y_1 \notin \dom(\Gamma_1)$ and $y_2 \in \dom(\Gamma_1)$:\\
            
            Analogous to the previous case.\\

            \item $y_1 \in \dom(\Gamma_1)$, $y_2 \notin \dom(\Gamma_1)$, $y_1 \notin \dom(\Gamma_2)$ and $y_2 \in \dom(\Gamma_2)$:\\
            
            Since $y_1 \in \dom(\Gamma_1)$, by \autoref{xfreealg}, $y_1 \in \fv{M_1}$.
            
            So by \autoref{eqsubst1}, we have $\msf{T}(M_1[y/y_1]) = (\Gamma_1[y/y_1], \alpha_1)$.\\
            
            And since $y_2 \notin \dom(\Gamma_1)$, by \autoref{xfreealg}, $y_2 \notin \fv{M_1}$,
            
            so $M_1[y/y_1, y/y_2] = M_1[y/y_1]$
            
            and then
            \begin{equation}
                \msf{T}(M_1[y/y_1, y/y_2]) = \msf{T}(M_1[y/y_1]) = (\Gamma_1[y/y_1], \alpha_1). \tag{1}\label{eq:2.a.4.1}
            \end{equation}
            
            \hfill
            
            Since $y_2 \in \dom(\Gamma_2)$, by \autoref{xfreealg}, $y_2 \in \fv{M_2}$.
            
            So by \autoref{eqsubst1}, we have $\msf{T}(M_2[y/y_2]) = (\Gamma_2[y/y_2], \rz_2)$.\\
            
            And since $y_1 \notin \dom(\Gamma_2)$, by \autoref{xfreealg}, $y_1 \notin \fv{M_2}$,
            
            so $M_2[y/y_1, y/y_2] = M_2[y/y_2]$
            
            and then
            \begin{equation}
                \msf{T}(M_2[y/y_1, y/y_2]) = \msf{T}(M_2[y/y_2]) = (\Gamma_2[y/y_2], \rz_2). \tag{2}\label{eq:2.a.4.2}
            \end{equation}
            
            \hfill
            
            So by rule 3.(a) of the inference algorithm (and \eqref{eq:2.a.4.1}, \eqref{eq:2.a.4.2}),
        
            $\msf{T}(M_1[y/y_1, y/y_2] M_2[y/y_1, y/y_2]) = (\ts(\Gamma_1[y/y_1] + \Gamma_2[y/y_2]), \ts(\alpha_3))$.\\
            
            Since $\ts(\Gamma_1 + \Gamma_2) \equiv (\Gamma', y_1:\ro_1, y_2:\ro_2)$,
            
            we have $\ts(\Gamma_1[y/y_1] + \Gamma_2[y/y_2]) \equiv (\Gamma', y:\ro_1 \cap \ro_2)$.\\
            
            Let $\Gamma'' = \ts(\Gamma_1[y/y_1] + \Gamma_2[y/y_2])$.\\
            
            Also, $M[y/y_1, y/y_2] = M_1[y/y_1, y/y_2] M_2[y/y_1, y/y_2]$, so
            \begin{align*}
                \msf{T}(M[y/y_1, y/y_2])
                &= \msf{T}(M_1[y/y_1, y/y_2] M_2[y/y_1, y/y_2])\\
                &= (\Gamma'', \ts(\alpha_3))\\
                &= (\Gamma'', \rt),
            \end{align*}
        
            with $\Gamma'' \equiv (\Gamma', y:\ro_1 \cap \ro_2)$.\\

            \item $y_1 \notin \dom(\Gamma_1)$, $y_2 \in \dom(\Gamma_1)$, $y_1 \in \dom(\Gamma_2)$ and $y_2 \notin \dom(\Gamma_2)$:\\
            
            Analogous to the previous case.\\
        \end{enumerate}

        \item $(\Gamma, \rt) = (\ts(\Gamma_1' + \sum_{i=1}^n \Gamma_i), \ts(\rt_1'))$, with $\Gamma \equiv (\Gamma', y_1:\ro_1, y_2:\ro_2)$, where $\msf{T}(M_1) = (\Gamma_1', \rz_1' \cap \cdots \cap \rz_n' \rightarrow \rt_1')$, with $n \geq 2$, $\msf{T}(M_2) = (\Gamma_i, \rz_i)$ for $1 \leq i \leq n$, and $\ts = \msf{UNIFY}(\{\rz_i = \rz_i' \mid 1 \leq i \leq n\})$.\\
        
        Because $y$ does not occur in $M$, then $y$ does not occur in $M_1$ nor in $M_2$.\\
        
        Since $\ts(\Gamma_1' + \sum_{i=1}^n \Gamma_i) \equiv (\Gamma', y_1:\ro_1, y_2:\ro_2)$, then there are nine possible cases regarding the presence of $y_1$ and $y_2$ in $\dom(\Gamma_1')$ and $\dom(\Gamma_i)$ (for all $1 \leq i \leq n$):
        
        \begin{enumerate}
            \item $y_1, y_2 \in \dom(\Gamma_1')$ and $y_1, y_2 \notin \dom(\Gamma_i)$:\\
        
            So $\Gamma_1' \equiv (\Gamma_1'', y_1:\ro_3, y_2:\ro_4)$ (for some $\ro_3, \ro_4$ such that $\ts(\ro_3) = \ro_1$ and $\ts(\ro_4) = \ro_2$).\\
            
            By induction,
            \begin{equation}
                \msf{T}(M_1[y/y_1, y/y_2]) = (\Gamma_1''', \rz_1' \cap \cdots \cap \rz_n' \rightarrow \rt_1'), \tag{1}\label{eq:2.b.1.1}
            \end{equation}
            with $\Gamma_1''' \equiv (\Gamma_1'', y:\ro_3 \cap \ro_4)$.\\
            
            And since $y_1, y_2 \notin \dom(\Gamma_i)$, by \autoref{xfreealg}, $y_1, y_2 \notin \fv{M_2}$,
            
            so $M_2[y/y_1, y/y_2] = M_2$
            
            and then
            \begin{equation}
                \msf{T}(M_2[y/y_1, y/y_2]) = \msf{T}(M_2) = (\Gamma_i, \rz_i). \tag{2}\label{eq:2.b.1.2}
            \end{equation}
            
            \hfill
            
            So by rule 3.(b) of the inference algorithm (and \eqref{eq:2.b.1.1}, \eqref{eq:2.b.1.2}),
        
            $\msf{T}(M_1[y/y_1, y/y_2] M_2[y/y_1, y/y_2]) = (\ts(\Gamma_1''' + \sum_{i=1}^n \Gamma_i), \ts(\rt_1'))$.\\
            
            Since $\Gamma_1''' \equiv (\Gamma_1'', y:\ro_3 \cap \ro_4)$, $\Gamma_1' \equiv (\Gamma_1'', y_1:\ro_3, y_2:\ro_4)$ and $\ts(\Gamma_1' + \sum_{i=1}^n \Gamma_i) \equiv (\Gamma', y_1:\ro_1, y_2:\ro_2)$,
            
            we have $\ts(\Gamma_1''' + \sum_{i=1}^n \Gamma_i) \equiv (\Gamma', y:\ro_1 \cap \ro_2)$.\\
            
            Let $\Gamma'' = \ts(\Gamma_1''' + \sum_{i=1}^n \Gamma_i)$.\\
            
            Also, $M[y/y_1, y/y_2] = M_1[y/y_1, y/y_2] M_2[y/y_1, y/y_2]$, so
            \begin{align*}
                \msf{T}(M[y/y_1, y/y_2])
                &= \msf{T}(M_1[y/y_1, y/y_2] M_2[y/y_1, y/y_2])\\
                &= (\Gamma'', \ts(\rt_1'))\\
                &= (\Gamma'', \rt),
            \end{align*}
        
            with $\Gamma'' \equiv (\Gamma', y:\ro_1 \cap \ro_2)$.\\

            \item $y_1, y_2 \notin \dom(\Gamma_1')$ and $y_1, y_2 \in \dom(\Gamma_i)$:\\
            
            Analogous to the previous case.\\

            \item $y_1, y_2 \in \dom(\Gamma_1')$ and $y_1, y_2 \in \dom(\Gamma_i)$:\\
            
            So $\Gamma_1' \equiv (\Gamma_1'', y_1:\ro_3, y_2:\ro_4)$ and $\Gamma_i \equiv (\Gamma_i', y_1:\ro_{3_i}, y_2:\ro_{4_i})$ (for some $\ro_3, \ro_4, \ro_{3_i}, \ro_{4_i}$ such that $\ts(\ro_3 \cap \ro_{3_1} \cap \cdots \cap \ro_{3_n}) = \ro_1$ and $\ts(\ro_4 \cap \ro_{4_1} \cap \cdots \cap \ro_{4_n}) = \ro_2$).\\
            
            By induction,
            \begin{equation}
                \msf{T}(M_1[y/y_1, y/y_2]) = (\Gamma_1''', \rz_1' \cap \cdots \cap \rz_n' \rightarrow \rt_1'), \tag{1}\label{eq:2.b.3.1}
            \end{equation}
            with $\Gamma_1''' \equiv (\Gamma_1'', y:\ro_3 \cap \ro_4)$;
            
            \begin{equation}
                \msf{T}(M_2[y/y_1, y/y_2]) = (\Gamma_i'', \rz_i), \tag{2}\label{eq:2.b.3.2}
            \end{equation}
            with $\Gamma_i'' \equiv (\Gamma_i', y:\ro_{3_i} \cap \ro_{4_i})$.\\
            
            So by rule 3.(b) of the inference algorithm (and \eqref{eq:2.b.3.1}, \eqref{eq:2.b.3.2}),
        
            $\msf{T}(M_1[y/y_1, y/y_2] M_2[y/y_1, y/y_2]) = (\ts(\Gamma_1''' + \sum_{i=1}^n \Gamma_i''), \ts(\rt_1'))$.\\
            
            Since $\Gamma_1''' \equiv (\Gamma_1'', y:\ro_3 \cap \ro_4)$, $\Gamma_1' \equiv (\Gamma_1'', y_1:\ro_3, y_2:\ro_4)$, $\Gamma_i'' \equiv (\Gamma_i', y:\ro_{3_i} \cap \ro_{4_i})$, $\Gamma_i \equiv (\Gamma_i', y_1:\ro_{3_i}, y_2:\ro_{4_i})$, $\ts(\Gamma_1' + \sum_{i=1}^n \Gamma_i) \equiv (\Gamma', y_1:\ro_1, y_2:\ro_2)$ and $(\ro_3 \cap \ro_4) \cap (\ro_{3_1} \cap \ro_{4_1}) \cap \cdots \cap (\ro_{3_n} \cap \ro_{4_n}) = (\ro_3 \cap \ro_{3_1} \cap \cdots \cap \ro_{3_n}) \cap (\ro_4 \cap \ro_{4_1} \cap \cdots \cap \ro_{4_n})$,
            
            we have $\ts(\Gamma_1''' + \sum_{i=1}^n \Gamma_i'') \equiv (\Gamma', y:\ro_1 \cap \ro_2)$.\\
            
            Let $\Gamma'' = \ts(\Gamma_1''' + \sum_{i=1}^n \Gamma_i'')$.\\
            
            Also, $M[y/y_1, y/y_2] = M_1[y/y_1, y/y_2] M_2[y/y_1, y/y_2]$, so
            \begin{align*}
                \msf{T}(M[y/y_1, y/y_2])
                &= \msf{T}(M_1[y/y_1, y/y_2] M_2[y/y_1, y/y_2])\\
                &= (\Gamma'', \ts(\rt_1'))\\
                &= (\Gamma'', \rt),
            \end{align*}
        
            with $\Gamma'' \equiv (\Gamma', y:\ro_1 \cap \ro_2)$.\\

            \item $y_1, y_2 \in \dom(\Gamma_1')$, $y_1 \in \dom(\Gamma_i)$ and $y_2 \notin \dom(\Gamma_i)$:\\
            
            So $\Gamma_1' \equiv (\Gamma_1'', y_1:\ro_3, y_2:\ro_4)$ and $\Gamma_i \equiv (\Gamma_i', y_1:\ro_{3_i})$ (for some $\ro_3, \ro_4, \ro_{3_i}$ such that $\ts(\ro_3 \cap \ro_{3_1} \cap \cdots \cap \ro_{3_n}) = \ro_1$ and $\ts(\ro_4) = \ro_2$).\\
            
            By induction,
            \begin{equation}
                \msf{T}(M_1[y/y_1, y/y_2]) = (\Gamma_1''', \rz_1' \cap \cdots \cap \rz_n' \rightarrow \rt_1'), \tag{1}\label{eq:2.b.4.1}
            \end{equation}
            with $\Gamma_1''' \equiv (\Gamma_1'', y:\ro_3 \cap \ro_4)$.\\
            
            Since $y_1 \in \dom(\Gamma_i)$, by \autoref{xfreealg}, $y_1 \in \fv{M_2}$.
            
            So by \autoref{eqsubst1}, we have $\msf{T}(M_2[y/y_1]) = (\Gamma_i[y/y_1], \rz_i)$.\\
            
            And since $y_2 \notin \dom(\Gamma_i)$, by \autoref{xfreealg}, $y_2 \notin \fv{M_2}$,
            
            so $M_2[y/y_1, y/y_2] = M_2[y/y_1]$
            
            and then
            \begin{equation}
                \msf{T}(M_2[y/y_1, y/y_2]) = \msf{T}(M_2[y/y_1]) = (\Gamma_i[y/y_1], \rz_i). \tag{2}\label{eq:2.b.4.2}
            \end{equation}
            
            \hfill
            
            So by rule 3.(b) of the inference algorithm (and \eqref{eq:2.b.4.1}, \eqref{eq:2.b.4.2}),
        
            $\msf{T}(M_1[y/y_1, y/y_2] M_2[y/y_1, y/y_2]) = (\ts(\Gamma_1''' + \sum_{i=1}^n \Gamma_i[y/y_1]), \ts(\rt_1'))$.\\
            
            Since $\Gamma_1''' \equiv (\Gamma_1'', y:\ro_3 \cap \ro_4)$, $\Gamma_1' \equiv (\Gamma_1'', y_1:\ro_3, y_2:\ro_4)$, $\Gamma_i[y/y_1] \equiv (\Gamma_i', y:\ro_{3_i})$, $\Gamma_i \equiv (\Gamma_i', y_1:\ro_{3_i})$, $\ts(\Gamma_1' + \sum_{i=1}^n \Gamma_i) \equiv (\Gamma', y_1:\ro_1, y_2:\ro_2)$ and $(\ro_3 \cap \ro_4) \cap \ro_{3_1} \cap \cdots \cap \ro_{3_n} = (\ro_3 \cap \ro_{3_1} \cap \cdots \cap \ro_{3_n}) \cap \ro_4$,
            
            we have $\ts(\Gamma_1''' + \sum_{i=1}^n \Gamma_i[y/y_1]) \equiv (\Gamma', y:\ro_1 \cap \ro_2)$.\\
            
            Let $\Gamma'' = \ts(\Gamma_1''' + \sum_{i=1}^n \Gamma_i[y/y_1])$.\\
            
            Also, $M[y/y_1, y/y_2] = M_1[y/y_1, y/y_2] M_2[y/y_1, y/y_2]$, so
            \begin{align*}
                \msf{T}(M[y/y_1, y/y_2])
                &= \msf{T}(M_1[y/y_1, y/y_2] M_2[y/y_1, y/y_2])\\
                &= (\Gamma'', \ts(\rt_1'))\\
                &= (\Gamma'', \rt),
            \end{align*}
        
            with $\Gamma'' \equiv (\Gamma', y:\ro_1 \cap \ro_2)$.\\

            \item $y_1, y_2 \in \dom(\Gamma_1')$, $y_1 \notin \dom(\Gamma_i)$ and $y_2 \in \dom(\Gamma_i)$:\\
            
            Analogous to the previous case.\\

            \item $y_1, y_2 \in \dom(\Gamma_i)$, $y_1 \in \dom(\Gamma_1')$ and $y_2 \notin \dom(\Gamma_1')$:\\
            
            Analogous to the previous case.\\

            \item $y_1, y_2 \in \dom(\Gamma_i)$, $y_1 \notin \dom(\Gamma_1')$ and $y_2 \in \dom(\Gamma_1')$:\\
            
            Analogous to the previous case.\\

            \item $y_1 \in \dom(\Gamma_1')$, $y_2 \notin \dom(\Gamma_1')$, $y_1 \notin \dom(\Gamma_i)$ and $y_2 \in \dom(\Gamma_i)$:\\
            
            Since $y_1 \in \dom(\Gamma_1')$, by \autoref{xfreealg}, $y_1 \in \fv{M_1}$.
            
            So by \autoref{eqsubst1}, we have $\msf{T}(M_1[y/y_1]) = (\Gamma_1'[y/y_1], \rz_1' \cap \cdots \cap \rz_n' \rightarrow \rt_1')$.\\
            
            And since $y_2 \notin \dom(\Gamma_1')$, by \autoref{xfreealg}, $y_2 \notin \fv{M_1}$,
            
            so $M_1[y/y_1, y/y_2] = M_1[y/y_1]$
            
            and then
            \begin{equation}
                \msf{T}(M_1[y/y_1, y/y_2]) = \msf{T}(M_1[y/y_1]) = (\Gamma_1'[y/y_1], \rz_1' \cap \cdots \cap \rz_n' \rightarrow \rt_1'). \tag{1}\label{eq:2.b.8.1}
            \end{equation}
            
            \hfill
            
            Since $y_2 \in \dom(\Gamma_i)$, by \autoref{xfreealg}, $y_2 \in \fv{M_2}$.
            
            So by \autoref{eqsubst1}, we have $\msf{T}(M_2[y/y_2]) = (\Gamma_i[y/y_2], \rz_i)$.\\
            
            And since $y_1 \notin \dom(\Gamma_i)$, by \autoref{xfreealg}, $y_1 \notin \fv{M_2}$,
            
            so $M_2[y/y_1, y/y_2] = M_2[y/y_2]$
            
            and then
            \begin{equation}
                \msf{T}(M_2[y/y_1, y/y_2]) = \msf{T}(M_2[y/y_2]) = (\Gamma_i[y/y_2], \rz_i). \tag{2}\label{eq:2.b.8.2}
            \end{equation}
            
            \hfill
            
            So by rule 3.(b) of the inference algorithm (and \eqref{eq:2.b.8.1}, \eqref{eq:2.b.8.2}),
        
            $\msf{T}(M_1[y/y_1, y/y_2] M_2[y/y_1, y/y_2]) = (\ts(\Gamma_1'[y/y_1] + \sum_{i=1}^n \Gamma_i[y/y_2]), \ts(\rt_1'))$.\\
            
            Since $\ts(\Gamma_1' + \sum_{i=1}^n \Gamma_i) \equiv (\Gamma', y_1:\ro_1, y_2:\ro_2)$,
            
            we have $\ts(\Gamma_1'[y/y_1] + \sum_{i=1}^n \Gamma_i[y/y_2]) \equiv (\Gamma', y:\ro_1 \cap \ro_2)$.\\
            
            Let $\Gamma'' = \ts(\Gamma_1'[y/y_1] + \sum_{i=1}^n \Gamma_i[y/y_2])$.\\
            
            Also, $M[y/y_1, y/y_2] = M_1[y/y_1, y/y_2] M_2[y/y_1, y/y_2]$, so
            \begin{align*}
                \msf{T}(M[y/y_1, y/y_2])
                &= \msf{T}(M_1[y/y_1, y/y_2] M_2[y/y_1, y/y_2])\\
                &= (\Gamma'', \ts(\rt_1'))\\
                &= (\Gamma'', \rt),
            \end{align*}
        
            with $\Gamma'' \equiv (\Gamma', y:\ro_1 \cap \ro_2)$.\\

            \item $y_1 \notin \dom(\Gamma_1')$, $y_2 \in \dom(\Gamma_1')$, $y_1 \in \dom(\Gamma_i)$ and $y_2 \notin \dom(\Gamma_i)$:\\
            
            Analogous to the previous case.\\
        \end{enumerate}

        \item $(\Gamma, \rt) = (\ts(\Gamma_1 + \Gamma_2), \ts(\rt_1))$, with $\Gamma \equiv (\Gamma', y_1:\ro_1, y_2:\ro_2)$, where $\msf{T}(M_1) = (\Gamma_1, \rz \multimap \rt_1)$, $\msf{T}(M_2) = (\Gamma_2, \rz_2)$ and $\ts = \msf{UNIFY}(\{\rz_2 = \rz\})$.\\
        
        Because $y$ does not occur in $M$, then $y$ does not occur in $M_1$ nor in $M_2$.\\
        
        Since $\ts(\Gamma_1 + \Gamma_2) \equiv (\Gamma', y_1:\ro_1, y_2:\ro_2)$, then there are nine possible cases regarding the presence of $y_1$ and $y_2$ in $\dom(\Gamma_1)$ and $\dom(\Gamma_2)$:
        
        \begin{enumerate}
            \item $y_1, y_2 \in \dom(\Gamma_1)$ and $y_1, y_2 \notin \dom(\Gamma_2)$:\\
        
            So $\Gamma_1 \equiv (\Gamma_1', y_1:\ro_3, y_2:\ro_4)$ (for some $\ro_3, \ro_4$ such that $\ts(\ro_3) = \ro_1$ and $\ts(\ro_4) = \ro_2$).\\
            
            By induction,
            \begin{equation}
                \msf{T}(M_1[y/y_1, y/y_2]) = (\Gamma_1'', \rz \multimap \rt_1), \tag{1}\label{eq:2.c.1.1}
            \end{equation}
            with $\Gamma_1'' \equiv (\Gamma_1', y:\ro_3 \cap \ro_4)$.\\
            
            And since $y_1, y_2 \notin \dom(\Gamma_2)$, by \autoref{xfreealg}, $y_1, y_2 \notin \fv{M_2}$,
            
            so $M_2[y/y_1, y/y_2] = M_2$
            
            and then
            \begin{equation}
                \msf{T}(M_2[y/y_1, y/y_2]) = \msf{T}(M_2) = (\Gamma_2, \rz_2). \tag{2}\label{eq:2.c.1.2}
            \end{equation}
            
            \hfill
            
            So by rule 3.(c) of the inference algorithm (and \eqref{eq:2.c.1.1}, \eqref{eq:2.c.1.2}),
        
            $\msf{T}(M_1[y/y_1, y/y_2] M_2[y/y_1, y/y_2]) = (\ts(\Gamma_1'' + \Gamma_2), \ts(\rt_1))$.\\
            
            Since $\Gamma_1'' \equiv (\Gamma_1', y:\ro_3 \cap \ro_4)$, $\Gamma_1 \equiv (\Gamma_1', y_1:\ro_3, y_2:\ro_4)$ and $\ts(\Gamma_1 + \Gamma_2) \equiv (\Gamma', y_1:\ro_1, y_2:\ro_2)$,
            
            we have $\ts(\Gamma_1'' + \Gamma_2) \equiv (\Gamma', y:\ro_1 \cap \ro_2)$.\\
            
            Let $\Gamma'' = \ts(\Gamma_1'' + \Gamma_2)$.\\
            
            Also, $M[y/y_1, y/y_2] = M_1[y/y_1, y/y_2] M_2[y/y_1, y/y_2]$, so
            \begin{align*}
                \msf{T}(M[y/y_1, y/y_2])
                &= \msf{T}(M_1[y/y_1, y/y_2] M_2[y/y_1, y/y_2])\\
                &= (\Gamma'', \ts(\rt_1))\\
                &= (\Gamma'', \rt),
            \end{align*}
        
            with $\Gamma'' \equiv (\Gamma', y:\ro_1 \cap \ro_2)$.\\

            \item $y_1, y_2 \notin \dom(\Gamma_1)$ and $y_1, y_2 \in \dom(\Gamma_2)$:\\
            
            Analogous to the previous case.\\

            \item $y_1, y_2 \in \dom(\Gamma_1)$ and $y_1, y_2 \in \dom(\Gamma_2)$:\\
            
            So $\Gamma_1 \equiv (\Gamma_1', y_1:\ro_3, y_2:\ro_4)$ and $\Gamma_2 \equiv (\Gamma_2', y_1:\ro_3', y_2:\ro_4')$ (for some $\ro_3, \ro_4, \ro_3', \ro_4'$ such that $\ts(\ro_3 \cap \ro_3') = \ro_1$ and $\ts(\ro_4 \cap \ro_4') = \ro_2$).\\
            
            By induction,
            \begin{equation}
                \msf{T}(M_1[y/y_1, y/y_2]) = (\Gamma_1'', \rz \multimap \rt_1), \tag{1}\label{eq:2.c.3.1}
            \end{equation}
            with $\Gamma_1'' \equiv (\Gamma_1', y:\ro_3 \cap \ro_4)$;
            
            \begin{equation}
                \msf{T}(M_2[y/y_1, y/y_2]) = (\Gamma_2'', \rz_2), \tag{2}\label{eq:2.c.3.2}
            \end{equation}
            with $\Gamma_2'' \equiv (\Gamma_2', y:\ro_3' \cap \ro_4')$.\\
            
            So by rule 3.(c) of the inference algorithm (and \eqref{eq:2.c.3.1}, \eqref{eq:2.c.3.2}),
        
            $\msf{T}(M_1[y/y_1, y/y_2] M_2[y/y_1, y/y_2]) = (\ts(\Gamma_1'' + \Gamma_2''), \ts(\rt_1))$.\\
            
            Since $\Gamma_1'' \equiv (\Gamma_1', y:\ro_3 \cap \ro_4)$, $\Gamma_1 \equiv (\Gamma_1', y_1:\ro_3, y_2:\ro_4)$, $\Gamma_2'' \equiv (\Gamma_2', y:\ro_3' \cap \ro_4')$, $\Gamma_2 \equiv (\Gamma_2', y_1:\ro_3', y_2:\ro_4')$, $\ts(\Gamma_1 + \Gamma_2) \equiv (\Gamma', y_1:\ro_1, y_2:\ro_2)$ and $(\ro_3 \cap \ro_4) \cap (\ro_3' \cap \ro_4') = (\ro_3 \cap \ro_3') \cap (\ro_4 \cap \ro_4')$,
            
            we have $\ts(\Gamma_1'' + \Gamma_2'') \equiv (\Gamma', y:\ro_1 \cap \ro_2)$.\\
            
            Let $\Gamma'' = \ts(\Gamma_1'' + \Gamma_2'')$.\\
            
            Also, $M[y/y_1, y/y_2] = M_1[y/y_1, y/y_2] M_2[y/y_1, y/y_2]$, so
            \begin{align*}
                \msf{T}(M[y/y_1, y/y_2])
                &= \msf{T}(M_1[y/y_1, y/y_2] M_2[y/y_1, y/y_2])\\
                &= (\Gamma'', \ts(\rt_1))\\
                &= (\Gamma'', \rt),
            \end{align*}
        
            with $\Gamma'' \equiv (\Gamma', y:\ro_1 \cap \ro_2)$.\\

            \item $y_1, y_2 \in \dom(\Gamma_1)$, $y_1 \in \dom(\Gamma_2)$ and $y_2 \notin \dom(\Gamma_2)$:\\
            
            So $\Gamma_1 \equiv (\Gamma_1', y_1:\ro_3, y_2:\ro_4)$ and $\Gamma_2 \equiv (\Gamma_2', y_1:\ro_3')$ (for some $\ro_3, \ro_4, \ro_3'$ such that $\ts(\ro_3 \cap \ro_3') = \ro_1$ and $\ts(\ro_4) = \ro_2$).\\
            
            By induction,
            \begin{equation}
                \msf{T}(M_1[y/y_1, y/y_2]) = (\Gamma_1'', \rz \multimap \rt_1), \tag{1}\label{eq:2.c.4.1}
            \end{equation}
            with $\Gamma_1'' \equiv (\Gamma_1', y:\ro_3 \cap \ro_4)$.\\
            
            Since $y_1 \in \dom(\Gamma_2)$, by \autoref{xfreealg}, $y_1 \in \fv{M_2}$.
            
            So by \autoref{eqsubst1}, we have $\msf{T}(M_2[y/y_1]) = (\Gamma_2[y/y_1], \rz_2)$.\\
            
            And since $y_2 \notin \dom(\Gamma_2)$, by \autoref{xfreealg}, $y_2 \notin \fv{M_2}$,
            
            so $M_2[y/y_1, y/y_2] = M_2[y/y_1]$
            
            and then
            \begin{equation}
                \msf{T}(M_2[y/y_1, y/y_2]) = \msf{T}(M_2[y/y_1]) = (\Gamma_2[y/y_1], \rz_2). \tag{2}\label{eq:2.c.4.2}
            \end{equation}
            
            \hfill
            
            So by rule 3.(c) of the inference algorithm (and \eqref{eq:2.c.4.1}, \eqref{eq:2.c.4.2}),
        
            $\msf{T}(M_1[y/y_1, y/y_2] M_2[y/y_1, y/y_2]) = (\ts(\Gamma_1'' + \Gamma_2[y/y_1]), \ts(\rt_1))$.\\
            
            Since $\Gamma_1'' \equiv (\Gamma_1', y:\ro_3 \cap \ro_4)$, $\Gamma_1 \equiv (\Gamma_1', y_1:\ro_3, y_2:\ro_4)$, $\Gamma_2[y/y_1] \equiv (\Gamma_2', y:\ro_3')$, $\Gamma_2 \equiv (\Gamma_2', y_1:\ro_3')$, $\ts(\Gamma_1 + \Gamma_2) \equiv (\Gamma', y_1:\ro_1, y_2:\ro_2)$ and $(\ro_3 \cap \ro_4) \cap \ro_3' = (\ro_3 \cap \ro_3') \cap \ro_4$,
            
            we have $\ts(\Gamma_1'' + \Gamma_2[y/y_1]) \equiv (\Gamma', y:\ro_1 \cap \ro_2)$.\\
            
            Let $\Gamma'' = \ts(\Gamma_1'' + \Gamma_2[y/y_1])$.\\
            
            Also, $M[y/y_1, y/y_2] = M_1[y/y_1, y/y_2] M_2[y/y_1, y/y_2]$, so
            \begin{align*}
                \msf{T}(M[y/y_1, y/y_2])
                &= \msf{T}(M_1[y/y_1, y/y_2] M_2[y/y_1, y/y_2])\\
                &= (\Gamma'', \ts(\rt_1))\\
                &= (\Gamma'', \rt),
            \end{align*}
        
            with $\Gamma'' \equiv (\Gamma', y:\ro_1 \cap \ro_2)$.\\

            \item $y_1, y_2 \in \dom(\Gamma_1)$, $y_1 \notin \dom(\Gamma_2)$ and $y_2 \in \dom(\Gamma_2)$:\\
            
            Analogous to the previous case.\\

            \item $y_1, y_2 \in \dom(\Gamma_2)$, $y_1 \in \dom(\Gamma_1)$ and $y_2 \notin \dom(\Gamma_1)$:\\
            
            Analogous to the previous case.\\

            \item $y_1, y_2 \in \dom(\Gamma_2)$, $y_1 \notin \dom(\Gamma_1)$ and $y_2 \in \dom(\Gamma_1)$:\\
            
            Analogous to the previous case.\\

            \item $y_1 \in \dom(\Gamma_1)$, $y_2 \notin \dom(\Gamma_1)$, $y_1 \notin \dom(\Gamma_2)$ and $y_2 \in \dom(\Gamma_2)$:\\
            
            Since $y_1 \in \dom(\Gamma_1)$, by \autoref{xfreealg}, $y_1 \in \fv{M_1}$.
            
            So by \autoref{eqsubst1}, we have $\msf{T}(M_1[y/y_1]) = (\Gamma_1[y/y_1], \rz \multimap \rt_1)$.\\
            
            And since $y_2 \notin \dom(\Gamma_1)$, by \autoref{xfreealg}, $y_2 \notin \fv{M_1}$,
            
            so $M_1[y/y_1, y/y_2] = M_1[y/y_1]$
            
            and then
            \begin{equation}
                \msf{T}(M_1[y/y_1, y/y_2]) = \msf{T}(M_1[y/y_1]) = (\Gamma_1[y/y_1], \rz \multimap \rt_1). \tag{1}\label{eq:2.c.8.1}
            \end{equation}
            
            \hfill
            
            Since $y_2 \in \dom(\Gamma_2)$, by \autoref{xfreealg}, $y_2 \in \fv{M_2}$.
            
            So by \autoref{eqsubst1}, we have $\msf{T}(M_2[y/y_2]) = (\Gamma_2[y/y_2], \rz_2)$.\\
            
            And since $y_1 \notin \dom(\Gamma_2)$, by \autoref{xfreealg}, $y_1 \notin \fv{M_2}$,
            
            so $M_2[y/y_1, y/y_2] = M_2[y/y_2]$
            
            and then
            \begin{equation}
                \msf{T}(M_2[y/y_1, y/y_2]) = \msf{T}(M_2[y/y_2]) = (\Gamma_2[y/y_2], \rz_2). \tag{2}\label{eq:2.c.8.2}
            \end{equation}
            
            \hfill
            
            So by rule 3.(c) of the inference algorithm (and \eqref{eq:2.c.8.1}, \eqref{eq:2.c.8.2}),
        
            $\msf{T}(M_1[y/y_1, y/y_2] M_2[y/y_1, y/y_2]) = (\ts(\Gamma_1[y/y_1] + \Gamma_2[y/y_2]), \ts(\rt_1))$.\\
            
            Since $\ts(\Gamma_1 + \Gamma_2) \equiv (\Gamma', y_1:\ro_1, y_2:\ro_2)$,
            
            we have $\ts(\Gamma_1[y/y_1] + \Gamma_2[y/y_2]) \equiv (\Gamma', y:\ro_1 \cap \ro_2)$.\\
            
            Let $\Gamma'' = \ts(\Gamma_1[y/y_1] + \Gamma_2[y/y_2])$.\\
            
            Also, $M[y/y_1, y/y_2] = M_1[y/y_1, y/y_2] M_2[y/y_1, y/y_2]$, so
            \begin{align*}
                \msf{T}(M[y/y_1, y/y_2])
                &= \msf{T}(M_1[y/y_1, y/y_2] M_2[y/y_1, y/y_2])\\
                &= (\Gamma'', \ts(\rt_1))\\
                &= (\Gamma'', \rt),
            \end{align*}
        
            with $\Gamma'' \equiv (\Gamma', y:\ro_1 \cap \ro_2)$.\\

            \item $y_1 \notin \dom(\Gamma_1)$, $y_2 \in \dom(\Gamma_1)$, $y_1 \in \dom(\Gamma_2)$ and $y_2 \notin \dom(\Gamma_2)$:\\
            
            Analogous to the previous case.
        \end{enumerate}
    \end{enumerate}
\end{enumerate}

Any other possible case makes the left side of the implication ($\msf{T}(M) = (\Gamma, \rt)$, with $\Gamma \equiv (\Gamma', y_1:\ro_1, y_2:\ro_2)$, and $y$ does not occur in  $M$) false, which makes the statement true.
\end{proof}
\end{lemma}

\begin{theorem}[Completeness]\label{completeness} 
If $\Phi \rhd \inftwo{ \Gamma }{ M }{ \rt }$, then $\msf{T}(M) = (\Gamma', \rt')$ (for some environment $\Gamma'$ and type $\rt'$) and there is a substitution $\ts$ such that $\ts(\rt') = \rt$ and $\ts(\Gamma') \equiv \Gamma$.

\begin{proof}
By induction on $|\Phi|$.

\begin{enumerate}
    \item \tul{(Axiom)}: Then $\Gamma = [x:\rz]$, $M = x$ and $\rt = \rz$.\\
    
    $\msf{T}(M) = ([x:\alpha], \alpha)$ and let $\ts = [\rz/\alpha]$.\\
    
    Then $\ts(\rt') = \ts(\alpha) = \tsub{\alpha}{\rz/\alpha} = \rz = \rt$
    
    and $\ts(\Gamma') = \ts([x:\alpha]) = [x:\tsub{\alpha}{\rz/\alpha}] = [x:\rz] = \Gamma \equiv \Gamma$.\\

    \item \tul{(Exchange)}: Then $\Gamma = (\Gamma_1, y:\ro_2, x:\ro_1, \Gamma_2)$, $M = M_1$, $\rt = \rt_1$, and assuming that the premise $\inftwo{ \Gamma_1, x:\ro_1, y:\ro_2, \Gamma_2 }{ M_1 }{ \rt_1 }$ holds.\\
    
    By the induction hypothesis, $\msf{T}(M_1) = (\Gamma'', \rt'')$
    
    and there is a substitution $\ts'$ such that $\ts'(\rt'') = \rt_1$ and $\ts'(\Gamma'') \equiv (\Gamma_1, x:\ro_1, y:\ro_2, \Gamma_2)$.\\
    
    By definition, $(\Gamma_1, x:\ro_1, y:\ro_2, \Gamma_2) \equiv (\Gamma_1, y:\ro_2, x:\ro_1, \Gamma_2)$.
    
    So $\ts'(\Gamma'') \equiv (\Gamma_1, x:\ro_1, y:\ro_2, \Gamma_2) \equiv (\Gamma_1, y:\ro_2, x:\ro_1, \Gamma_2) = \Gamma$.\\
    
    And $\msf{T}(M) = \msf{T}(M_1)$ and $\ts'(\rt'') = \rt_1 = \rt$.\\
    
    So for $\Gamma' = \Gamma''$, $\rt' = \rt''$ and $\ts = \ts'$,
    
    we have $\msf{T}(M) = (\Gamma', \rt')$, $\ts(\rt') = \rt$ and $\ts(\Gamma') \equiv \Gamma$.\\

    \item \tul{(Contraction)}: Then $\Gamma = (\Gamma_1, x:\ro_1 \cap \ro_2, \Gamma_2)$, $M = M_1 [x/x_1, x/x_2]$, $\rt = \rt_1$, and assuming that the premise $\inftwo{ \Gamma_1, x_1:\ro_1, x_2:\ro_2, \Gamma_2 }{ M_1 }{ \rt_1 }$ holds.\\
    
    By the induction hypothesis, $\msf{T}(M_1) = (\Gamma'', \rt'')$
    
    and there is a substitution $\ts'$ such that $\ts'(\rt'') = \rt_1$ and $\ts'(\Gamma'') \equiv (\Gamma_1, x_1:\ro_1, x_2:\ro_2, \Gamma_2)$.\\
    
    Because $\ts'(\Gamma'') \equiv (\Gamma_1, x_1:\ro_1, x_2:\ro_2, \Gamma_2)$, then $\dom(\Gamma'') = \dom(\Gamma_1, x_1:\ro_1, x_2:\ro_2, \Gamma_2)$.
    
    So $\Gamma'' \equiv (\Gamma_3, x_1:\ro_1', x_2:\ro_2')$ for some environment $\Gamma_3$ and types $\ro_1'$, $\ro_2'$ such that $\ts'(\ro_1') = \ro_1$, $\ts'(\ro_2') = \ro_2$ and $\ts'(\Gamma_3) \equiv (\Gamma_1, \Gamma_2)$.\\
    
    Then by \autoref{eqsubst2} ($x$ does not occur in $M_1$),
    
    $\msf{T}(M_1[x/x_1, x/x_2]) = (\Gamma_1'', \rt'')$, with $\Gamma_1'' \equiv (\Gamma_3, x:\ro_1' \cap \ro_2')$.
    
    And $\ts'(\Gamma_3, x:\ro_1' \cap \ro_2') \equiv (\Gamma_1, \Gamma_2, x:\ro_1 \cap \ro_2) \equiv (\Gamma_1, x:\ro_1 \cap \ro_2, \Gamma_2) = \Gamma$.\\
    
    Also, $M = M_1[x/x_1, x/x_2]$, so $\msf{T}(M) = \msf{T}(M_1[x/x_1, x/x_2]) = (\Gamma_1'', \rt'')$.\\
    
    So for $\Gamma' = \Gamma_1''$, $\rt' = \rt''$ and $\ts = \ts'$, we have $\msf{T}(M) = (\Gamma', \rt')$, $\ts(\rt') = \rt$ and $\ts(\Gamma') \equiv \Gamma$.\\

    \item \tul{($\rightarrow$ Intro)}: Then $\Gamma = \Gamma_1$, $M = \lambda x.M_1$, $\rt = \rz_1 \cap \cdots \cap \rz_n \rightarrow \rt_1 $, and assuming that the premise $\inftwo{ \Gamma_1, x:\rz_1 \cap \cdots \cap \rz_n }{ M_1 }{ \rt_1 }$ (with $n \geq 2$) holds.\\
    
    By the induction hypothesis, $\msf{T}(M_1) = (\Gamma'', \rt'')$
    
    and there is a substitution $\ts'$ such that $\ts'(\rt'') = \rt_1$ and $\ts'(\Gamma'') \equiv (\Gamma_1, x:\rz_1 \cap \cdots \cap \rz_n)$.\\
    
    There is only one possible case for $\msf{T}(M)$:
    \begin{itemize}
        \item $(x:\rz_1' \cap \cdots \cap \rz_{m}') \in \Gamma''$ (with $m \geq 2$). Then $\msf{T}(M) = ({\Gamma''}_x, \rz_1' \cap \cdots \cap \rz_{m}' \rightarrow \rt'')$.\\
        
        By $\ts'(\Gamma'') \equiv (\Gamma_1, x:\rz_1 \cap \cdots \cap \rz_n)$ and the assumption that $(x:\rz_1' \cap \cdots \cap \rz_{m}') \in \Gamma''$,
        
        we have $\ts'(\rz_1' \cap \cdots \cap \rz_{m}') = \rz_1 \cap \cdots \cap \rz_n$.\\
        
        Then by that and by $\ts'(\rt'') = \rt_1$,
        
        we have $\ts'(\rz_1' \cap \cdots \cap \rz_{m}' \rightarrow \rt'') = \rz_1 \cap \cdots \cap \rz_n \rightarrow \rt_1$.\\
        
        By $\ts'(\Gamma'') \equiv (\Gamma_1, x:\rz_1 \cap \cdots \cap \rz_n)$ and the definition of environment, ${\ts'(\Gamma''}_x) \equiv \Gamma_1$.\\
        
        So for $\Gamma' = {\Gamma''}_x$, $\rt' = \rz_1' \cap \cdots \cap \rz_{m}' \rightarrow \rt''$ and $\ts = \ts'$, we have $\msf{T}(M) = (\Gamma', \rt')$, $\ts(\rt') = \rt$ and $\ts(\Gamma') \equiv \Gamma$.
    \end{itemize}
    
    Note that there is not the case where $x \notin \dom(\Gamma'')$ because by $\inftwo{ \Gamma_1, x:\rz_1 \cap \cdots \cap \rz_n }{ M_1 }{ \rt_1 }$ (with $n \geq 2$) and \autoref{equaldom}, $x \in \dom(\Gamma'')$.
    
    There is also not the case where $(x:\rz) \in \Gamma''$, as the substitution $\ts'$ could not exist (because there is no substitution $\ts'$ such that $\ts'(\rz) = \rz_1 \cap \cdots \cap \rz_n$).\\

    \item \tul{($\rightarrow$ Elim)}: Then $\Gamma = (\Gamma_0, \sum_{i=1}^n \Gamma_i)$, $M = M_1 M_2$, $\rt = \rt_1$, and assuming that the premises $\inftwo{ \Gamma_0 }{ M_1 }{ \rz_1 \cap \cdots \cap \rz_n \rightarrow \rt_1 }$ and $\inftwo{ \Gamma_i }{ M_2 }{ \rz_i }$, for $1 \leq i \leq n$ (with $n \geq 2$), hold.\\
    
    By the induction hypothesis,
    \begin{itemize}
        \item $\msf{T}(M_1) = (\Gamma_0', \rt_0')$ and there is a substitution $\ts_0'$ such that $\ts_0'(\rt_0') = \rz_1 \cap \cdots \cap \rz_n \rightarrow \rt_1$ and $\ts_0'(\Gamma_0') \equiv \Gamma_0$;
        
        \item $\msf{T}(M_2) = (\Gamma_i', \rt_i')$ and there are substitutions $\ts_i'$ such that $\ts_i'(\rt_i') = \rz_i$ and $\ts_i'(\Gamma_i') \equiv \Gamma_i$, for $1 \leq i \leq n$.
    \end{itemize}
    
    There is only one possible case for $\msf{T}(M)$:
    \begin{itemize}
        \item $\rt_0' = \rz_1' \cap \cdots \cap \rz_n' \rightarrow \rt_3$ and, for each $1 \leq i \leq n$, $\rt_i' = \rz_i''$:\\
        
        Let $P = \{\rz_i' = \rz_i'' \mid 1 \leq i \leq n\}$.\\
        
        Let us assume, without loss of generality, that $\Gamma_0', \rt_0', \ts_0'$ and all $\Gamma_i', \rt_i', \ts_i'$ do not have type variables in common (if they did, we could simply rename the type variables in each of the $\Gamma_i', \rt_i', \ts_i'$ to fresh type variables and we would have the same result, as we consider types equal up to renaming of variables).\\
        
        We have $\ts_i'(\rt_i') = \rz_i$ and $\rt_i' = \rz_i''$, so $\ts_i'(\rz_i'') = \rz_i$, for each $1 \leq i \leq n$.\\
        
        And $\ts_0'(\rt_0') = \rz_1 \cap \cdots \cap \rz_n \rightarrow \rt_1$ and $\rt_0' = \rz_1' \cap \cdots \cap \rz_n' \rightarrow \rt_3$,
        
        so $\ts_0'(\rz_1' \cap \cdots \cap \rz_n' \rightarrow \rt_3) = \rz_1 \cap \cdots \cap \rz_n \rightarrow \rt_1$.
        
        Equivalently, $\ts_0'(\rz_1') \cap \cdots \cap \ts_0'(\rz_n') \rightarrow \ts_0'(\rt_3) = \rz_1 \cap \cdots \cap \rz_n \rightarrow \rt_1$.
        
        So $\ts_0'(\rt_3) = \rt_1$ and $\ts_0'(\rz_i') = \rz_i$, for each $1 \leq i \leq n$.\\
        
        Then $\ts_3 = \ts_0' \cup \ts_1' \cup \cdots \cup \ts_n'$ is a solution to $P$:
        
        for all $1 \leq i \leq n$, $\ts_3(\rz_i') = \ts_0'(\rz_i') = \rz_i = \ts_i'(\rz_i'') = \ts_3(\rz_i'')$.\\
        
        Let $\ts' = \msf{UNIFY}(P)$.
        
        Then we have $\msf{T}(M) = (\ts'(\Gamma_0' + \sum_{i=1}^n \Gamma_i'), \ts'(\rt_3))$, given by the algorithm.\\
        
        By \autoref{mgu} of most general unifier, there exists an $\ts$ such that
        \begin{equation}
            (\ts(\ts'(\Gamma_0' + \sum_{i=1}^n \Gamma_i')), \ts(\ts'(\rt_3))) = (\ts_3(\Gamma_0' + \sum_{i=1}^n \Gamma_i'), \ts_3(\rt_3)). \tag{1}\label{eq:comp.5}
        \end{equation}
        And $(\ts(\ts'(\Gamma_0' + \sum_{i=1}^n \Gamma_i')), \ts(\ts'(\rt_3)))$ is also a solution to $\msf{T}(M)$.\\
        
        We have $\dom(\Gamma_0) \cap \dom(\Gamma_i) = \emptyset$, for all $1 \leq i \leq n$ (otherwise $\Gamma = \Gamma_0, \sum_{i=1}^n \Gamma_i$ would be inconsistent),
        
        so $\Gamma_0, \sum_{i=1}^n \Gamma_i \equiv \Gamma_0 + \sum_{i=1}^n \Gamma_i$.\\
        
        Because of that and our initial assumption that $\Gamma_0', \rt_0', \ts_0'$ and all $\Gamma_i', \rt_i', \ts_i'$ do not have type variables in common, we have $\ts_0'(\Gamma_0') + \sum_{i=1}^n \ts_i'(\Gamma_i') \equiv \Gamma_0, \sum_{i=1}^n \Gamma_i$.\\
        
        And $\ts_3(\Gamma_0' + \sum_{i=1}^n \Gamma_i') = \ts_0'(\Gamma_0') + \sum_{i=1}^n \ts_i'(\Gamma_i')$,
        
        so $\ts_3(\Gamma_0' + \sum_{i=1}^n \Gamma_i') \equiv \Gamma_0, \sum_{i=1}^n \Gamma_i$,
        
        which, by \eqref{eq:comp.5}, is equivalent to $\ts(\ts'(\Gamma_0' + \sum_{i=1}^n \Gamma_i')) \equiv \Gamma_0, \sum_{i=1}^n \Gamma_i$.\\
        
        Finally, we have $\ts_3(\rt_3) = \ts_0'(\rt_3)$ and $\ts_0'(\rt_3) = \rt_1$,
        
        so $\ts_3(\rt_3) = \rt_1$,
        
        which, by \eqref{eq:comp.5}, is equivalent to $\ts(\ts'(\rt_3)) = \rt_1$.\\
        
        So for $\Gamma' = \ts'(\Gamma_0' + \sum_{i=1}^n \Gamma_i')$ and $\rt' = \ts'(\rt_3)$, we have $\msf{T}(M) = (\Gamma', \rt')$ and there is an $\ts$ such that $\ts(\rt') = \rt$ and $\ts(\Gamma') \equiv \Gamma$.
    \end{itemize}
    
    Note that there is not the case where $\rt_0' = \rz_1' \cap \cdots \cap \rz_m' \rightarrow \rt_3$ with $m \neq n$, as the substitution $\ts_0'$ could not exist (because there is no substitution $\ts_0'$ such that $\ts_0'(\rz_1' \cap \cdots \cap \rz_m' \rightarrow \rt_3) = \rz_1 \cap \cdots \cap \rz_n \rightarrow \rt_1$, for $m \neq n$).
    
    There is also not the case where $\rt_0' = \rz' \multimap \rt_3$ nor the case where $\rt_0' = \alpha$ as the substitution $\ts_0'$ (such that $\ts_0'(\rt_0') = \rz_1 \cap \cdots \cap \rz_n \rightarrow \rt_1$) could not exist.\\

    \item \tul{($\multimap$ Intro)}: Then $\Gamma = \Gamma_1$, $M = \lambda x.M_1$, $\rt = \rz \multimap \rt_1 $, and assuming that the premise $\inftwo{ \Gamma_1, x:\rz }{ M_1 }{ \rt_1 }$ holds.\\
    
    By the induction hypothesis, $\msf{T}(M_1) = (\Gamma'', \rt'')$
    
    and there is a substitution $\ts'$ such that $\ts'(\rt'') = \rt_1$ and $\ts'(\Gamma'') \equiv (\Gamma_1, x:\rz)$.\\
    
    There is only one possible case for $\msf{T}(M)$:
    \begin{itemize}
        \item $(x:\rz') \in \Gamma''$. Then $\msf{T}(M) = ({\Gamma''}_x, \rz' \multimap \rt'')$.\\
        
        By $\ts'(\Gamma'') \equiv (\Gamma_1, x:\rz)$ and the assumption that $(x:\rz') \in \Gamma''$,
        
        we have $\ts'(\rz') = \rz$.\\
        
        Then by that and by $\ts'(\rt'') = \rt_1$,
        
        we have $\ts'(\rz' \multimap \rt'') = \rz \multimap \rt_1$.\\
        
        By $\ts'(\Gamma'') \equiv (\Gamma_1, x:\rz)$ and the definition of environment, ${\ts'(\Gamma''}_x) \equiv \Gamma_1$.\\
        
        So for $\Gamma' = {\Gamma''}_x$, $\rt' = \rz' \multimap \rt''$ and $\ts = \ts'$, we have $\msf{T}(M) = (\Gamma', \rt')$, $\ts(\rt') = \rt$ and $\ts(\Gamma') \equiv \Gamma$.
    \end{itemize}
    
    Note that there is not the case where $x \notin \dom(\Gamma'')$ because by $\inftwo{ \Gamma_1, x:\rz }{ M_1 }{ \rt_1 }$ and \autoref{equaldom}, $x \in \dom(\Gamma'')$.
    
    There is also not the case where $(x:\rz_1 \cap \cdots \cap \rz_n) \in \Gamma''$, as the substitution $\ts'$ could not exist (because there is no substitution $\ts'$ such that $\ts'(\rz_1 \cap \cdots \cap \rz_n) = \rz$).\\

    \item \tul{($\multimap$ Elim)}: Then $\Gamma = (\Gamma_1, \Gamma_2)$, $M = M_1 M_2$, $\rt = \rt_1$, and assuming that the premises $\inftwo{ \Gamma_1 }{ M_1 }{ \rz \multimap \rt_1 }$ and $\inftwo{ \Gamma_2 }{ M_2 }{ \rz }$ hold.\\
    
    By the induction hypothesis,
    \begin{itemize}
        \item $\msf{T}(M_1) = (\Gamma_1', \rt_1')$ and there is a substitution $\ts_1'$ such that $\ts_1'(\rt_1') = \rz \multimap \rt_1$ and $\ts_1'(\Gamma_1') \equiv \Gamma_1$;
        
        \item $\msf{T}(M_2) = (\Gamma_2', \rt_2')$ and there is a substitution $\ts_2'$ such that $\ts_2'(\rt_2') = \rz$ and $\ts_2'(\Gamma_2') \equiv \Gamma_2$.
    \end{itemize}
    
    There are two possible cases for $\msf{T}(M)$:
    \begin{itemize}
        \item $\rt_1' = \alpha_1$ and $\rt_2' = \rz_2$:\\
        
        Let $P = \{\alpha_1 = \alpha_2 \multimap \alpha_3, \rz_2 = \alpha_2\}$, where $\alpha_2,\alpha_3$ are fresh variables.\\
        
        Let us assume, without loss of generality, that $\Gamma_1', \rt_1', \ts_1'$ and $\Gamma_2', \rt_2', \ts_2'$ do not have type variables in common (if they did, we could simply rename the type variables in $\Gamma_2', \rt_2', \ts_2'$ to fresh type variables and we would have the same result, as we consider types equal up to renaming of variables).\\
        
        We have $\ts_2'(\rt_2') = \rz$ and $\rt_2' = \rz_2$, so $\ts_2'(\rz_2) = \rz$.\\
        
        And $\ts_1'(\rt_1') = \rz \multimap \rt_1$ and $\rt_1' = \alpha_1$, so $\ts_1'(\alpha_1) = \rz \multimap \rt_1$.\\
        
        Then $\ts_3 = \ts_1' \cup \ts_2' \cup [\rz/\alpha_2, \rt_1/\alpha_3]$ is a solution to $P$:
        
        \begin{itemize}
            \item $\ts_3(\alpha_1) = \ts_1'(\alpha_1) = \rz \multimap \rt_1 = \tsub{(\alpha_2 \multimap \alpha_3)}{\rz/\alpha_2, \rt_1/\alpha_3} = \ts_3(\alpha_2 \multimap \alpha_3)$;
            
            \item $\ts_3(\rz_2) = \ts_2'(\rz_2) = \rz = \tsub{\alpha_2}{\rz/\alpha_2, \rt_1/\alpha_3} = \ts_3(\alpha_2)$.\\
        \end{itemize}
        
        Let $\ts' = \msf{UNIFY}(P)$.
        
        Then we have $\msf{T}(M) = (\ts'(\Gamma_1' + \Gamma_2'), \ts'(\alpha_3))$, given by the algorithm.\\
        
        By \autoref{mgu} of most general unifier, there exists an $\ts$ such that
        \begin{equation}
            (\ts(\ts'(\Gamma_1' + \Gamma_2')), \ts(\ts'(\alpha_3))) = (\ts_3(\Gamma_1' + \Gamma_2'), \ts_3(\alpha_3)). \tag{1}\label{eq:comp.7.1}
        \end{equation}
        And $(\ts(\ts'(\Gamma_1' + \Gamma_2')), \ts(\ts'(\alpha_3)))$ is also a solution to $\msf{T}(M)$.\\
        
        We have $\dom(\Gamma_1) \cap \dom(\Gamma_2) = \emptyset$ (otherwise $\Gamma = \Gamma_1, \Gamma_2$ would be inconsistent),
        
        so $\Gamma_1, \Gamma_2 \equiv \Gamma_1 + \Gamma_2$.\\
        
        Because of that and our initial assumption that $\Gamma_1', \rt_1', \ts_1'$ and $\Gamma_2', \rt_2', \ts_2'$ do not have type variables in common, we have $\ts_1'(\Gamma_1') +  \ts_2'(\Gamma_2') \equiv \Gamma_1, \Gamma_2$.\\
        
        And $\ts_3(\Gamma_1' + \Gamma_2') = \ts_1'(\Gamma_1') +  \ts_2'(\Gamma_2')$,
        
        so $\ts_3(\Gamma_1' + \Gamma_2') \equiv \Gamma_1, \Gamma_2$,
        
        which, by \eqref{eq:comp.7.1}, is equivalent to $\ts(\ts'(\Gamma_1' + \Gamma_2')) \equiv \Gamma_1, \Gamma_2$.\\
        
        Finally, we have $\ts_3(\alpha_3) = \rt_1$,
        
        which, by \eqref{eq:comp.7.1}, is equivalent to $\ts(\ts'(\alpha_3)) = \rt_1$.\\
        
        So for $\Gamma' = \ts'(\Gamma_1' + \Gamma_2')$ and $\rt' = \ts'(\alpha_3)$, we have $\msf{T}(M) = (\Gamma', \rt')$ and there is an $\ts$ such that $\ts(\rt') = \rt$ and $\ts(\Gamma') \equiv \Gamma$.\\

        \item $\rt_1' = \rz' \multimap \rt_3$ and $\rt_2' = \rz_2$:\\
        
        Let $P = \{\rz_2 = \rz'\}$.\\
        
        Let us assume, without loss of generality, that $\Gamma_1', \rt_1', \ts_1'$ and $\Gamma_2', \rt_2', \ts_2'$ do not have type variables in common (if they did, we could simply rename the type variables in $\Gamma_2', \rt_2', \ts_2'$ to fresh type variables and we would have the same result, as we consider types equal up to renaming of variables).\\
        
        We have $\ts_2'(\rt_2') = \rz$ and $\rt_2' = \rz_2$, so $\ts_2'(\rz_2) = \rz$.\\
        
        And $\ts_1'(\rt_1') = \rz \multimap \rt_1$ and $\rt_1' = \rz' \multimap \rt_3$,
        
        so $\ts_1'(\rz' \multimap \rt_3) = \rz \multimap \rt_1$.
        
        Equivalently, $(\ts_1'(\rz')) \multimap (\ts_1'(\rt_3)) = \rz \multimap \rt_1$.
        
        So $\ts_1'(\rz') = \rz$ and $\ts_1'(\rt_3) = \rt_1$.\\
        
        Then $\ts_3 = \ts_1' \cup \ts_2'$ is a solution to $P$:
        
        $\ts_3(\rz_2) = \ts_2'(\rz_2) = \rz = \ts_1'(\rz') = \ts_3(\rz')$.\\
        
        Let $\ts' = \msf{UNIFY}(P)$.
        
        Then we have $\msf{T}(M) = (\ts'(\Gamma_1' + \Gamma_2'), \ts'(\rt_3))$, given by the algorithm.\\
        
        By \autoref{mgu} of most general unifier, there exists an $\ts$ such that
        \begin{equation}
            (\ts(\ts'(\Gamma_1' + \Gamma_2')), \ts(\ts'(\rt_3))) = (\ts_3(\Gamma_1' + \Gamma_2'), \ts_3(\rt_3)). \tag{1}\label{eq:comp.7.2}
        \end{equation}
        And $(\ts(\ts'(\Gamma_1' + \Gamma_2')), \ts(\ts'(\rt_3)))$ is also a solution to $\msf{T}(M)$.\\
        
        We have $\dom(\Gamma_1) \cap \dom(\Gamma_2) = \emptyset$ (otherwise $\Gamma = \Gamma_1, \Gamma_2$ would be inconsistent),
        
        so $\Gamma_1, \Gamma_2 \equiv \Gamma_1 + \Gamma_2$.\\
        
        Because of that and our initial assumption that $\Gamma_1', \rt_1', \ts_1'$ and $\Gamma_2', \rt_2', \ts_2'$ do not have type variables in common, we have $\ts_1'(\Gamma_1') +  \ts_2'(\Gamma_2') \equiv \Gamma_1, \Gamma_2$.\\
        
        And $\ts_3(\Gamma_1' + \Gamma_2') = \ts_1'(\Gamma_1') +  \ts_2'(\Gamma_2')$,
        
        so $\ts_3(\Gamma_1' + \Gamma_2') \equiv \Gamma_1, \Gamma_2$,
        
        which, by \eqref{eq:comp.7.2}, is equivalent to $\ts(\ts'(\Gamma_1' + \Gamma_2')) \equiv \Gamma_1, \Gamma_2$.\\
        
        Finally, we have $\ts_3(\rt_3) = \ts_1'(\rt_3)$ and $\ts_1'(\rt_3) = \rt_1$,
        
        so $\ts_3(\rt_3) = \rt_1$,
        
        which, by \eqref{eq:comp.7.2}, is equivalent to $\ts(\ts'(\rt_3)) = \rt_1$.\\
        
        So for $\Gamma' = \ts'(\Gamma_1' + \Gamma_2')$ and $\rt' = \ts'(\rt_3)$, we have $\msf{T}(M) = (\Gamma', \rt')$ and there is an $\ts$ such that $\ts(\rt') = \rt$ and $\ts(\Gamma') \equiv \Gamma$.
    \end{itemize}
    
    Note that there is not the case where $\rt_1' = \rz_1' \cap \cdots \cap \rz_n' \rightarrow \rt_3$, as the substitution $\ts_1'$ could not exist (because there is no substitution $\ts_1'$ such that $\ts_1'(\rz_1' \cap \cdots \cap \rz_n' \rightarrow \rt_3) = \rz \multimap \rt_1$).
\end{enumerate}
\end{proof}
\end{theorem}

Hence, we end up with a sound and complete type inference algorithm for the Linear Rank~2 Intersection Type System.

\subsection{Remarks}\label{sec:linearfinalremarks}

A $\lambda$-term $M$ is called a $\lambda I$-term if and only if, for each subterm of the form $\lambda x.N$ in $M$, $x$ occurs free in $N$ at least once. Note that our type system and type inference algorithm only type $\lambda I$-terms, but we could have extended them for the \emph{affine terms} -- a $\lambda$-term $M$ is affine if and only if, for each subterm of the form $\lambda x.N$ in $M$, $x$ occurs free in $N$ at most once, and if each free variable of $M$ has just one occurrence free in $M$.

There is no unique and final way of typing affine terms. For instance, in the systems in \cite{accattoli2018tight}, arguments that do not occur in the body of the function get the empty type $[\;]$. Since we do not allow the empty sequence in our definition and adding it would make the system more complex, we decided to only work with $\lambda I$-terms.\\

Regarding our choice of defining environments as lists and having the rules (Exchange) and (Contraction) in the type system, instead of defining environments as sets and using the $(+)$ operation for concatenation, that decision had to do with the fact that, this way, the system is closer to a linear type system. In the Linear Rank~2 Intersection Type System, a term is linear until we need to contract variables, so using these definitions makes us have more control over linearity and non-linearity. Also, it makes the system more easily extensible for other algebraic properties of intersection. We could also have rewritten the rule ($\rightarrow$ Elim) in order not to use the $(+)$ operation, which is something we might do in the future.

The downside of choosing these definitions is that it makes the proofs (in \autoref{chap:linearrank} and \autoref{chap:resource}) more complex, as they are not syntax directed because of the rules (Exchange) and (Contraction).

\section{Resource Inference}\label{chap:resource}

Given the quantitative properties of the Linear Rank 2 Intersection Types, we now aim to redefine the type system and the type inference algorithm, in order to infer not only the type of a $\lambda$-term, but also parameters related to resource usage. In this case, we are interested in obtaining the number of evaluation steps of the $\lambda$-term to its normal form, for the leftmost-outermost strategy.

\subsection{Type System}\label{sec:tighttypesystem}

The new type system defined in this chapter results from an adaptation and merge between our Linear Rank~2 Intersection Type System (\autoref{lineartypesystem}) and the system for the leftmost-outermost evaluation strategy presented in \cite{accattoli2018tight}, as that system is able to derive a measure related to the number of evaluation steps for the leftmost-outermost strategy. We then begin by adapting some definitions from \cite{accattoli2018tight} and others that were already introduced in \autoref{chap:linearrank}.

The predicates $\msf{normal}$ and $\msf{neutral}$ defining, respectively, the leftmost-outermost normal terms and neutral terms, are in \autoref{normalneutral2}. The predicate $\msf{abs}(M)$ is true if and only if $M$ is an abstraction; $\msf{normal(M)}$ means that $M$ is in normal form; and $\msf{neutral}(M)$ means that $M$ is in normal form and can never behave as an abstraction, i.e., it does not create a redex when applied to an argument.

\begin{definition}[Leftmost-outermost normal forms]\label{normalneutral2} 
\hfill
\begin{center}
    \AxiomC{}
    \UnaryInfC{$\msf{neutral}(x)$}
    \DisplayProof
    \quad
    \AxiomC{$\msf{neutral}(M)$}
    \AxiomC{$\msf{normal}(N)$}
    \BinaryInfC{$\msf{neutral}(M N)$}
    \DisplayProof
    \quad
    \AxiomC{$\msf{neutral}(M)$}
    \UnaryInfC{$\msf{normal}(M)$}
    \DisplayProof
    \quad
    \AxiomC{$\msf{normal}(M)$}
    \UnaryInfC{$\msf{normal}(\lambda x.M)$}
    \DisplayProof
\end{center}
\end{definition}

\begin{definition} [Leftmost-outermost evaluation strategy] 
\hfill
\begin{center}
    \AxiomC{}
    \UnaryInfC{$(\lambda x.M) N \longrightarrow \sub{M}{N/x}$}
    \DisplayProof
    \quad
    \AxiomC{$M \longrightarrow M'$}
    \UnaryInfC{$\lambda x.M \longrightarrow \lambda x.M'$}
    \DisplayProof
    \quad
    \AxiomC{$M \longrightarrow M'$}
    \AxiomC{$\neg \msf{abs}(M)$}
    \BinaryInfC{$M N \longrightarrow M' N$}
    \DisplayProof
    
    \hfill
    
    \AxiomC{$\msf{neutral}(N)$}
    \AxiomC{$M \longrightarrow M'$}
    \BinaryInfC{$N M \longrightarrow N M'$}
    \DisplayProof
\end{center}
\end{definition}

\begin{definition}[Finite rank multi-types]
We define the finite rank multi-types by the following grammar:
\begin{Dequation}
    \begin{align*}
        \msf{tight} &\Coloneqq \msf{Neutral} \mid \msf{Abs}   \tag{Tight constants}\\
        \trz &\Coloneqq \msf{tight} \mid \alpha \mid \trz \multimap \trz   \tag{Rank 0 multi-types}\\
        \tro &\Coloneqq \trz \mid \tro \cap \tro   \tag{Rank 1 multi-types}\\
        \trt &\Coloneqq \trz \mid \tro \rightarrow \trt   \tag{Rank 2 multi-types}
    \end{align*}
\end{Dequation}
\end{definition}

\begin{definition}
    ~\begin{itemize}
        \item Here, a \emph{statement} is an expression of the form $M:(\ro, \tro)$, where the pair $(\ro, \tro)$ is called the \emph{predicate}, and the term $M$ is called the \emph{subject} of the statement.
        
        \item A \emph{declaration} is a statement where the subject is a term variable.
        
        \item The comma operator (,) appends a declaration to the end of a list (of declarations). The list $(\Gamma_1, \Gamma_2)$ is the list that results from appending the  list $\Gamma_2$ to the end of the list $\Gamma_1$.
        
        \item A finite list of declarations is \emph{consistent} if and only if the term variables are all distinct.
        
        \item An \emph{environment} is a consistent finite list of declarations which predicates are pairs with a sequence from $\tl{1}$ as the first element and a rank 1 multi-type as the second element of the pair (i.e., the declarations are of the form $x:(\ro, \tro)$), and we use $\Gamma$ (possibly with single quotes and/or number subscripts) to range over environments.
        
        \item An environment $\Gamma = [x_1:(\ro_1, \tro_1), \dots, x_n:(\ro_n, \tro_n)]$ induces a partial function $\Gamma$ with domain $\dom(\Gamma) = \{x_1, \dots, x_n\}$, and $\Gamma(x_i) = (\ro_i, \tro_i)$.
        
        \item We write $\Gamma_x$ for the resulting environment of eliminating the declaration of $x$ from $\Gamma$ (if there is no declaration of $x$ in $\Gamma$, then $\Gamma_x = \Gamma$).
        
        \item We write $\Gamma_1 \equiv \Gamma_2$ if the environments $\Gamma_1$ and $\Gamma_2$ are equal up to the order of the declarations.
        
        \item If $\Gamma_1$ and $\Gamma_2$ are environments, the environment $\Gamma_1 + \Gamma_2$ is defined as follows:
        
        for each $x \in \dom(\Gamma_1) \cup \dom(\Gamma_2)$,
            \[
                (\Gamma_1 + \Gamma_2)(x) = \left\{
                \begin{array}{ll}
                    \Gamma_1(x) &\text{if } x \notin \dom(\Gamma_2)\\
                    \Gamma_2(x) &\text{if } x \notin \dom(\Gamma_1)\\
                    (\ro_1 \cap \ro_2, \tro_1 \cap \tro_2) &\text{if } \Gamma_1(x) = (\ro_1, \tro_1) \text{ and } \Gamma_2(x) = (\ro_2, \tro_2)
                \end{array}
                \right.
            \]
        with the declarations of the variables in $\dom(\Gamma_1)$ in the beginning of the list, by the same order they appear in $\Gamma_1$, followed by the declarations of the variables in $\dom(\Gamma_2) \setminus \dom(\Gamma_1)$, by the order they appear in $\Gamma_2$.
        
        \item We write $\msf{tight}(\trt)$ if $\trt$ is of the form $\msf{tight}$ and $\msf{tight}(\trz_1 \cap \cdots \cap \trz_n)$ if $\msf{tight}(\trz_i)$ for all $1 \leq i \leq n$.
        
        For $\Gamma = [x_1:(\ro_1, \tro_1), \dots, x_n:(\ro_n, \tro_n)]$, we write $\msf{tight}(\Gamma)$ if $\msf{tight}(\tro_i)$ for all $1 \leq i \leq n$, in which case we also say that $\Gamma$ is tight.
    \end{itemize}
\end{definition}

\begin{definition}[Linear Rank~2 Quantitative Type System]\label{tightlinearsystem}
In the Linear Rank~2 Quantitative Type System, we say that $M$ has type $\rt$ and multi-type $\trt$ given the environment $\Gamma$, with index~$b$, and write $\infertight{ \Gamma }{ b }{ M }{ \rt }{ \trt }$, if it can be obtained from the \emph{derivation rules} in \autoref{fig:lr2qts}.
\begin{figure}[!ht]
\begin{equation}\tag{Axiom}
\begin{mathprooftree}

    \AxiomC{$\infertight{ [x:(\rz, \trz)] }{ 0 }{ x }{ \rz }{ \trz }$}
\end{mathprooftree}
\end{equation}

\begin{equation}\tag{Exchange}
\begin{mathprooftree}

    \AxiomC{$\infertight{ \Gamma_1, x:(\ro_1, \tro_1), y:(\ro_2, \tro_2), \Gamma_2 }{ b }{ M }{ \rt }{ \trt }$}
    
    \UnaryInfC{$\infertight{ \Gamma_1, y:(\ro_2, \tro_2), x:(\ro_1, \tro_1), \Gamma_2 }{ b }{ M }{ \rt }{ \trt }$}
    
\end{mathprooftree}
\end{equation}

\begin{equation}\tag{Contraction}
\begin{mathprooftree}

    \AxiomC{$\infertight{ \Gamma_1, x_1:(\ro_1, \tro_1), x_2:(\ro_2, \tro_2), \Gamma_2 }{ b }{ M }{ \rt }{ \trt }$}
    
    \UnaryInfC{$\infertight{ \Gamma_1, x:(\ro_1 \cap \ro_2, \tro_1 \cap \tro_2), \Gamma_2 }{ b }{ M [x/x_1, x/x_2] }{ \rt }{ \trt }$}
    
\end{mathprooftree}
\end{equation}

\begin{equation}\tag{$\multimap$ Intro}
\begin{mathprooftree}

    \AxiomC{$\infertight{ \Gamma, x:(\rz, \trz) }{ \color{azulciencias} b }{ M }{ \rt }{ \trt }$}
    
    \UnaryInfC{$\infertight{ \Gamma }{ \color{azulciencias} b+1 }{ \lambda x.M }{ \rz \multimap \rt }{ \trz \multimap \trt }$}
    
\end{mathprooftree}
\end{equation}

\begin{equation}\tag{$\multimap$ Intro\textsubscript{t}}
\begin{mathprooftree}

    \AxiomC{$\infertight{ \Gamma, x:(\rz, \msf{tight}) }{ b }{ M }{ \rt }{ \msf{tight} }$}
    
    \UnaryInfC{$\infertight{ \Gamma }{ b }{ \lambda x.M }{ \rz \multimap \rt }{ \msf{Abs} }$}
    
\end{mathprooftree}
\end{equation}

\begin{equation}\tag{$\rightarrow$ Intro}
\begin{mathprooftree}

    \AxiomC{$\infertight{ \Gamma, x:(\rz_1 \cap \cdots \cap \rz_n, \trz_1 \cap \cdots \cap \trz_n) }{ \color{azulciencias} b }{ M }{ \rt }{ \trt }$}
    
    \AxiomC{$n \geq 2$}
    
    \BinaryInfC{$\infertight{ \Gamma }{ \color{azulciencias} b+1 }{ \lambda x.M }{ \rz_1 \cap \cdots \cap \rz_n \rightarrow \rt }{ \trz_1 \cap \cdots \cap \trz_n \rightarrow \trt }$}
    
\end{mathprooftree}
\end{equation}

    \begin{equation}\tag{$\rightarrow$ Intro\textsubscript{t}}
\begin{mathprooftree}

    \AxiomC{$\infertight{ \Gamma, x:(\rz_1 \cap \cdots \cap \rz_n, \tro) }{ b }{ M }{ \rt }{ \msf{tight} }$}
    
    \AxiomC{$\msf{tight}(\tro)$}
    
    \AxiomC{$n \geq 2$}
    
    \TrinaryInfC{$\infertight{ \Gamma }{ b }{ \lambda x.M }{ \rz_1 \cap \cdots \cap \rz_n \rightarrow \rt }{ \msf{Abs} }$}
    
\end{mathprooftree}
\end{equation}

\begin{equation}\tag{$\multimap$ Elim}
\begin{mathprooftree}

    \AxiomC{$\infertight{ \Gamma_1 }{ b_1 }{ M_1 }{ \rz \multimap \rt }{ \trz \multimap \trt }$}
    
    \AxiomC{$\infertight{ \Gamma_2 }{ b_2 }{ M_2 } { \rz }{ \trz }$}
    
    \BinaryInfC{$\infertight{ \Gamma_1, \Gamma_2 }{ b_1 + b_2 }{ M_1 M_2 }{ \rt }{ \trt }$}
    
\end{mathprooftree}
\end{equation}

\begin{equation}\tag{$\multimap$ Elim\textsubscript{t}}
\begin{mathprooftree}

    \AxiomC{$\infertight{ \Gamma_1 }{ b_1 }{ M_1 }{ \rz \multimap \rt }{ \msf{Neutral} }$}
    
    \AxiomC{$\infertight{ \Gamma_2 }{ b_2 }{ M_2 } { \rz }{ \msf{tight} }$}
    
    \BinaryInfC{$\infertight{ \Gamma_1, \Gamma_2 }{ b_1 + b_2 }{ M_1 M_2 }{ \rt }{ \msf{Neutral} }$}
    
\end{mathprooftree}
\end{equation}

\begin{equation}\tag{$\rightarrow$ Elim}
\begin{mathprooftree}

    \AxiomC{$\infertight{ \Gamma }{ b }{ M_1 }{ \rz_1 \cap \cdots \cap \rz_n \rightarrow \rt }{ \trz_1 \cap \cdots \cap \trz_n \rightarrow \trt }$}
    
    \noLine
    
    \UnaryInfC{$\infertight{ \Gamma_1 }{ b_1 }{ M_2 } { \rz_1 }{ \trz_1 }  \text { } \cdots \text{ }  \infertight{ \Gamma_n }{ b_n }{ M_2 } { \rz_n }{ \trz_n }$}
    
    \AxiomC{$n \geq 2$}
    
    \BinaryInfC{$\infertight{ \Gamma, \sum_{i=1}^n \Gamma_i }{ b + b_1 + \dots + b_n }{ M_1 M_2 }{ \rt }{ \trt }$}
    
\end{mathprooftree}
\end{equation}

\begin{equation}\tag{$\rightarrow$ Elim\textsubscript{t}}
\begin{mathprooftree}

    \AxiomC{$\infertight{ \Gamma }{ b }{ M_1 }{ \rz_1 \cap \cdots \cap \rz_n \rightarrow \rt }{ \msf{Neutral} }$}
    
    \noLine
    
    \UnaryInfC{$\infertight{ \Gamma_1 }{ b_1 }{ M_2 } { \rz_1 }{ \msf{tight} }  \text { } \cdots \text{ }  \infertight{ \Gamma_n }{ {b_n} }{ M_2 } { \rz_n }{ \msf{tight} }$}
    
    \AxiomC{$n \geq 2$}
    
    \BinaryInfC{$\infertight{ \Gamma, \sum_{i=1}^n \Gamma_i }{ b + b_1 + \dots + b_n }{ M_1 M_2 }{ \rt }{ \msf{Neutral} }$}
    
\end{mathprooftree}
\end{equation}
    \caption{Linear Rank~2 Quantitative Type System}
    \label{fig:lr2qts}
\end{figure}
\end{definition}

The tight rules (the t-indexed ones) are used to introduce the tight constants $\msf{Neutral}$ and $\msf{Abs}$, and they are related to minimal typings. Note that the index is only incremented in rules ($\multimap$ Intro) and ($\rightarrow$ Intro), as these are used to type abstractions that will be applied, contrary to the abstractions typed with the constant $\msf{Abs}$.

\begin{notation}
We write $\Phi \rhd \infertight{ \Gamma }{ b }{ M }{ \rt }{ \trt }$ if $\Phi$ is a derivation tree ending with $\infertight{ \Gamma }{ b }{ M }{ \rt }{ \trt }$. In this case, $|\Phi|$ is the depth of the derivation tree $\Phi$.
\end{notation}

\begin{definition}[Tight derivations]
A derivation $\Phi \rhd \infertight{ \Gamma }{ b }{ M }{ \rt }{ \trt }$ is tight if $\msf{tight}(\trt)$ and $\msf{tight}(\Gamma)$.
\end{definition}

Similarly to what has been done in \cite{accattoli2018tight}, in this section we prove that, in the Linear Rank~2 Quantitative Type System, whenever a term is tightly typable with index $b$, then $b$ is exactly the number of evaluations steps to leftmost-outermost normal form.

\begin{example}
Let $M = (\lambda x_1.(\lambda x_2.x_2 x_1) x_1) I$, where $I$ is the  identity function $\lambda y.y$.

Let us first consider the leftmost-outermost evaluation of $M$ to normal form:
\[
(\lambda x_1.(\lambda x_2.x_2 x_1) x_1) I \longrightarrow (\lambda x_2.x_2 I) I \longrightarrow I I \longrightarrow I
\]
\indent So the evaluation sequence has length $3$.

Let us write $\overmultimap{\alpha}$ for the type $(\alpha \multimap \alpha)$ and $\overlongmultimap{\msf{Abs}}$ for the type $\msf{Abs} \multimap \msf{Abs}$.

To make the derivation tree easier to read, let us first get the following derivation $\Phi$ for the term $\lambda x_1.(\lambda x_2.x_2 x_1) x_1$:

\scalebox{0.7}{
\begin{mathprooftree}
    \AxiomC{$\infertight{ [x_2:(\overmultimap{\alpha} \multimap \overmultimap{\alpha}, \overlongmultimap{\msf{Abs}})] }{ 0 }{ x_2 }{ \overmultimap{\alpha} \multimap \overmultimap{\alpha} }{ \overlongmultimap{\msf{Abs}} }$}
    
    \AxiomC{$\infertight{ [x_3:(\overmultimap{\alpha}, \msf{Abs})] }{ 0 }{ x_3 }{ \overmultimap{\alpha} }{ \msf{Abs} }$}
    
    \BinaryInfC{$\infertight{ [x_2:(\overmultimap{\alpha} \multimap \overmultimap{\alpha}, \overlongmultimap{\msf{Abs}}), x_3:(\overmultimap{\alpha}, \msf{Abs})] }{ 0 }{ x_2 x_3 }{ \overmultimap{\alpha} }{ \msf{Abs} }$}
    \UnaryInfC{$\infertight{ [x_3:(\overmultimap{\alpha}, \msf{Abs})] }{ 1 }{ \lambda x_2.x_2 x_3 }{ (\overmultimap{\alpha} \multimap \overmultimap{\alpha}) \multimap \; \overmultimap{\alpha} }{ \overlongmultimap{\msf{Abs}} \multimap \msf{Abs} }$}
    
    \AxiomC{$\infertight{ [x_4:(\overmultimap{\alpha} \multimap \overmultimap{\alpha}, \overlongmultimap{\msf{Abs}})] }{ 0 }{ x_4 }{ \overmultimap{\alpha} \multimap \overmultimap{\alpha} }{ \overlongmultimap{\msf{Abs}}) }$}
    
    \BinaryInfC{$\infertight{ [x_3:(\overmultimap{\alpha}, \msf{Abs}), x_4:(\overmultimap{\alpha} \multimap \overmultimap{\alpha}, \overlongmultimap{\msf{Abs}})] }{ 1 }{ (\lambda x_2.x_2 x_3) x_4 }{ \overmultimap{\alpha} }{ \msf{Abs} }$}
    \UnaryInfC{$\infertight{ [x_1:(\overmultimap{\alpha} \cap \; (\overmultimap{\alpha} \multimap \overmultimap{\alpha}), \msf{Abs} \; \cap \overlongmultimap{\msf{Abs}})] }{ 1 }{ (\lambda x_2.x_2 x_1) x_1 }{ \overmultimap{\alpha} }{ \msf{Abs} }$}
    \UnaryInfC{$\infertight{ [\;] }{ 2 }{ \lambda x_1.(\lambda x_2.x_2 x_1) x_1 }{ (\overmultimap{\alpha} \cap \; (\overmultimap{\alpha} \multimap \overmultimap{\alpha})) \rightarrow \; \overmultimap{\alpha} }{ (\msf{Abs} \; \cap \overlongmultimap{\msf{Abs}}) \rightarrow \msf{Abs} }$}
\end{mathprooftree}
}\\

Then for the $\lambda$-term $M$, the following tight derivation is obtained:

\begin{center}
\begin{mathprooftree}
    \AxiomC{$\Phi$}
    
    \AxiomC{$\infertight{ [y:(\alpha, \msf{Neutral})] }{ 0 }{ y }{ \alpha }{ \msf{Neutral} }$}
    \UnaryInfC{$\infertight{ [\;] }{ 0 }{ I }{ \overmultimap{\alpha} }{ \msf{Abs} }$}
    
    \AxiomC{$\infertight{ [y:(\overmultimap{\alpha}, \msf{Abs})] }{ 0 }{ y }{ \overmultimap{\alpha} }{ \msf{Abs} }$}
    \UnaryInfC{$\infertight{ [\;] }{ 1 }{ I }{ \overmultimap{\alpha} \multimap \overmultimap{\alpha} }{ \overlongmultimap{\msf{Abs}} }$}
    
    \TrinaryInfC{$\infertight{ [\;] }{ 3 }{ (\lambda x_1.(\lambda x_2.x_2 x_1) x_1) I }{ \overmultimap{\alpha} }{ \msf{Abs} }$}
\end{mathprooftree}
\end{center}

So indeed, the index $3$ represents the number of evaluation steps to leftmost-outermost normal form.
\end{example}

We now show several properties of the type system, adapted from \cite{accattoli2018tight}, in order to prove the \emph{tight correctness} (\autoref{tightcorrect}).

\begin{lemma}[Tight spreading on neutral terms]\label{spreading} 
If $M$ is a term such that $\msf{neutral}(M)$ and $\Phi \rhd \infertight{ \Gamma }{ b }{ M } { \rt }{ \trt }$ is a typing derivation such that $\msf{tight}(\Gamma)$, then $\msf{tight}(\trt)$.

\begin{proof}
By induction on $|\Phi|$.

Note that the last rule in $\Phi$ cannot be any of the $\multimap$ and $\rightarrow$ intro ones, because $M = \lambda x.M_1$ is not neutral.

\begin{enumerate}
    \item \tul{(Axiom)}: Then $\Gamma = [x:(\rz, \trz)]$, $M = x$ and $\trt = \trz$.\\
    
    Since by hypothesis $\msf{tight}(\Gamma)$, then $\trz$ is tight. So $\msf{tight}(\trt)$.\\

    \item \tul{(Exchange)}: Then $\Gamma = (\Gamma_1, y:(\ro_2, \tro_2), x:(\ro_1, \tro_1), \Gamma_2)$, $M = M_1$, $\trt = \trt_1$, and assuming that the premise $\infertight{ \Gamma_1, x:(\ro_1, \tro_1), y:(\ro_2, \tro_2), \Gamma_2 }{ b }{ M_1 }{ \rt }{ \trt_1 }$ holds.\\
    
    Since by hypothesis $\msf{tight}(\Gamma)$, and $(\Gamma_1, x:(\ro_1, \tro_1), y:(\ro_2, \tro_2), \Gamma_2) \equiv \Gamma$, then $\msf{tight}(\Gamma_1, x:(\ro_1, \tro_1), y:(\ro_2, \tro_2), \Gamma_2)$. And since $\msf{neutral}(M)$ and $M = M_1$, then $\msf{neutral}(M_1)$.\\
    
    So by induction we get $\msf{tight}(\trt_1)$. And because $\trt = \trt_1$, we have $\msf{tight}(\trt)$.\\

    \item \tul{(Contraction)}: Then $\Gamma = (\Gamma_1, x:(\ro_1 \cap \ro_2, \tro_1 \cap \tro_2), \Gamma_2)$, $M = M_1 [x/x_1, x/x_2]$, $\trt = \trt_1$, and assuming that the premise $\infertight{ \Gamma_1, x_1:(\ro_1, \tro_1), x_2:(\ro_2, \tro_2), \Gamma_2 }{ b }{ M_1 }{ \rt }{ \trt_1 }$ holds.\\
    
    Since by hypothesis $\msf{tight}(\Gamma)$, and all types in $(\Gamma_1, x_1:(\ro_1, \tro_1), x_2:(\ro_2, \tro_2), \Gamma_2)$ appear in $\Gamma$, then $\msf{tight}(\Gamma_1, x_1:(\ro_1, \tro_1), x_2:(\ro_2, \tro_2), \Gamma_2)$. And $\msf{neutral}(M_1)$ because $\msf{neutral}(M_1 [x/x_1, x/x_2])$.\\
    
    So by induction we get $\msf{tight}(\trt_1)$. And because $\trt = \trt_1$, we have $\msf{tight}(\trt)$.\\

    \item \tul{($\multimap$ Elim)}: Then $\Gamma = (\Gamma_1, \Gamma_2)$, $M = M_1 M_2$, $\trt = \trt_1$, and assuming that the premises $\infertight{ \Gamma_1 }{ b_1 }{ M_1 }{ \rz \multimap \rt }{ \trz \multimap \trt_1 }$ and $\infertight{ \Gamma_2 }{ b_2 }{ M_2 } { \rz }{ \trz }$ hold.\\
    
    Since by hypothesis $\msf{neutral}(M_1 M_2)$, then $\msf{neutral}(M_1)$ and $\msf{normal}(M_2)$
    
    All types in $\Gamma_1$ appear in $\Gamma$. Then since by hypothesis $\msf{tight}(\Gamma)$, we have $\msf{tight}(\Gamma_1)$.\\
    
    Then we could apply the induction hypothesis to obtain $\msf{tight}(\trz \multimap \trt_1)$, which is false.
    
    So ($\multimap$ Elim) cannot be the last rule in $\Phi$.\\

    \item \tul{($\rightarrow$ Elim)}: Similarly to ($\multimap$ Elim), here we would obtain $\msf{tight}(\trz_1 \cap \cdots \cap \trz_n \rightarrow \trt_1)$, so ($\rightarrow$~Elim) also cannot be the last rule in $\Phi$.\\

    \item \tul{($\multimap$ Elim\textsubscript{t})}: Then $\Gamma = (\Gamma_1, \Gamma_2)$, $M = M_1 M_2$ and $\trt = \msf{Neutral}$.\\
    
    Since $\trt = \msf{Neutral}$, we already have $\msf{tight}(\trt)$.\\

    \item \tul{($\rightarrow$ Elim\textsubscript{t})}: Then $\Gamma = (\Gamma, \sum_{i=1}^n \Gamma_i)$, $M = M_1 M_2$ and $\trt = \msf{Neutral}$.\\
    
    Since $\trt = \msf{Neutral}$, we already have $\msf{tight}(\trt)$.
\end{enumerate}
\end{proof}
\end{lemma}

\begin{lemma}[Properties of tight typings for normal forms]\label{tightproperties} 
Let $M$ be such that $\msf{normal}(M)$ and $\Phi \rhd \infertight{ \Gamma }{ b }{ M } { \rt }{ \trt }$ be a typing derivation.

\begin{enumerate}[(i)]
    \item \emph{Tightness}: if $\Phi$ is tight, then $b = 0$.
    
    \item \emph{Neutrality}: if $\trt = \msf{Neutral}$ then $\msf{neutral}(M)$.
\end{enumerate}

\begin{proof}
By induction on $|\Phi|$.

\begin{enumerate}
    \item \tul{(Axiom)}: Then $\Gamma = [x:(\rz, \trz)]$, $M = x$, $b = 0$ and $\trt = \trz$.\\
    
    Clearly, both properties of the statement are verified in this case.\\

    \item \tul{(Exchange)}: Then $\Gamma = (\Gamma_1, y:(\ro_2, \tro_2), x:(\ro_1, \tro_1), \Gamma_2)$, $M = M_1$, $b = b_1$, $\trt = \trt_1$, and assuming $\Phi_1 \rhd \infertight{ \Gamma_1, x:(\ro_1, \tro_1), y:(\ro_2, \tro_2), \Gamma_2 }{ b_1 }{ M_1 }{ \rt }{ \trt_1 }$.\\
    
    Since by hypothesis $\msf{normal}(M)$ and $M = M_1$, then $\msf{normal}(M_1)$.
    
    \begin{enumerate}[(i)]
        \item \emph{Tightness}: if $\Phi$ is tight, then $\Phi_1$ is tight and by induction, $b = b_1 = 0$.
        
        \item \emph{Neutrality}: if $\trt = \msf{Neutral}$, since $\trt = \trt_1$, $\trt_1 = \msf{Neutral}$ and by induction, $\msf{neutral}(M_1)$. So since $M = M_1$, $\msf{neutral}(M)$.\\
    \end{enumerate}

    \item \tul{(Contraction)}: Then $\Gamma = (\Gamma_1, x:(\ro_1 \cap \ro_2, \tro_1 \cap \tro_2), \Gamma_2)$, $M = M_1 [x/x_1, x/x_2]$, $b = b_1$, $\trt = \trt_1$, and assuming $\Phi_1 \rhd \infertight{ \Gamma_1, x_1:(\ro_1, \tro_1), x_2:(\ro_2, \tro_2), \Gamma_2 }{ b_1 }{ M_1 }{ \rt }{ \trt_1 }$.\\
    
    Since by hypothesis $\msf{normal}(M)$ and $M = M_1 [x/x_1, x/x_2]$, then $\msf{normal}(M_1)$.
    
    \begin{enumerate}[(i)]
        \item \emph{Tightness}: if $\Phi$ is tight, then $\Phi_1$ is tight and by induction, $b = b_1 = 0$.
        
        \item \emph{Neutrality}: if $\trt = \msf{Neutral}$, since $\trt = \trt_1$, $\trt_1 = \msf{Neutral}$ and by induction, $\msf{neutral}(M_1)$. So since $M = M_1 [x/x_1, x/x_2]$, $\msf{neutral}(M)$.\\
    \end{enumerate}

    \item \tul{($\multimap$ Intro)}: Then $\Gamma = \Gamma_1$, $M = \lambda x.M_1$, $b = b_1+1$, $\trt = \trz \multimap \trt_1$, and assuming $\Phi_1 \rhd \infertight{ \Gamma_1, x:(\rz, \trz) }{ b_1 }{ M_1 }{ \rt }{ \trt_1 }$.\\
    
    Since by hypothesis $\msf{normal}(M)$ and $M = \lambda x.M_1$, then $\msf{normal}(M_1)$.
    
    \begin{enumerate}[(i)]
        \item \emph{Tightness}: $\Phi$ is not tight, so the statement trivially holds.
        
        \item \emph{Neutrality}: $\trt \neq \msf{Neutral}$, so the statement trivially holds.\\
    \end{enumerate}

    \item \tul{($\rightarrow$ Intro)}: Then $\Gamma = \Gamma_1$, $M = \lambda x.M_1$, $b = b_1+1$, $\trt = \trz_1 \cap \cdots \cap \trz_n \rightarrow \trt_1$, and assuming $\Phi_1 \rhd \infertight{ \Gamma_1, x:(\rz_1 \cap \cdots \cap \rz_n, \trz_1 \cap \cdots \cap \trz_n) }{ b_1 }{ M_1 }{ \rt }{ \trt_1 }$, with $n \geq 2$.\\
    
    Since by hypothesis $\msf{normal}(M)$ and $M = \lambda x.M_1$, then $\msf{normal}(M_1)$.
    
    \begin{enumerate}[(i)]
        \item \emph{Tightness}: $\Phi$ is not tight, so the statement trivially holds.
        
        \item \emph{Neutrality}: $\trt \neq \msf{Neutral}$, so the statement trivially holds.\\
    \end{enumerate}

    \item \tul{($\multimap$ Intro\textsubscript{t})}: Then $\Gamma = \Gamma_1$, $M = \lambda x.M_1$, $b = b_1$, $\trt = \msf{Abs}$, and assuming $\Phi_1 \rhd \infertight{ \Gamma_1, x:(\rz, \msf{tight}) }{ b_1 }{ M_1 }{ \rt }{ \msf{tight} }$.\\
    
    Since by hypothesis $\msf{normal}(M)$ and $M = \lambda x.M_1$, then $\msf{normal}(M_1)$.
    
    \begin{enumerate}[(i)]
        \item \emph{Tightness}: if $\Phi$ is tight, then $\Phi_1$ is tight and by induction, $b = b_1 = 0$.
        
        \item \emph{Neutrality}: $\trt \neq \msf{Neutral}$, so the statement trivially holds.\\
    \end{enumerate}

    \item \tul{($\rightarrow$ Intro\textsubscript{t})}: Then $\Gamma = \Gamma_1$, $M = \lambda x.M_1$, $b = b_1$, $\trt = \msf{Abs}$, and assuming $\Phi_1 \rhd \infertight{ \Gamma_1, x:(\rz_1 \cap \cdots \cap \rz_n, \tro) }{ b_1 }{ M_1 }{ \rt }{ \msf{tight} }$, with $\msf{tight}(\tro)$ and $n \geq 2$.\\
    
    Since by hypothesis $\msf{normal}(M)$ and $M = \lambda x.M_1$, then $\msf{normal}(M_1)$.
    
    \begin{enumerate}[(i)]
        \item \emph{Tightness}: if $\Phi$ is tight, then $\Phi_1$ is tight and by induction, $b = b_1 = 0$.
        
        \item \emph{Neutrality}: $\trt \neq \msf{Neutral}$, so the statement trivially holds.\\
    \end{enumerate}

    \item \tul{($\multimap$ Elim)}: Then $\Gamma = (\Gamma_1, \Gamma_2)$, $M = M_1 M_2$, $b = b_1 + b_2$, $\trt = \trt_1$, and assuming $\Phi_1 \rhd \infertight{ \Gamma_1 }{ b_1 }{ M_1 }{ \rz \multimap \rt }{ \trz \multimap \trt_1 }$ and $\Phi_2 \rhd \infertight{ \Gamma_2 }{ b_2 }{ M_2 } { \rz }{ \trz }$.\\
    
    Since by hypothesis $\msf{normal}(M)$ and $M = M_1 M_2$, then $\msf{neutral}(M_1 M_2)$. So $\msf{neutral}(M_1)$ (and then $\msf{normal}(M_1)$) and $\msf{normal}(M_2)$.
    
    \begin{enumerate}[(i)]
        \item \emph{Tightness}: this case is impossible. If $\Phi$ is tight, then $\Gamma = (\Gamma_1, \Gamma_2)$ is tight, and so is $\Gamma_1$. And since $\msf{neutral}(M_1)$, \autoref{spreading} implies that the type of $M_1$ in $\Phi_1$ has to be tight, which is absurd.
        
        \item \emph{Neutrality}: $\msf{neutral}(M)$ holds by hypothesis.\\
    \end{enumerate}

    \item \tul{($\rightarrow$ Elim)}: Then $\Gamma = (\Gamma', \sum_{i=1}^n \Gamma_i)$, $M = M_1 M_2$, $b = b' + b_1 + \dots + b_n$, $\trt = \trt_1$, and assuming $\Phi' \rhd \infertight{ \Gamma' }{ b' }{ M_1 }{ \rz_1 \cap \cdots \cap \rz_n \rightarrow \rt }{ \trz_1 \cap \cdots \cap \trz_n \rightarrow \trt_1 }$ and $\Phi_i \rhd \infertight{ \Gamma_i }{ b_i }{ M_2 } { \rz_i }{ \trz_i }$, for $1 \leq i \leq n$, with $n \geq 2$.\\
    
    Since by hypothesis $\msf{normal}(M)$ and $M = M_1 M_2$, then $\msf{neutral}(M_1 M_2)$. So $\msf{neutral}(M_1)$ (and then $\msf{normal}(M_1)$) and $\msf{normal}(M_2)$.
    
    \begin{enumerate}[(i)]
        \item \emph{Tightness}: this case is impossible. If $\Phi$ is tight, then $\Gamma = (\Gamma', \sum_{i=1}^n \Gamma_i)$ is tight, and so is $\Gamma'$. And since $\msf{neutral}(M_1)$, \autoref{spreading} implies that the type of $M_1$ in $\Phi'$ has to be tight, which is absurd.
        
        \item \emph{Neutrality}: $\msf{neutral}(M)$ holds by hypothesis.\\
    \end{enumerate}

    \item \tul{($\multimap$ Elim\textsubscript{t})}: Then $\Gamma = (\Gamma_1, \Gamma_2)$, $M = M_1 M_2$, $b = b_1 + b_2$, $\trt = \msf{Neutral}$, and assuming $\Phi_1 \rhd \infertight{ \Gamma_1 }{ b_1 }{ M_1 }{ \rz \multimap \rt }{ \msf{Neutral} }$ and $\Phi_2 \rhd \infertight{ \Gamma_2 }{ b_2 }{ M_2 } { \rz }{ \msf{tight} }$.\\
    
    Since by hypothesis $\msf{normal}(M)$ and $M = M_1 M_2$, then $\msf{neutral}(M_1 M_2)$. So $\msf{neutral}(M_1)$ (and then $\msf{normal}(M_1)$) and $\msf{normal}(M_2)$.
    
    \begin{enumerate}[(i)]
        \item \emph{Tightness}: if $\Phi$ is tight, then $\Phi_1$ and $\Phi_2$ are tight and by induction, $b_1 = 0$ and $b_2 = 0$. So $b = b_1 + b_2 = 0$.
        
        \item \emph{Neutrality}: $\msf{neutral}(M)$ holds by hypothesis.\\
    \end{enumerate}

    \item \tul{($\rightarrow$ Elim\textsubscript{t})}: Then $\Gamma = (\Gamma', \sum_{i=1}^n \Gamma_i)$, $M = M_1 M_2$, $b = b' + b_1 + \dots + b_n$, $\trt = \msf{Neutral}$, and assuming $\Phi' \rhd \infertight{ \Gamma' }{ b' }{ M_1 }{ \rz_1 \cap \cdots \cap \rz_n \rightarrow \rt }{ \msf{Neutral} }$ and $\Phi_i \rhd \infertight{ \Gamma_i }{ b_i }{ M_2 } { \rz_i }{ \msf{tight} }$, for $1 \leq i \leq n$, with $n \geq 2$.\\
    
    Since by hypothesis $\msf{normal}(M)$ and $M = M_1 M_2$, then $\msf{neutral}(M_1 M_2)$. So $\msf{neutral}(M_1)$ (and then $\msf{normal}(M_1)$) and $\msf{normal}(M_2)$.
    
    \begin{enumerate}[(i)]
        \item \emph{Tightness}: if $\Phi$ is tight, then $\Phi'$ and $\Phi_i$ (for all $1 \leq i \leq n$) are tight and by induction, $b' = 0$ and $b_i = 0$ (for all $1 \leq i \leq n$). So $b = b' + b_1 + \dots + b_n = 0$.
        
        \item \emph{Neutrality}: $\msf{neutral}(M)$ holds by hypothesis.
    \end{enumerate}
\end{enumerate}

\end{proof}
\end{lemma}

\begin{lemma}[Relevance]\label{xfreesys2}
If $\Phi \rhd \infertight{ \Gamma }{ b }{ M }{ \rt }{ \trt }$, then $x \in \dom(\Gamma)$ if and only if $x \in \fv{M}$.

\begin{proof}
Easy induction on $|\Phi|$.
\end{proof}
\end{lemma}

\begin{lemma}[Substitution and typings]\label{substlema} 

Let $\Phi \rhd \infertight{ \Gamma }{ b }{ M_1 }{ \rt }{ \trt }$ be a derivation with $x \in \dom(\Gamma)$ and $\Gamma(x) = (\rz_1 \cap \cdots \cap \rz_n, \trz_1 \cap \cdots \cap \trz_n)$, for $n \geq 1$. And, for each $1 \leq i \leq n$, let $\Phi_i \rhd \infertight{ \Gamma_i }{ b_i }{ M_2 } { \rz_i }{ \trz_i }$.

Then there exists a derivation $\Phi' \rhd \infertight{ \Gamma_x, \sum_{i=1}^n \Gamma_i }{ b + b_1 + \dots + b_n }{ \sub{M_1}{M_2/x} }{ \rt }{ \trt }$. Moreover, if the derivations $\Phi, \Phi_1, \dots, \Phi_n$ are tight, then so is the derivation $\Phi'$.

\begin{proof}

The proof follows by induction on $|\Phi|$.

Without loss of generality, we assume that $\fv{M_1} \cap \fv{M_2} = \emptyset$, so that $\Gamma_x, \sum_{i=1}^n \Gamma_i$ is consistent. Otherwise, we could simply rename the free variables in $M_1$ to get $M_1'$ (and the same derivation $\Phi$, with the variables renamed) such that $\fv{M_1'} \cap \fv{M_2} = \emptyset$. Then, by a weaker form of the lemma (for $M_1, M_2$ such that $\fv{M_1} \cap \fv{M_2} = \emptyset$), we would get the derivation $\Phi'$ (with the renamed variables) and finally we could apply the rule (Contraction) (and (Exchange), when necessary) to the variables that were renamed in $M_1$, in order to end up with the proper derivation $\Phi'$.

\begin{enumerate}
    \item \tul{(Axiom)}:\\
    
    Then we have
    \[
        \Phi \rhd \infertight{ [x:(\rz_1, \trz_1)] }{ 0 }{ x }{ \rz_1 }{ \trz_1 }.
    \]
    
    So $\Gamma = [x:(\rz_1, \trz_1)]$, $\Gamma_x = [\;]$, $M_1 = x$, $b = 0$, $\rt = \rz_1$ and $\trt = \trz_1$.\\
    
    By hypothesis we also have:
    \[
        \Phi_1 \rhd \infertight{ \Gamma_1 }{ b_1 }{ M_2 } { \rz_1 }{ \trz_1 }
    \]
    
    Given that $(\Gamma_x, \Gamma_1) = ([\;], \Gamma_1) = \Gamma_1$, $\sub{M_1}{M_2/x} = \sub{x}{M_2/x} = M_2$, $b+b_1 = 0+b_1 = b_1$, $\rt = \rz_1$ and $\trt = \trz_1$, then we already have the derivation $\Phi' = \Phi_1$, as we wanted.\\
    
    \item \tul{(Exchange)}:\\
    
    Then we have
    \[
        \Phi \rhd \infertight{ \Gamma_1', y_2:(\ro_2, \tro_2), y_1:(\ro_1, \tro_1), \Gamma_2' }{ b }{ M_1 }{ \rt }{ \trt },
    \]
    
    which follows from $\Phi_1' \rhd \infertight{ \Gamma_1', y_1:(\ro_1, \tro_1), y_2:(\ro_2, \tro_2), \Gamma_2' }{ b }{ M_1 }{ \rt }{ \trt }$.
    
    And there are three different cases depending on $x$:
    
    \begin{enumerate}
        \item $x \neq y_1$ and $x \neq y_2$:\\
    
        So $\Gamma = (\Gamma_1', y_2:(\ro_2, \tro_2), y_1:(\ro_1, \tro_1), \Gamma_2')$ and $\Gamma_x = ({\Gamma_1'}_x, y_2:(\ro_2, \tro_2), y_1:(\ro_1, \tro_1), {\Gamma_2'}_x)$.\\
        
        Since $x:(\rz_1 \cap \cdots \cap \rz_n, \trz_1 \cap \cdots \cap \trz_n) \in \Gamma$, then either that declaration is in $\Gamma_1'$ or in $\Gamma_2'$, and $x:(\rz_1 \cap \cdots \cap \rz_n, \trz_1 \cap \cdots \cap \trz_n) \in (\Gamma_1', y_1:(\ro_1, \tro_1), y_2:(\ro_2, \tro_2), \Gamma_2')$.\\
        
        By hypothesis we also have:
        \[
            \Phi_i \rhd \infertight{ \Gamma_i }{ b_i }{ M_2 } { \rz_i }{ \trz_i }
        \]
        for $1 \leq i \leq n$.\\
        
        Given $\Phi_1'$ and $\Phi_i$, by the induction hypothesis, there is a derivation ending with
        \[
            \infertight{ {\Gamma_1'}_x, y_1:(\ro_1, \tro_1), y_2:(\ro_2, \tro_2), {\Gamma_2'}_x, \sum_{i=1}^n \Gamma_i }{ b + b_1 + \dots + b_n }{ \sub{M_1}{M_2/x} }{ \rt }{ \trt }.
        \]
        
        By rule (Exchange), we get the final judgment we wanted:
        
        \begin{center}
        \begin{mathprooftree}
            \AxiomC{$\infertight{ {\Gamma_1'}_x, y_1:(\ro_1, \tro_1), y_2:(\ro_2, \tro_2), {\Gamma_2'}_x, \sum_{i=1}^n \Gamma_i }{ b + b_1 + \dots + b_n }{ \sub{M_1}{M_2/x} }{ \rt }{ \trt }$}
            \UnaryInfC{$\infertight{ {\Gamma_1'}_x, y_2:(\ro_2, \tro_2), y_1:(\ro_1, \tro_1), {\Gamma_2'}_x, \sum_{i=1}^n \Gamma_i }{ b + b_1 + \dots + b_n }{ \sub{M_1}{M_2/x} }{ \rt }{ \trt }$}
        \end{mathprooftree}
        \end{center}
        
        So there is indeed $\Phi' \rhd \infertight{ \Gamma_x, \sum_{i=1}^n \Gamma_i }{ b + b_1 + \dots + b_n }{ \sub{M_1}{M_2/x} }{ \rt }{ \trt }$.\\
        
        \item $x = y_1$:\\
        
        So $\Gamma = (\Gamma_1', y_2:(\ro_2, \tro_2), y_1:(\ro_1, \tro_1), \Gamma_2') = (\Gamma_1', y_2:(\ro_2, \tro_2), x:(\ro_1, \tro_1), \Gamma_2')$ and $\Gamma_x = (\Gamma_1', y_2:(\ro_2, \tro_2), \Gamma_2')$.\\
        
        By hypothesis we also have:
        \[
            \Phi_i \rhd \infertight{ \Gamma_i }{ b_i }{ M_2 } { \rz_i }{ \trz_i }
        \]
        for $1 \leq i \leq n$.\\
        
        Given $\Phi_1' \rhd \infertight{ \Gamma_1', x:(\ro_1, \tro_1), y_2:(\ro_2, \tro_2), \Gamma_2' }{ b }{ M_1 }{ \rt }{ \trt }$ and $\Phi_i$, by the induction hypothesis, there is a derivation
        \[
            \Phi' \rhd \infertight{ \Gamma_1', y_2:(\ro_2, \tro_2), \Gamma_2', \sum_{i=1}^n \Gamma_i }{ b + b_1 + \dots + b_n }{ \sub{M_1}{M_2/x} }{ \rt }{ \trt },
        \]
        
        which is the derivation we wanted.\\
        
        \item $x = y_2$:\\
        
        Analogous to the case where $x = y_1$.\\
    \end{enumerate}

    \item \tul{(Contraction)}:\\
    
    Then we have
    \[
        \Phi \rhd \infertight{ \Gamma_1', y:(\ro_1 \cap \ro_2, \tro_1 \cap \tro_2), \Gamma_2' }{ b }{ M_1' [y/y_1, y/y_2] }{ \rt }{ \trt },
    \]
    
    which follows from $\Phi_1' \rhd \infertight{ \Gamma_1', y_1:(\ro_1, \tro_1), y_2:(\ro_2, \tro_2), \Gamma_2' }{ b }{ M_1' }{ \rt }{ \trt }$.
    
    And there are two different cases depending on $x$:
    
    \begin{enumerate}
        \item $x \neq y$:\\
    
        So $\Gamma = (\Gamma_1', y:(\ro_1 \cap \ro_2, \tro_1 \cap \tro_2), \Gamma_2')$, $\Gamma_x = ({\Gamma_1'}_x, y:(\ro_1 \cap \ro_2, \tro_1 \cap \tro_2), {\Gamma_2'}_x)$, $M_1 = M_1' [y/y_1, y/y_2]$, $x \neq y_1$ and $x \neq y_2$.\\
        
        Since $x:(\rz_1 \cap \cdots \cap \rz_n, \trz_1 \cap \cdots \cap \trz_n) \in \Gamma$, then either that declaration is in $\Gamma_1'$ or in $\Gamma_2'$, and $x:(\rz_1 \cap \cdots \cap \rz_n, \trz_1 \cap \cdots \cap \trz_n) \in (\Gamma_1', y_1:(\ro_1, \tro_1), y_2:(\ro_2, \tro_2), \Gamma_2')$.\\
        
        By hypothesis we also have:
        \[
            \Phi_i \rhd \infertight{ \Gamma_i }{ b_i }{ M_2 } { \rz_i }{ \trz_i }
        \]
        for $1 \leq i \leq n$.\\
        
        Given $\Phi_1'$ and $\Phi_i$, by the induction hypothesis, there is a derivation ending with
        \[
            \infertight{ {\Gamma_1'}_x, y_1:(\ro_1, \tro_1), y_2:(\ro_2, \tro_2), {\Gamma_2'}_x, \sum_{i=1}^n \Gamma_i }{ b + b_1 + \dots + b_n }{ \sub{M_1'}{M_2/x} }{ \rt }{ \trt }.
        \]
        Note that this implies that $y_1$ and $y_2$ do not occur free in $M_2$, otherwise, by \autoref{xfreesys2}, $y_1, y_2 \in \dom(\Gamma_i)$ and so ${\Gamma_1'}_x, y_1:(\ro_1, \tro_1), y_2:(\ro_2, \tro_2), {\Gamma_2'}_x, \sum_{i=1}^n \Gamma_i$ would not be consistent.\\
        
        By rule (Contraction), we get the final judgment we wanted:
        
        \begin{center}
        \begin{mathprooftree}
            \AxiomC{$\infertight{ {\Gamma_1'}_x, y_1:(\ro_1, \tro_1), y_2:(\ro_2, \tro_2), {\Gamma_2'}_x, \sum_{i=1}^n \Gamma_i }{ b + b_1 + \dots + b_n }{ \sub{M_1'}{M_2/x} }{ \rt }{ \trt }$}
            \UnaryInfC{$\infertight{ {\Gamma_1'}_x, y:(\ro_1 \cap \ro_2, \tro_1 \cap \tro_2), {\Gamma_2'}_x, \sum_{i=1}^n \Gamma_i }{ b + b_1 + \dots + b_n }{ \sub{(\sub{M_1'}{M_2/x})}{y/y_1, y/y_2} }{ \rt }{ \trt }$}
        \end{mathprooftree}
        \end{center}

        \hfill
        
        Since $x \neq y_1$, $x \neq y_2$, $x \neq y$ and $y_1, y_2$ do not occur free in $M_2$,
        
        then $\sub{(\sub{M_1'}{M_2/x})}{y/y_1, y/y_2} = \sub{(\sub{M_1'}{y/y_1, y/y_2})}{M_2/x}$.
        
        And as $M_1 = M_1' [y/y_1, y/y_2]$, we have $\sub{(\sub{M_1'}{y/y_1, y/y_2})}{M_2/x} = \sub{M_1}{M_2/x}$.\\
        
        So there is indeed $\Phi' \rhd \infertight{ \Gamma_x, \sum_{i=1}^n \Gamma_i }{ b + b_1 + \dots + b_n }{ \sub{M_1}{M_2/x} }{ \rt }{ \trt }$.\\

        \item $x = y$:\\
        
        So $\Gamma = (\Gamma_1', y:(\ro_1 \cap \ro_2, \tro_1 \cap \tro_2), \Gamma_2') = (\Gamma_1', x:(\ro_1 \cap \ro_2, \tro_1 \cap \tro_2), \Gamma_2')$, $\Gamma_x = (\Gamma_1', \Gamma_2')$, $M_1 = M_1' [y/y_1, y/y_2] = M_1' [x/y_1, x/y_2]$ and $x \neq y_1$ and $x \neq y_2$ (assuming that $y \neq y_1$ and $y \neq y_2$, without loss of generality). Let us also assume, without loss of generality, that $y_1, y_2$ do not occur in $M_2$.\\
        
        By hypothesis we have $x:(\rz_1 \cap \cdots \cap \rz_n, \trz_1 \cap \cdots \cap \trz_n) \in \Gamma$ and $y:(\ro_1 \cap \ro_2, \tro_1 \cap \tro_2) \in \Gamma$.
        
        So since $x = y$, we have $\rz_1 \cap \cdots \cap \rz_n = \ro_1 \cap \ro_2$ and $\trz_1 \cap \cdots \cap \trz_n = \tro_1 \cap \tro_2$.
        
        So for some $1 \leq k < n$, we have $\ro_1 = \rz_1 \cap \cdots \cap \rz_k$, $\ro_2 = \rz_{k+1} \cap \cdots \cap \rz_n$, $\tro_1 = \trz_1 \cap \cdots \cap \trz_k$ and $\tro_2 = \trz_{k+1} \cap \cdots \cap \trz_n$.\\
        
        By hypothesis we also have:
        \[
            \Phi_i \rhd \infertight{ \Gamma_i }{ b_i }{ M_2 } { \rz_i }{ \trz_i }
        \]
        for $1 \leq i \leq n$.\\
        
        Given $\Phi_1' \rhd \infertight{ \Gamma_1', y_1:(\rz_1 \cap \cdots \cap \rz_k, \trz_1 \cap \cdots \cap \trz_k), y_2:(\rz_{k+1} \cap \cdots \cap \rz_n, \trz_{k+1} \cap \cdots \cap \trz_n), \Gamma_2' }{ b }{ M_1' }{ \rt }{ \trt }$ and $\Phi_j$ for $1 \leq j \leq k$, by the induction hypothesis, there is a derivation ending with
        \[
            \infertight{ \Gamma_1', y_2:(\rz_{k+1} \cap \cdots \cap \rz_n, \trz_{k+1} \cap \cdots \cap \trz_n), \Gamma_2', \sum_{j=1}^k \Gamma_j }{ b + b_1 + \dots + b_k }{ \sub{M_1'}{M_2/y_1} }{ \rt }{ \trt }.
        \]
        
        Now given that derivation and $\Phi_j$ for $k+1 \leq j \leq n$, by the induction hypothesis, there is a derivation
        \[
            \Phi' \rhd \infertight{ \Gamma_1', \Gamma_2', \sum_{j=k+1}^n \Gamma_j }{ b + b_1 + \dots + b_k + b_{k+1} + \dots + b_n }{ \sub{(\sub{M_1'}{M_2/y_1})}{M_2/y_2} }{ \rt }{ \trt },
        \]
        
        which is the derivation we wanted.\\
        
        Since $x \neq y_1$, $x \neq y_2$, $y_1 \neq y_2$ and $y_1, y_2, x$ do not occur in $M_2$ and $x$ does not occur free in $M_1'$, then:
        \begin{align*}
                \sub{(\sub{M_1'}{M_2/y_1})}{M_2/y_2}
                &= \sub{(\sub{(\sub{M_1'}{x/y_1})}{M_2/x})}{M_2/y_2}\\
                &= \sub{(\sub{(\sub{(\sub{M_1'}{x/y_1})}{M_2/x})}{x/y_2})}{M_2/x}\\
                &= \sub{(\sub{(\sub{M_1'}{x/y_1})}{x/y_2})}{M_2/x}\\
                &= \sub{(\sub{M_1'}{x/y_1,x/y_2})}{M_2/x}
        \end{align*}
        
        And as $M_1 = M_1' [x/y_1, x/y_2]$, we have $\sub{(\sub{M_1'}{M_2/y_1})}{M_2/y_2} = \sub{M_1}{M_2/x}$.\\
        
        Also $(\Gamma_1', \Gamma_2') = \Gamma_x$ and $b + b_1 + \dots + b_k + b_{k+1} + \dots + b_n = b + b_1 + \dots + b_n$.\\
        
        So there is indeed $\Phi' \rhd \infertight{ \Gamma_x, \sum_{i=1}^n \Gamma_i }{ b + b_1 + \dots + b_n }{ \sub{M_1}{M_2/x} }{ \rt }{ \trt }$.\\
    \end{enumerate}
    
    \item \tul{($\multimap$ Intro)}:\\
    
    Then we have
    \[
        \Phi \rhd \infertight{ \Gamma }{ b'+1 }{ \lambda y.M }{ \rz \multimap \rt' }{ \trz \multimap \trt' },
    \]
    
    which follows from $\Phi_1' \rhd \infertight{ \Gamma, y:(\rz, \trz) }{ b' }{ M }{ \rt' }{ \trt' }$.
    
    So $M_1 = \lambda y.M$, $b = b'+1$, $\rt = \rz \multimap \rt'$, $\trt = \trz \multimap \trt'$ and $x \neq y$.\\
    
    Since $x:(\rz_1 \cap \cdots \cap \rz_n, \trz_1 \cap \cdots \cap \trz_n) \in \Gamma$, then $x:(\rz_1 \cap \cdots \cap \rz_n, \trz_1 \cap \cdots \cap \trz_n) \in (\Gamma, y:(\rz, \trz))$.\\
    
    By hypothesis we also have:
    \[
        \Phi_i \rhd \infertight{ \Gamma_i }{ b_i }{ M_2 } { \rz_i }{ \trz_i }
    \]
    for $1 \leq i \leq n$.\\
    
    Given $\Phi_1'$ and $\Phi_i$, by the induction hypothesis, there is a derivation ending with
    \[
        \infertight{ \Gamma_x, y:(\rz, \trz), \sum_{i=1}^n \Gamma_i }{ b' + b_1 + \dots + b_n }{ \sub{M}{M_2/x} }{ \rt' }{ \trt' }.
    \]
    
    We can now perform consecutive applications of (Exchange) in order to get
    \[
        \infertight{ \Gamma_x, \sum_{i=1}^n \Gamma_i, y:(\rz, \trz) }{ b' + b_1 + \dots + b_n }{ \sub{M}{M_2/x} }{ \rt' }{ \trt' }.
    \]
    
    Finally, by rule ($\multimap$ Intro), we get the final judgment we wanted:
    
    \begin{center}
    \begin{mathprooftree}
        \AxiomC{$\infertight{ \Gamma_x, \sum_{i=1}^n \Gamma_i, y:(\rz, \trz) }{ b' + b_1 + \dots + b_n }{ \sub{M}{M_2/x} }{ \rt' }{ \trt' }$}
        \UnaryInfC{$\infertight{ \Gamma_x, \sum_{i=1}^n \Gamma_i }{ b' + b_1 + \dots + b_n + 1 }{ \lambda y.(\sub{M}{M_2/x}) }{ \rz \multimap \rt' }{ \trz \multimap \trt' }$}
    \end{mathprooftree}
    \end{center}
    
    Since $M_1 = \lambda y.M$ and $x \neq y$, then $\lambda y.(\sub{M}{M_2/x}) = \sub{(\lambda y.M)}{M_2/x} = \sub{M_1}{M_2/x}$.
    
    Also $b = b'+1$, so $b' + b_1 + \dots + b_n + 1 = b + b_1 + \dots + b_n$.\\
    
    So there is indeed $\Phi' \rhd \infertight{ \Gamma_x, \sum_{i=1}^n \Gamma_i }{ b + b_1 + \dots + b_n }{ \sub{M_1}{M_2/x} }{ \rt }{ \trt }$.\\

    \item \tul{($\rightarrow$ Intro)}:\\
    
    Similar to the previous case.\\

    \item \tul{($\multimap$ Intro\textsubscript{t})}:\\
    
    Then we have
    \[
        \Phi \rhd \infertight{ \Gamma }{ b' }{ \lambda y.M }{ \rz \multimap \rt' }{ \msf{Abs} },
    \]
    
    which follows from $\Phi_1' \rhd \infertight{ \Gamma, y:(\rz, \msf{tight}) }{ b' }{ M }{ \rt' }{ \msf{tight} }$.
    
    So $M_1 = \lambda y.M$, $b = b'$, $\rt = \rz \multimap \rt'$, $\trt = \msf{Abs}$ and $x \neq y$.\\
    
    Since $x:(\rz_1 \cap \cdots \cap \rz_n, \trz_1 \cap \cdots \cap \trz_n) \in \Gamma$, then $x:(\rz_1 \cap \cdots \cap \rz_n, \trz_1 \cap \cdots \cap \trz_n) \in (\Gamma, y:(\rz, \msf{tight}))$.\\
    
    By hypothesis we also have:
    \[
        \Phi_i \rhd \infertight{ \Gamma_i }{ b_i }{ M_2 } { \rz_i }{ \trz_i }
    \]
    for $1 \leq i \leq n$.\\
    
    Given $\Phi_1'$ and $\Phi_i$, by the induction hypothesis, there is a derivation ending with
    \[
        \infertight{ \Gamma_x, y:(\rz, \msf{tight}), \sum_{i=1}^n \Gamma_i }{ b' + b_1 + \dots + b_n }{ \sub{M}{M_2/x} }{ \rt' }{ \msf{tight} }.
    \]
    
    We can now perform consecutive applications of (Exchange) in order to get
    \[
        \infertight{ \Gamma_x, \sum_{i=1}^n \Gamma_i, y:(\rz, \msf{tight}) }{ b' + b_1 + \dots + b_n }{ \sub{M}{M_2/x} }{ \rt' }{ \msf{tight} }.
    \]
    
    Finally, by rule ($\multimap$ Intro\textsubscript{t}), we get the final judgment we wanted:
    
    \begin{center}
    \begin{mathprooftree}
        \AxiomC{$\infertight{ \Gamma_x, \sum_{i=1}^n \Gamma_i, y:(\rz, \msf{tight}) }{ b' + b_1 + \dots + b_n }{ \sub{M}{M_2/x} }{ \rt' }{ \msf{tight} }$}
        \UnaryInfC{$\infertight{ \Gamma_x, \sum_{i=1}^n \Gamma_i }{ b' + b_1 + \dots + b_n }{ \lambda y.(\sub{M}{M_2/x}) }{ \rz \multimap \rt' }{ \msf{Abs} }$}
    \end{mathprooftree}
    \end{center}
    
    Since $M_1 = \lambda y.M$ and $x \neq y$, then $\lambda y.(\sub{M}{M_2/x}) = \sub{(\lambda y.M)}{M_2/x} = \sub{M_1}{M_2/x}$.
    
    Also $b = b'$, so $b' + b_1 + \dots + b_n = b + b_1 + \dots + b_n$.\\
    
    So there is indeed $\Phi' \rhd \infertight{ \Gamma_x, \sum_{i=1}^n \Gamma_i }{ b + b_1 + \dots + b_n }{ \sub{M_1}{M_2/x} }{ \rt }{ \trt }$.\\

    \item \tul{($\rightarrow$ Intro\textsubscript{t})}:\\
    
    Similar to the previous case.\\

    \item \tul{($\multimap$ Elim)}:\\
    
    Then we have
    \[
        \Phi \rhd \infertight{ \Gamma_1', \Gamma_2' }{ b_1' + b_2' }{ N_1 N_2 }{ \rt }{ \trt },
    \]
    
    which follows from $\Phi_1' \rhd \infertight{ \Gamma_1' }{ b_1' }{ N_1 }{ \rz \multimap \rt }{ \trz \multimap \trt }$ and $\Phi_2' \rhd \infertight{ \Gamma_2' }{ b_2' }{ N_2 } { \rz }{ \trz }$.
    
    Since $x \in \dom(\Gamma)$ and $\Gamma = (\Gamma_1', \Gamma_2')$, either $x \in \dom(\Gamma_1')$ (and $x \in \fv{N_1}$, by \autoref{xfreesys2}) or $x \in \dom(\Gamma_2')$ (and $x \in \fv{N_2}$, by \autoref{xfreesys2}). Note that there is not the case where $x \in \dom(\Gamma_1')$ and $x \in \dom(\Gamma_2')$ simultaneously, otherwise $(\Gamma_1', \Gamma_2')$ would be inconsistent.
    
    So there are two different cases depending on $x$:
    
    \begin{enumerate}
        \item $x \in \dom(\Gamma_1')$ and $x \notin \dom(\Gamma_2')$:\\
        
        So $\Gamma = (\Gamma_1', \Gamma_2')$, $\Gamma_x = ({\Gamma_1'}_x, \Gamma_2')$, $M_1 = N_1 N_2$ and $b = b_1' + b_2'$.\\
        
        And $x:(\rz_1 \cap \cdots \cap \rz_n, \trz_1 \cap \cdots \cap \trz_n) \in \Gamma_1'$.\\
        
        By hypothesis we also have:
        \[
            \Phi_i \rhd \infertight{ \Gamma_i }{ b_i }{ M_2 } { \rz_i }{ \trz_i }
        \]
        for $1 \leq i \leq n$.\\
        
        Given $\Phi_1'$ and $\Phi_i$, by the induction hypothesis, there is a derivation ending with
        \[
            \infertight{ {\Gamma_1'}_x, \sum_{i=1}^n \Gamma_i }{ b_1' + b_1 + \dots + b_n }{ \sub{N_1}{M_2/x} }{ \rz \multimap \rt }{ \trz \multimap \trt }.
        \]
        
        By rule ($\multimap$ Elim), with $\infertight{ \Gamma_2' }{ b_2' }{ N_2 } { \rz }{ \trz }$ from $\Phi_2'$, we get:
        
        \begin{center}
        \begin{mathprooftree}
            \AxiomC{$\infertight{ {\Gamma_1'}_x, \sum_{i=1}^n \Gamma_i }{ b_1' + b_1 + \dots + b_n }{ \sub{N_1}{M_2/x} }{ \rz \multimap \rt }{ \trz \multimap \trt }$}
            \AxiomC{$\infertight{ \Gamma_2' }{ b_2' }{ N_2 } { \rz }{ \trz }$}
            \BinaryInfC{$\infertight{ {\Gamma_1'}_x, \sum_{i=1}^n \Gamma_i, \Gamma_2' }{ b_1' + b_1 + \dots + b_n + b_2' }{ (\sub{N_1}{M_2/x}) N_2 }{ \rt }{ \trt }$}
        \end{mathprooftree}
        \end{center}
        
        Note that $({\Gamma_1'}_x, \sum_{i=1}^n \Gamma_i, \Gamma_2')$ is consistent because of our initial assumption that $\fv{M_1} \cap \fv{M_2} = \emptyset$.\\
        
        We can now perform consecutive applications of (Exchange) in order to get the final judgment we wanted:
        \[
            \infertight{ {\Gamma_1'}_x, \Gamma_2', \sum_{i=1}^n \Gamma_i }{ b_1' + b_1 + \dots + b_n + b_2' }{ (\sub{N_1}{M_2/x}) N_2 }{ \rt }{ \trt }
        \]
        
        Since $M_1 = N_1 N_2$ and $x \notin \fv{N_2}$ (by \autoref{xfreesys2}, since $x \notin \dom(\Gamma_2')$), then $(\sub{N_1}{M_2/x}) N_2 = (\sub{N_1}{M_2/x}) (\sub{N_2}{M_2/x}) = \sub{(N_1 N_2)}{M_2/x} = \sub{M_1}{M_2/x}$.
    
        Also $b = b_1' + b_2'$, so $b_1' + b_1 + \dots + b_n + b_2' = b + b_1 + \dots + b_n$.\\
    
        So there is indeed $\Phi' \rhd \infertight{ \Gamma_x, \sum_{i=1}^n \Gamma_i }{ b + b_1 + \dots + b_n }{ \sub{M_1}{M_2/x} }{ \rt }{ \trt }$.\\

        \item $x \in \dom(\Gamma_2')$ and $x \notin \dom(\Gamma_1')$:\\
        
        So $\Gamma = (\Gamma_1', \Gamma_2')$, $\Gamma_x = (\Gamma_1', {\Gamma_2'}_x)$, $M_1 = N_1 N_2$ and $b = b_1' + b_2'$.\\
        
        And $x:(\rz_1 \cap \cdots \cap \rz_n, \trz_1 \cap \cdots \cap \trz_n) \in \Gamma_2'$.\\
        
        By hypothesis we also have:
        \[
            \Phi_i \rhd \infertight{ \Gamma_i }{ b_i }{ M_2 } { \rz_i }{ \trz_i }
        \]
        for $1 \leq i \leq n$.\\
        
        Given $\Phi_2'$ and $\Phi_i$, by the induction hypothesis, there is a derivation ending with
        \[
            \infertight{ {\Gamma_2'}_x, \sum_{i=1}^n \Gamma_i }{ b_2' + b_1 + \dots + b_n }{ \sub{N_2}{M_2/x} }{ \rz }{ \trz }.
        \]
        
        By rule ($\multimap$ Elim), with $\infertight{ \Gamma_1' }{ b_1' }{ N_1 }{ \rz \multimap \rt }{ \trz \multimap \trt }$ from $\Phi_1'$, we get the final judgment we wanted:
        
        \begin{center}
        \begin{mathprooftree}
            \AxiomC{$\infertight{ \Gamma_1' }{ b_1' }{ N_1 }{ \rz \multimap \rt }{ \trz \multimap \trt }$}
            \AxiomC{$\infertight{ {\Gamma_2'}_x, \sum_{i=1}^n \Gamma_i }{ b_2' + b_1 + \dots + b_n }{ \sub{N_2}{M_2/x} }{ \rz }{ \trz }$}
            \BinaryInfC{$\infertight{ \Gamma_1', {\Gamma_2'}_x, \sum_{i=1}^n \Gamma_i }{ b_1' + b_2' + b_1 + \dots + b_n }{ N_1 (\sub{N_2}{M_2/x}) }{ \rt }{ \trt }$}
        \end{mathprooftree}
        \end{center}
        
        Note that $(\Gamma_1', {\Gamma_2'}_x, \sum_{i=1}^n \Gamma_i)$ is consistent because of our initial assumption that $\fv{M_1} \cap \fv{M_2} = \emptyset$.\\
        
        Since $M_1 = N_1 N_2$ and $x \notin \fv{N_1}$ (by \autoref{xfreesys2}, since $x \notin \dom(\Gamma_1')$), then $N_1 (\sub{N_2}{M_2/x}) = (\sub{N_1}{M_2/x}) (\sub{N_2}{M_2/x}) = \sub{(N_1 N_2)}{M_2/x} = \sub{M_1}{M_2/x}$.
    
        Also $b = b_1' + b_2'$, so $b_1' + b_2' + b_1 + \dots + b_n = b + b_1 + \dots + b_n$.\\
    
        So there is indeed $\Phi' \rhd \infertight{ \Gamma_x, \sum_{i=1}^n \Gamma_i }{ b + b_1 + \dots + b_n }{ \sub{M_1}{M_2/x} }{ \rt }{ \trt }$.\\
    \end{enumerate}
    
    \item \tul{($\multimap$ Elim\textsubscript{t})}:\\
    
    Similar to the previous case.\\
    
    \item \tul{($\rightarrow$ Elim)}:\\
    
    Then we have
    \[
        \Phi \rhd \infertight{ \Gamma', \sum_{j=1}^m \Gamma_j' }{ b' + b_1' + \dots + b_m' }{ N_1 N_2 }{ \rt }{ \trt }
    \]
    
    which follows from $\Phi_1'' \rhd \infertight{ \Gamma' }{ b' }{ N_1 }{ \rz_1' \cap \cdots \cap \rz_m' \rightarrow \rt }{ \trz_1' \cap \cdots \cap \trz_m' \rightarrow \trt }$,
    
    $\Phi_1' \rhd \infertight{ \Gamma_1' }{ b_1' }{ N_2 } { \rz_1' }{ \trz_1' }$, \dots, $\Phi_m' \rhd \infertight{ \Gamma_m' }{ b_m' }{ N_2 } { \rz_m' }{ \trz_m' }$ and $m \geq 2$.
    
    Since $x \in \dom(\Gamma)$ and $\Gamma = (\Gamma', \sum_{j=1}^m \Gamma_j')$, either $x \in \dom(\Gamma')$ (and $x \in \fv{N_1}$, by \autoref{xfreesys2}) or $x \in \dom(\Gamma_j')$, for $1 \leq j \leq m$ (and $x \in \fv{N_2}$, by \autoref{xfreesys2}). Note that there is not the case where $x \in \dom(\Gamma')$ and $x \in \dom(\Gamma_j')$ (for $1 \leq j \leq m$) simultaneously, otherwise $(\Gamma', \sum_{j=1}^m \Gamma_j')$ would be inconsistent.
    
    So there are two different cases depending on $x$:
    
    \begin{enumerate}
        \item $x \in \dom(\Gamma')$ and $x \notin \dom(\Gamma_j')$, for $1 \leq j \leq m$:\\
        
        So $\Gamma = (\Gamma', \sum_{j=1}^m \Gamma_j')$, $\Gamma_x = ({\Gamma'}_x, \sum_{j=1}^m \Gamma_j')$, $M_1 = N_1 N_2$ and $b = b' + b_1' + \dots + b_m'$.\\
        
        And $x:(\rz_1 \cap \cdots \cap \rz_n, \trz_1 \cap \cdots \cap \trz_n) \in \Gamma'$.\\
        
        By hypothesis we also have:
        \[
            \Phi_i \rhd \infertight{ \Gamma_i }{ b_i }{ M_2 } { \rz_i }{ \trz_i }
        \]
        for $1 \leq i \leq n$.\\
        
        Given $\Phi_1''$ and $\Phi_i$, by the induction hypothesis, there is a derivation ending with
        \[
            \infertight{ {\Gamma'}_x, \sum_{i=1}^n \Gamma_i }{ b' + b_1 + \dots + b_n }{ \sub{N_1}{M_2/x} }{ \rz_1' \cap \cdots \cap \rz_m' \rightarrow \rt }{ \trz_1' \cap \cdots \cap \trz_m' \rightarrow \trt }.
        \]
        
        By rule ($\rightarrow$ Elim), with $\infertight{ \Gamma_1' }{ b_1' }{ N_2 } { \rz_1' }{ \trz_1' }$, \dots, $\infertight{ \Gamma_m' }{ b_m' }{ N_2 } { \rz_m' }{ \trz_m' }$ from $\Phi_1'$, \dots, $\Phi_m'$, respectively, we get:
        
        \begin{center}
        \begin{mathprooftree}
            \AxiomC{$\infertight{ {\Gamma'}_x, \sum_{i=1}^n \Gamma_i }{ b' + b_1 + \dots + b_n }{ \sub{N_1}{M_2/x} }{ \rz_1' \cap \cdots \cap \rz_m' \rightarrow \rt }{ \trz_1' \cap \cdots \cap \trz_m' \rightarrow \trt }$}
            \noLine
            \UnaryInfC{$\infertight{ \Gamma_1' }{ b_1' }{ N_2 } { \rz_1' }{ \trz_1' }  \text { } \cdots \text{ }  \infertight{ \Gamma_m' }{ b_m' }{ N_2 } { \rz_m' }{ \trz_m' }$}
            \UnaryInfC{$\infertight{ {\Gamma'}_x, \sum_{i=1}^n \Gamma_i, \sum_{j=1}^m \Gamma_j' }{ b' + b_1 + \dots + b_n + b_1' + \dots + b_m' }{ (\sub{N_1}{M_2/x}) N_2 }{ \rt }{ \trt }$}
        \end{mathprooftree}
        \end{center}
        
        \hfill\\
        Note that $({\Gamma'}_x, \sum_{i=1}^n \Gamma_i, \sum_{j=1}^m \Gamma_j')$ is consistent because of our initial assumption that $\fv{M_1} \cap \fv{M_2} = \emptyset$.\\
        
        We can now perform consecutive applications of (Exchange) in order to get the final judgment we wanted:
        \[
            \infertight{ {\Gamma'}_x, \sum_{j=1}^m \Gamma_j', \sum_{i=1}^n \Gamma_i }{ b' + b_1 + \dots + b_n + b_1' + \dots + b_m' }{ (\sub{N_1}{M_2/x}) N_2 }{ \rt }{ \trt }
        \]
        
        Since $M_1 = N_1 N_2$ and $x \notin \fv{N_2}$ (by \autoref{xfreesys2}, since $x \notin \dom(\Gamma_j')$ for $1 \leq j \leq m$), then $(\sub{N_1}{M_2/x}) N_2 = (\sub{N_1}{M_2/x}) (\sub{N_2}{M_2/x}) = \sub{(N_1 N_2)}{M_2/x} = \sub{M_1}{M_2/x}$.
    
        Also $b = b' + b_1' + \dots + b_m'$, so $b' + b_1 + \dots + b_n + b_1' + \dots + b_m' = b + b_1 + \dots + b_n$.\\
    
        So there is indeed $\Phi' \rhd \infertight{ \Gamma_x, \sum_{i=1}^n \Gamma_i }{ b + b_1 + \dots + b_n }{ \sub{M_1}{M_2/x} }{ \rt }{ \trt }$.\\

        \item $x \in \dom(\Gamma_j')$, for $1 \leq j \leq m$, and $x \notin \dom(\Gamma')$:\\
        
        So $\Gamma = (\Gamma', \sum_{j=1}^m \Gamma_j')$, $\Gamma_x = (\Gamma', \sum_{j=1}^m {\Gamma_j'}_x)$, $M_1 = N_1 N_2$ and $b = b' + b_1' + \dots + b_m'$.\\
        
        And $x:(\rz_1 \cap \cdots \cap \rz_n, \trz_1 \cap \cdots \cap \trz_n) \in \sum_{j=1}^m \Gamma_j'$, which means that the sequences $(\rz_1 \cap \cdots \cap \rz_n, \trz_1 \cap \cdots \cap \trz_n)$ can be split between the environments $\Gamma_1', \dots, \Gamma_m'$. Also, note that this implies $n \geq m$ (because by \autoref{xfreesys2}, $x \in \dom(\Gamma_j')$ for all $1 \leq j \leq m$).\\
        
        Then for $1 = k_1 < \dots < k_m < k_{m+1} = n+1$ and $1 \leq j \leq m$,
        
        let $x:(\rz_{k_j} \cap \cdots \cap \rz_{k_{(j+1)}-1}, \trz_{k_j} \cap \cdots \cap \trz_{k_{(j+1)}-1}) \in \Gamma_j'$.\\
        
        By hypothesis we also have:
        \[
            \Phi_i \rhd \infertight{ \Gamma_i }{ b_i }{ M_2 } { \rz_i }{ \trz_i }
        \]
        for $1 \leq i \leq n$.\\
        
        So for each $1 \leq j \leq m$, given $\Phi_j'$ and $\Phi_{k_j}, \dots, \Phi_{k_{(j+1)}-1}$, by the induction hypothesis, there is a derivation ending with
        \[
            \infertight{ {\Gamma_j'}_x, \sum_{i={k_j}}^{k_{(j+1)}-1} \Gamma_i }{ b_j' + b_{k_j} + \dots + b_{k_{(j+1)}-1} }{ \sub{N_2}{M_2/x} }{ \rz_j' }{ \trz_j' }.
        \]
        
        By rule ($\rightarrow$ Elim), with $\infertight{ \Gamma' }{ b' }{ N_1 }{ \rz_1' \cap \cdots \cap \rz_m' \rightarrow \rt }{ \trz_1' \cap \cdots \cap \trz_m' \rightarrow \trt }$ from $\Phi_1''$, we get the final judgment we wanted:\\
        
        \scalebox{0.79}{
        \begin{mathprooftree}
            \AxiomC{$\infertight{ \Gamma' }{ b' }{ N_1 }{ \rz_1' \cap \cdots \cap \rz_m' \rightarrow \rt }{ \trz_1' \cap \cdots \cap \trz_m' \rightarrow \trt }$}
            \noLine
            \UnaryInfC{$\infertight{ {\Gamma_1'}_x, \sum_{i=1}^{k_{2}-1} \Gamma_i }{ b_1' + b_{1} + \dots + b_{k_{2}-1} }{ \sub{N_2}{M_2/x} }{ \rz_1' }{ \trz_1' }  \text { } \cdots \text{ }  \infertight{ {\Gamma_m'}_x, \sum_{i={k_m}}^{n} \Gamma_i }{ b_m' + b_{k_m} + \dots + b_{n} }{ \sub{N_2}{M_2/x} } { \rz_m' }{ \trz_m' }$}
            \UnaryInfC{$\infertight{ \Gamma', (({\Gamma_1'}_x, \sum_{i=1}^{k_{2}-1} \Gamma_i) + \dots + ({\Gamma_m'}_x, \sum_{i={k_m}}^{n} \Gamma_i)) }{ b'' }{ N_1 (\sub{N_2}{M_2/x}) }{ \rt }{ \trt }$}
        \end{mathprooftree}}\\
        
        where $b'' = b' + (b_1' + b_{1} + \dots + b_{k_{2}-1}) + \dots + (b_m' + b_{k_m} + \dots + b_{n})$.\\
        
        By our initial assumption that $\fv{M_1} \cap \fv{M_2} = \emptyset$, by \autoref{xfreesys2}, and by looking at the definition of (+), we have:
        \begin{align*}
            &\Gamma', (({\Gamma_1'}_x, \sum_{i=1}^{k_{2}-1} \Gamma_i) + \dots + ({\Gamma_m'}_x, \sum_{i={k_m}}^{n} \Gamma_i))\\
            =\; &\Gamma', (({\Gamma_1'}_x + \dots + {\Gamma_m'}_x) , (\sum_{i=1}^{k_{2}-1} \Gamma_i + \dots + \sum_{i={k_m}}^{n} \Gamma_i))\\
            =\; &\Gamma', (\sum_{j=1}^m {\Gamma_j'}_x , \sum_{i=1}^n \Gamma_i)\\
            =\; &(\Gamma', \sum_{j=1}^m {\Gamma_j'}_x) , \sum_{i=1}^n \Gamma_i\\
            =\; &\Gamma_x , \sum_{i=1}^n \Gamma_i.
        \end{align*}
        
        Since $M_1 = N_1 N_2$ and $x \notin \fv{N_1}$ (by \autoref{xfreesys2}, since $x \notin \dom(\Gamma')$), then $N_1 (\sub{N_2}{M_2/x}) = (\sub{N_1}{M_2/x}) (\sub{N_2}{M_2/x}) = \sub{(N_1 N_2)}{M_2/x} = \sub{M_1}{M_2/x}$.\\
    
        Also since $b = b' + b_1' + \dots + b_m'$, we have:
        \begin{align*}
            b''
            &= b' + (b_1' + b_{1} + \dots + b_{k_{2}-1}) + \dots + (b_m' + b_{k_m} + \dots + b_{n})\\
            &= b' + (b_1' + \dots + b_m') + (b_{1} + \dots + b_{k_{2}-1} + \dots + b_{k_m} + \dots + b_{n})\\
            &= b' + (b_1' + \dots + b_m') + (b_1 + \dots + b_n)\\
            &= b + b_1 + \dots + b_n.
        \end{align*}
    
        So there is indeed $\Phi' \rhd \infertight{ \Gamma_x, \sum_{i=1}^n \Gamma_i }{ b + b_1 + \dots + b_n }{ \sub{M_1}{M_2/x} }{ \rt }{ \trt }$.\\
    \end{enumerate}

    \item \tul{($\rightarrow$ Elim\textsubscript{t})}:\\
    
    Similar to the previous case.
\end{enumerate}
\end{proof}
\end{lemma}

We now show an important property that relates contracted terms with their linear counterpart. Basically, it says that the following diagram commutes (under the described conditions):

\begin{center}
\begin{tikzpicture}[line width=1pt,shorten >=2pt,shorten <=2pt]
\node (m) at (0,0) {$M$};
\node (n) at (2,0) {$N$};
\node (m') at (0,2) {$M'$};
\node (n') at (2,2) {$N'$};
\draw[-To] (m) to node[auto, swap] {$\beta$} (n);
\draw[->] (m') to node[auto, swap] {$\s(M')$} (m);
\draw[-To] (m') to node[auto] {$\beta$} (n');
\draw[->] (n') to node[auto] {$\s(N')$} (n);
\end{tikzpicture}
\end{center}

\begin{lemma}\label{substreduction}
Let $M \longrightarrow N$ and $M = \s(M')$ for some substitution $\s = \subst{x/x_1, x/x_2}$ where $x_1, x_2$ occur free in $M'$ and $x$ does not occur in $M'$. Then there exists a term $N'$ such that $N = \s(N')$ and $M' \longrightarrow N'$.

\begin{proof}
This trivially holds because $\s$ simply renames free variables -- suppose that in $M$ we annotate each occurrence of $x$ with the variable it substituted (so that each occurrence is either $x^{x_1}$ or $x^{x_2}$). Then $M$ is equal to $M'$, up to renaming of variables.
\end{proof}
\end{lemma}

\begin{convention}\label{contractedvars}
Without loss of generality, we assume that, in a derivation tree, all contracted variables (i.e., variables that, at some point in the derivation tree, disappear from the term and environment by an application of the (Contraction) rule) are different from any other variable in the derivation tree.\\
We also assume that when applying (Contraction), the new variables that substitute the contracted ones are also different from any other variable in the derivation tree.
\end{convention}

\begin{lemma}[Quantitative subject reduction]\label{subjectreduction} 
If $\Phi \rhd \infertight{ \Gamma }{ b }{ M } { \rt }{ \trt }$ is tight and $M \longrightarrow N$, then $b \geq 1$ and there exists a tight derivation $\Phi'$ such that $\Phi' \rhd \infertight{ \Gamma }{ b-1 }{ N } { \rt }{ \trt }$.

\begin{proof}
We prove the following stronger statement:

Assume $M \longrightarrow N$, $\Phi \rhd \infertight{ \Gamma }{ b }{ M } { \rt }{ \trt }$, $\msf{tight}(\Gamma)$, and either $\msf{tight}(\trt)$ or $\neg \msf{abs}(M)$.

Then there exists a derivation $\Phi' \rhd \infertight{ \Gamma }{ b-1 }{ N } { \rt }{ \trt }$.

We prove this statement by induction on $M \longrightarrow N$.

\begin{enumerate}
    \item Rule
    \AxiomC{}
    \UnaryInfC{$(\lambda x.M_1) N_1 \longrightarrow \sub{M_1}{N_1/x}$}
    \DisplayProof:\\
    
    Assume $\Phi \rhd \infertight{ \Gamma }{ b }{ (\lambda x.M_1) N_1 } { \rt }{ \trt }$ and $\msf{tight}(\Gamma)$.\\
    
    So at some point in the derivation $\Phi$, either the rule ($\multimap$ Elim) or ($\rightarrow$ Elim) is applied (not ($\multimap$ Elim\textsubscript{t}) nor ($\rightarrow$ Elim\textsubscript{t}) because it is not possible to derive the type $\msf{Neutral}$ for an abstraction) and is then followed by zero or more applications of the rules (Exchange) and/or (Contraction).
    
    Case where the last rule different from (Exchange) and (Contraction) applied in $\Phi$ is:
    
    \begin{enumerate}
        \item ($\multimap$ Elim):
        
        Then at some point in $\Phi$ we have:
        
        \begin{center}
        \begin{mathprooftree}
            \AxiomC{$\Phi_1 \rhd \infertight{ \Gamma_1 }{ b_1 }{ (\lambda x.M_1') }{ \rz \multimap \rt }{ \trz \multimap \trt }$}
            \AxiomC{$\Phi_2 \rhd \infertight{ \Gamma_2 }{ b_2 }{ N_1' } { \rz }{ \trz }$}
            \BinaryInfC{$\infertight{ \Gamma_1, \Gamma_2 }{ b_1 + b_2 }{ (\lambda x.M_1') N_1' }{ \rt }{ \trt }$}
        \end{mathprooftree}
        \end{center}
        
        \hfill
        
        Assume that after the application of this rule, the rule (Contraction) was applied $n$ times (with $n \geq 0$) and (Exchange) zero or more times, and let $\s = [y_n/x_n, y_n/x_n'] \circ [y_{n-1}/x_{n-1}, y_{n-1}/x_{n-1}'] \circ \cdots \circ [y_1/x_1, y_1/x_1']$ be the substitution that reflects the $n$ applications of the rule (Contraction). Then we have:
        
        \begin{itemize}
            \item $M_1 = \s(M_1')$;
            \item $N_1 = \s(N_1')$;
            \item $b = b_1 + b_2$.\\
        \end{itemize}
        
        Since $\Phi_1 \rhd \infertight{ \Gamma_1 }{ b_1 }{ (\lambda x.M_1') }{ \rz \multimap \rt }{ \trz \multimap \trt }$, at some point in the derivation $\Phi_1$, the rule ($\multimap$ Intro) is applied, followed by zero or more applications of the rules (Exchange) and/or (Contraction). So at some point in $\Phi_1$ we have:
        
        \begin{center}
        \begin{mathprooftree}
            \AxiomC{$\Phi_1' \rhd \infertight{ \Gamma_1', x:(\rz, \trz) }{ b_1' }{ M_1'' }{ \rt }{ \trt }$}
            \UnaryInfC{$\infertight{ \Gamma_1' }{ b_1'+1 }{ \lambda x.M_1'' }{ \rz \multimap \rt }{ \trz \multimap \trt }$}
        \end{mathprooftree}
        \end{center}
        
        \hfill
        
        Assume that after the application of this rule, the rule (Contraction) was applied $m$ times (with $m \geq 0$) and (Exchange) zero or more times, and let $\s_1 = [y_n'/z_m, y_n'/z_m'] \circ [y_{m-1}'/z_{m-1}, y_{m-1}'/z_{m-1}'] \circ \cdots \circ [y_1'/z_1, y_1'/z_1']$ be the substitution that reflects the $m$ applications of the rule (Contraction). Then we have:
        
        \begin{itemize}
            \item $M_1' = \s_1(M_1'')$;
            \item $b_1 = b_1' + 1$.\\
        \end{itemize}
        
        We can then apply \autoref{substlema} for the derivations $\Phi_1' \rhd \infertight{ \Gamma_1', x:(\rz, \trz) }{ b_1' }{ M_1'' }{ \rt }{ \trt }$ and $\Phi_2 \rhd \infertight{ \Gamma_2 }{ b_2 }{ N_1' } { \rz }{ \trz }$ to obtain
        \[
            \Phi'' \rhd \infertight{ \Gamma_1', \Gamma_2 }{ b_1' + b_2 }{ \sub{M_1''}{N_1'/x} }{ \rt }{ \trt }.
        \]
        
        Note that we can assume $\Gamma_1', \Gamma_2$ to be consistent by \autoref{contractedvars} and the fact that $\Gamma_1, \Gamma_2$ is consistent.\\
        
        If we perform the $m$ applications of (Contraction) (and the necessary applications of (Exchange)) over the same variables over which they were performed in $\Phi_1$ after the application of ($\multimap$ Intro), i.e., over the variables $z_1, z_1', z_2, z_2', \dots, z_m, z_m'$, we can obtain:
        \[
            \infertight{ \Gamma_3' }{ b_1' + b_2 }{ \s_1 (\sub{M_1''}{N_1'/x}) }{ \rt }{ \trt }
        \]
        
        where $\Gamma_3' \equiv (\Gamma_1, \Gamma_2)$. (By \autoref{contractedvars}, the variables in $\Gamma_2$ are not substituted.)\\
        
        If we now perform the $n$ applications of (Contraction) (and the necessary applications of (Exchange)) over the same variables over which they were performed in $\Phi$ after the application of ($\multimap$ Elim), i.e., over the variables $x_1, x_1', x_2, x_2', \dots, x_n, x_n'$, we can obtain:
        \[
            \infertight{ \Gamma_3 }{ b_1' + b_2 }{ \s (\s_1 (\sub{M_1''}{N_1'/x})) }{ \rt }{ \trt }
        \]
        
        where $\Gamma_3 \equiv \Gamma$.\\
        
        Finally, since $\Gamma_3 \equiv \Gamma$, we can apply (Exchange) as many times as needed to get the final judgment we wanted:
        \[
            \infertight{ \Gamma }{ b_1' + b_2 }{ \s (\s_1 (\sub{M_1''}{N_1'/x})) }{ \rt }{ \trt }
        \]
        
        \hfill
        
        Since $b = b_1 + b_2$ and $b_1 = b_1' + 1$, then
        \begin{align*}
            b_1' + b_2
            &= b_1 - 1 + b_2\\
            &= b - 1.
        \end{align*}
        
        By \autoref{contractedvars}, we have $\s_1 (\sub{M_1''}{N_1'/x}) = \sub{(\s_1 (M_1''))}{N_1'/x}$.
        
        Also, $M_1' = \s_1(M_1'')$, so
        \begin{align*}\label{eq:subred.1.a1}
            \s_1 (\sub{M_1''}{N_1'/x})
            &= \sub{(\s_1 (M_1''))}{N_1'/x}\\
            &= \sub{M_1'}{N_1'/x}. \tag{1}\\
        \end{align*}
        
        Because $x$ cannot be in $\s$, then $\s(\sub{M_1'}{N_1'/x}) = \sub{(\s(M_1'))}{\s(N_1')/x}$.
        
        And since $M_1 = \s(M_1')$ and $N_1 = \s(N_1')$, we have $\sub{(\s(M_1'))}{\s(N_1')/x} = \sub{M_1}{N_1/x}$, so
        \begin{align*}\label{eq:subred.1.a2}
            \s(\sub{M_1'}{N_1'/x})
            &= \sub{(\s(M_1'))}{\s(N_1')/x}\\
            &= \sub{M_1}{N_1/x}. \tag{2}\\
        \end{align*}
        
        Then, by \eqref{eq:subred.1.a1} and \eqref{eq:subred.1.a2} we have:
        \begin{align*}
            \s (\s_1 (\sub{M_1''}{N_1'/x}))
            &= \s (\sub{M_1'}{N_1'/x})\\
            &= \sub{M_1}{N_1/x}.
        \end{align*}
        
        So there is indeed $\Phi' \rhd \infertight{ \Gamma }{ b-1 }{ \sub{M_1}{N_1/x} } { \rt }{ \trt }$.\\

        \item ($\rightarrow$ Elim):
        
        Then at some point in $\Phi$ we have:
        
        \begin{center}
        \begin{mathprooftree}
            \AxiomC{$\Phi_1' \rhd \infertight{ \Gamma_1' }{ b_1' }{ (\lambda x.M_1') }{ \rz_1 \cap \cdots \cap \rz_k \rightarrow \rt }{ \trz_1 \cap \cdots \cap \trz_k \rightarrow \trt }$}
            \noLine
            \UnaryInfC{$\Phi_1 \rhd \infertight{ \Gamma_1 }{ b_1 }{ N_1' } { \rz_1 }{ \trz_1 }  \text { } \cdots \text{ }  \Phi_k \rhd \infertight{ \Gamma_k }{ b_k }{ N_1' } { \rz_k }{ \trz_k }$}
            \AxiomC{$k \geq 2$}
            \BinaryInfC{$\infertight{ \Gamma_1', \sum_{i=1}^k \Gamma_i }{ b_1' + b_1 + \dots + b_k }{ (\lambda x.M_1') N_1' }{ \rt }{ \trt }$}
        \end{mathprooftree}
        \end{center}
        
        \hfill
        
        Assume that after the application of this rule, the rule (Contraction) was applied $n$ times (with $n \geq 0$) and (Exchange) zero or more times, and let $\s = [y_n/x_n, y_n/x_n'] \circ [y_{n-1}/x_{n-1}, y_{n-1}/x_{n-1}'] \circ \cdots \circ [y_1/x_1, y_1/x_1']$ be the substitution that reflects the $n$ applications of the rule (Contraction). Then we have:
        
        \begin{itemize}
            \item $M_1 = \s(M_1')$;
            \item $N_1 = \s(N_1')$;
            \item $b = b_1' + b_1 + \dots + b_k$.\\
        \end{itemize}
        
        Since $\Phi_1' \rhd \infertight{ \Gamma_1' }{ b_1' }{ (\lambda x.M_1') }{ \rz_1 \cap \cdots \cap \rz_k \rightarrow \rt }{ \trz_1 \cap \cdots \cap \trz_k \rightarrow \trt }$, at some point in the derivation $\Phi_1'$, the rule ($\rightarrow$ Intro) is applied, followed by zero or more applications of the rules (Exchange) and/or (Contraction). So at some point in $\Phi_1'$ we have:
        
        \begin{center}
        \begin{mathprooftree}
            \AxiomC{$\Phi_1'' \rhd \infertight{ \Gamma_1'', x:(\rz_1 \cap \cdots \cap \rz_k, \trz_1 \cap \cdots \cap \trz_k) }{ b_1'' }{ M_1'' }{ \rt }{ \trt }$}
            \AxiomC{$k \geq 2$}
            \BinaryInfC{$\infertight{ \Gamma_1'' }{ b_1''+1 }{ \lambda x.M_1'' }{ \rz_1 \cap \cdots \cap \rz_k \rightarrow \rt }{ \trz_1 \cap \cdots \cap \trz_k \rightarrow \trt }$}
        \end{mathprooftree}
        \end{center}
        
        \hfill
        
        Assume that after the application of this rule, the rule (Contraction) was applied $m$ times (with $m \geq 0$) and (Exchange) zero or more times, and let $\s_1 = [y_n'/z_m, y_n'/z_m'] \circ [y_{m-1}'/z_{m-1}, y_{m-1}'/z_{m-1}'] \circ \cdots \circ [y_1'/z_1, y_1'/z_1']$ be the substitution that reflects the $m$ applications of the rule (Contraction). Then we have:
        
        \begin{itemize}
            \item $M_1' = \s_1(M_1'')$;
            \item $b_1' = b_1'' + 1$.\\
        \end{itemize}
        
        We can then apply \autoref{substlema} for the derivations $\Phi_1'' \rhd \infertight{ \Gamma_1'', x:(\rz_1 \cap \cdots \cap \rz_k, \trz_1 \cap \cdots \cap \trz_k) }{ b_1'' }{ M_1'' }{ \rt }{ \trt }$ and $\Phi_1 \rhd \infertight{ \Gamma_1 }{ b_1 }{ N_1' } { \rz_1 }{ \trz_1 }, \dots, \Phi_k \rhd \infertight{ \Gamma_k }{ b_k }{ N_1' } { \rz_k }{ \trz_k }$ to obtain
        \[
            \Phi'' \rhd \infertight{ \Gamma_1'', \sum_{i=1}^k \Gamma_i }{ b_1'' + b_1 + \dots + b_k }{ \sub{M_1''}{N_1'/x} }{ \rt }{ \trt }.
        \]
        
        Note that we can assume $\Gamma_1'', \sum_{i=1}^k \Gamma_i$ to be consistent by \autoref{contractedvars} and the fact that $\Gamma_1', \sum_{i=1}^k \Gamma_i$ is consistent.\\
        
        If we perform the $m$ applications of (Contraction) (and the necessary applications of (Exchange)) over the same variables over which they were performed in $\Phi_1'$ after the application of ($\rightarrow$ Intro), i.e., over the variables $z_1, z_1', z_2, z_2', \dots, z_m, z_m'$, we can obtain:
        \[
            \infertight{ \Gamma'' }{ b_1'' + b_1 + \dots + b_k }{ \s_1 (\sub{M_1''}{N_1'/x}) }{ \rt }{ \trt }
        \]
        
        where $\Gamma'' \equiv (\Gamma_1', \sum_{i=1}^k \Gamma_i)$. (By \autoref{contractedvars}, the variables in $\sum_{i=1}^k \Gamma_i$ are not substituted.)\\
        
        If we now perform the $n$ applications of (Contraction) (and the necessary applications of (Exchange)) over the same variables over which they were performed in $\Phi$ after the application of ($\rightarrow$ Elim), i.e., over the variables $x_1, x_1', x_2, x_2', \dots, x_n, x_n'$, we can obtain:
        \[
            \infertight{ \Gamma' }{ b_1'' + b_1 + \dots + b_k }{ \s (\s_1 (\sub{M_1''}{N_1'/x})) }{ \rt }{ \trt }
        \]
        
        where $\Gamma' \equiv \Gamma$.\\
        
        Finally, since $\Gamma' \equiv \Gamma$, we can apply (Exchange) as many times as needed to get the final judgment we wanted:
        \[
            \infertight{ \Gamma }{ b_1'' + b_1 + \dots + b_k }{ \s (\s_1 (\sub{M_1''}{N_1'/x})) }{ \rt }{ \trt }
        \]
        
        \hfill
        
        Since $b = b_1' + b_1 + \dots + b_k$ and $b_1' = b_1'' + 1$, then
        \begin{align*}
            b_1'' + b_1 + \dots + b_k
            &= b_1' - 1 + b_1 + \dots + b_k\\
            &= b - 1.
        \end{align*}
        
        By \autoref{contractedvars}, we have $\s_1 (\sub{M_1''}{N_1'/x}) = \sub{(\s_1 (M_1''))}{N_1'/x}$.
        
        Also, $M_1' = \s_1(M_1'')$, so
        \begin{align*}\label{eq:subred.1.b1}
            \s_1 (\sub{M_1''}{N_1'/x})
            &= \sub{(\s_1 (M_1''))}{N_1'/x}\\
            &= \sub{M_1'}{N_1'/x}. \tag{1}\\
        \end{align*}
        
        Because $x$ cannot be in $\s$, then $\s(\sub{M_1'}{N_1'/x}) = \sub{(\s(M_1'))}{\s(N_1')/x}$.
        
        And since $M_1 = \s(M_1')$ and $N_1 = \s(N_1')$, we have $\sub{(\s(M_1'))}{\s(N_1')/x} = \sub{M_1}{N_1/x}$, so
        \begin{align*}\label{eq:subred.1.b2}
            \s(\sub{M_1'}{N_1'/x})
            &= \sub{(\s(M_1'))}{\s(N_1')/x}\\
            &= \sub{M_1}{N_1/x}. \tag{2}\\
        \end{align*}
        
        Then, by \eqref{eq:subred.1.b1} and \eqref{eq:subred.1.b2} we have:
        \begin{align*}
            \s (\s_1 (\sub{M_1''}{N_1'/x}))
            &= \s (\sub{M_1'}{N_1'/x})\\
            &= \sub{M_1}{N_1/x}.
        \end{align*}
        
        So there is indeed $\Phi' \rhd \infertight{ \Gamma }{ b-1 }{ \sub{M_1}{N_1/x} } { \rt }{ \trt }$.\\
    \end{enumerate}

    \item Rule
    \AxiomC{$M_1 \longrightarrow M_2$}
    \UnaryInfC{$\lambda x.M_1 \longrightarrow \lambda x.M_2$}
    \DisplayProof:\\
    
    Assume $\Phi \rhd \infertight{ \Gamma }{ b }{ \lambda x.M_1 } { \rt }{ \trt }$, $\msf{tight}(\Gamma)$ and the premise $M_1 \longrightarrow M_2$.
    
    Since $\msf{abs}(\lambda x.M_1)$, we must have hypothesis $\msf{tight}(\trt)$.\\
    
    So at some point in the derivation $\Phi$, either the rule ($\multimap$ Intro\textsubscript{t}) or ($\rightarrow$ Intro\textsubscript{t}) is applied and is then followed by zero or more applications of the rules (Exchange) and/or (Contraction).
    
    As the two cases are similar, we will only show the case in which the last rule different from (Exchange) and (Contraction) applied in $\Phi$ is ($\multimap$ Intro\textsubscript{t}):\\
    
    Then at some point in $\Phi$ we have:
    
    \begin{center}
    \begin{mathprooftree}
        \AxiomC{$\Phi_1 \rhd \infertight{ \Gamma', x:(\rz, \msf{tight}) }{ b' }{ M_1' }{ \rt_1 }{ \msf{tight} }$}
        \UnaryInfC{$\infertight{ \Gamma' }{ b' }{ \lambda x.M_1' }{ \rz \multimap \rt_1 }{ \msf{Abs} }$}
    \end{mathprooftree}
    \end{center}
    
    where $\rt = \rz \multimap \rt_1$ and $\trt = \msf{Abs}$.\\
    
    Assume that after the application of this rule, the rule (Contraction) was applied $n$ times (with $n \geq 0$) and (Exchange) zero or more times, and let $\s = [y_n/x_n, y_n/x_n'] \circ [y_{n-1}/x_{n-1}, y_{n-1}/x_{n-1}'] \circ \cdots \circ [y_1/x_1, y_1/x_1']$ be the substitution that reflects the $n$ applications of the rule (Contraction). Then we have:
    
    \begin{itemize}
        \item $M_1 = \s(M_1')$;
        \item $b = b'$.\\
    \end{itemize}

    Since $\msf{tight}(\Gamma)$, we have $\msf{tight}(\Gamma', x:(\rz, \msf{tight}))$.
    
    Since $M_1 \longrightarrow M_2$ and $M_1 = \s(M_1')$, by applying \autoref{substreduction} $n$ times for the substitutions resulting from the $n$ contractions, we get a term $M_2'$ such that $M_2 = \s(M_2')$ and $M_1' \longrightarrow M_2'$.
    
    Then we can apply the induction hypothesis on $\Phi_1$ and get a derivation ending with
    \[
        \infertight{ \Gamma', x:(\rz, \msf{tight}) }{ b'-1 }{ M_2' }{ \rt_1 }{ \msf{tight} }.
    \]
    
    By rule ($\multimap$ Intro\textsubscript{t}), we have:
    
    \begin{center}
    \begin{mathprooftree}
        \AxiomC{$\infertight{ \Gamma', x:(\rz, \msf{tight}) }{ b'-1 }{ M_2' }{ \rt_1 }{ \msf{tight} }$}
        \UnaryInfC{$\infertight{ \Gamma' }{ b'-1 }{ \lambda x.M_2' }{ \rz \multimap \rt_1 }{ \msf{Abs} }$}
    \end{mathprooftree}
    \end{center}
    
    If we now perform the $n$ applications of (Contraction) (and the necessary applications of (Exchange)) over the same variables over which they were performed in $\Phi$ after the application of ($\multimap$ Intro\textsubscript{t}), i.e., over the variables $x_1, x_1', x_2, x_2', \dots, x_n, x_n'$, we can obtain:
    \[
            \infertight{ \Gamma_2 }{ b'-1 }{ \s (\lambda x.M_2') }{ \rz \multimap \rt_1 }{ \msf{Abs} }
    \]
        
    where $\Gamma_2 \equiv \Gamma$.\\
        
    Finally, since $\Gamma_2 \equiv \Gamma$, we can apply (Exchange) as many times as needed to get the final judgment we wanted:
    \[
        \infertight{ \Gamma }{ b'-1 }{ \s (\lambda x.M_2') }{ \rz \multimap \rt_1 }{ \msf{Abs} }
    \]
        
    \hfill
        
    Since $b = b'$, then $b'-1 = b-1$.
        
    And since $M_2 = \s(M_2')$ and $x$ cannot be in $\s$, we have $\s (\lambda x.M_2') = \lambda x.M_2$.\\
        
    So there is indeed $\Phi' \rhd \infertight{ \Gamma }{ b-1 }{ \lambda x.M_2 } { \rt }{ \trt }$.\\

    \item Rule
    \AxiomC{$M_1 \longrightarrow M_2$}
    \AxiomC{$\neg \msf{abs}(M_1)$}
    \BinaryInfC{$M_1 N_1 \longrightarrow M_2 N_1$}
    \DisplayProof:\\
    
    Assume $\Phi \rhd \infertight{ \Gamma }{ b }{ M_1 N_1 } { \rt }{ \trt }$, $\msf{tight}(\Gamma)$ and the premises $M_1 \longrightarrow M_2$ and $\neg \msf{abs}(M_1)$.\\
    
    So at some point in the derivation $\Phi$, either the rule ($\multimap$ Elim), or ($\multimap$ Elim\textsubscript{t}), or ($\rightarrow$ Elim), or ($\rightarrow$ Elim\textsubscript{t}) is applied and is then followed by zero or more applications of the rules (Exchange) and/or (Contraction).
    
    As the four cases are similar, we will only show the case in which the last rule different from (Exchange) and (Contraction) applied in $\Phi$ is ($\multimap$ Elim):\\
        
    Then at some point in $\Phi$ we have:
    
    \begin{center}
    \begin{mathprooftree}
        \AxiomC{$\Phi_1 \rhd \infertight{ \Gamma_1 }{ b_1 }{ M_1' }{ \rz \multimap \rt }{ \trz \multimap \trt }$}
        \AxiomC{$\Phi_2 \rhd \infertight{ \Gamma_2 }{ b_2 }{ N_1' } { \rz }{ \trz }$}
        \BinaryInfC{$\infertight{ \Gamma_1, \Gamma_2 }{ b_1 + b_2 }{ M_1' N_1' }{ \rt }{ \trt }$}
    \end{mathprooftree}
    \end{center}
    
    \hfill
    
    Assume that after the application of this rule, the rule (Contraction) was applied $n$ times (with $n \geq 0$) and (Exchange) zero or more times, and let $\s = [y_n/x_n, y_n/x_n'] \circ [y_{n-1}/x_{n-1}, y_{n-1}/x_{n-1}'] \circ \cdots \circ [y_1/x_1, y_1/x_1']$ be the substitution that reflects the $n$ applications of the rule (Contraction). Then we have:
    
    \begin{itemize}
        \item $M_1 = \s(M_1')$;
        \item $N_1 = \s(N_1')$;
        \item $b = b_1 + b_2$.\\
    \end{itemize}
    
    Since $\msf{tight}(\Gamma)$, we have $\msf{tight}(\Gamma_1)$. And also, as $\neg \msf{abs}(M_1)$, then $\neg \msf{abs}(M_1')$ ($\s$ simply renames free variables).

    Since $M_1 \longrightarrow M_2$ and $M_1 = \s(M_1')$, by applying \autoref{substreduction} $n$ times for the substitutions resulting from the $n$ contractions, we get a term $M_2'$ such that $M_2 = \s(M_2')$ and $M_1' \longrightarrow M_2'$.

    Then we can apply the induction hypothesis on $\Phi_1$ and get a derivation ending with
    \[
        \infertight{ \Gamma_1 }{ b_1-1 }{ M_2' }{ \rz \multimap \rt }{ \trz \multimap \trt }.
    \]

    By rule ($\multimap$ Elim), with $\infertight{ \Gamma_2 }{ b_2 }{ N_1' } { \rz }{ \trz }$ from $\Phi_2$, we have:
    
    \begin{center}
    \begin{mathprooftree}
        \AxiomC{$\infertight{ \Gamma_1 }{ b_1-1 }{ M_2' }{ \rz \multimap \rt }{ \trz \multimap \trt }$}
        \AxiomC{$\infertight{ \Gamma_2 }{ b_2 }{ N_1' } { \rz }{ \trz }$}
        \BinaryInfC{$\infertight{ \Gamma_1, \Gamma_2 }{ b_1 + b_2 - 1 }{ M_2' N_1' }{ \rt }{ \trt }$}
    \end{mathprooftree}
    \end{center}
    
    If we now perform the $n$ applications of (Contraction) (and the necessary applications of (Exchange)) over the same variables over which they were performed in $\Phi$ after the application of ($\multimap$ Elim), i.e., over the variables $x_1, x_1', x_2, x_2', \dots, x_n, x_n'$, we can obtain:
    \[
        \infertight{ \Gamma_3 }{ b_1 + b_2 - 1 }{ \s (M_2' N_1') }{ \rt }{ \trt }
    \]
    
    where $\Gamma_3 \equiv \Gamma$.\\
    
    Finally, since $\Gamma_3 \equiv \Gamma$, we can apply (Exchange) as many times as needed to get the final judgment we wanted:
    \[
        \infertight{ \Gamma }{ b_1 + b_2 - 1 }{ \s (M_2' N_1') }{ \rt }{ \trt }
    \]
    
    \hfill
    
    Since $b = b_1+b_2$, then $b_1+b_2-1 = b-1$.
    
    And since $M_2 = \s(M_2')$ and $N_1 = \s(N_1')$, we have $\s (M_2' N_1') = M_2 N_1$.\\
    
    So there is indeed $\Phi' \rhd \infertight{ \Gamma }{ b-1 }{ M_2 N_1 } { \rt }{ \trt }$.\\

    \item Rule
    \AxiomC{$\msf{neutral}(N_1)$}
    \AxiomC{$M_1 \longrightarrow M_2$}
    \BinaryInfC{$N_1 M_1 \longrightarrow N_1 M_2$}
    \DisplayProof:\\
    
    Assume $\Phi \rhd \infertight{ \Gamma }{ b }{ N_1 M_1 } { \rt }{ \trt }$, $\msf{tight}(\Gamma)$ and the premises $\msf{neutral}(N_1)$ and $M_1 \longrightarrow M_2$.\\
    
    So at some point in the derivation $\Phi$, either the rule ($\multimap$ Elim), or ($\multimap$ Elim\textsubscript{t}), or ($\rightarrow$ Elim), or ($\rightarrow$ Elim\textsubscript{t}) is applied, giving $\infertight{ \Gamma' }{ b }{ N_1' M_1' } { \rt }{ \trt }$, and is then followed by zero or more applications of the rules (Exchange) and/or (Contraction).\\
    
    Assume that the rule (Contraction) is applied $n$ times (with $n \geq 0$) and (Exchange) zero or more times, and let $\s = [y_n/x_n, y_n/x_n'] \circ [y_{n-1}/x_{n-1}, y_{n-1}/x_{n-1}'] \circ \cdots \circ [y_1/x_1, y_1/x_1']$ be the substitution that reflects the $n$ applications of the rule (Contraction). Then we have:
    
    \begin{itemize}
        \item $N_1 = \s(N_1')$;
        \item $M_1 = \s(M_1')$.\\
    \end{itemize}
    
    Then as a premise for any of those four rules, we have a derivation for $N_1'$ of the form:
    \[
        \Phi_1 \rhd \infertight{ \Gamma_1 }{ b_1 }{ N_1' }{ \rt' }{ \trt' }
    \]
    
    Since $\msf{neutral}(N_1)$, we have $\msf{neutral}(N_1')$ ($\s$ simply renames free variables).
    
    Also, since $\msf{tight}(\Gamma)$, independently from the elimination rule that was applied, we have $\msf{tight}(\Gamma_1)$.
    
    Then by \autoref{spreading}, we have $\msf{tight}(\trt')$.\\
    
    So this means that actually, the last rule different from (Exchange) and (Contraction) applied in $\Phi$ must be either ($\multimap$ Elim\textsubscript{t}) or ($\rightarrow$ Elim\textsubscript{t}), and not ($\multimap$ Elim) nor ($\rightarrow$ Elim).
    
    And as the two cases are similar, we will only show the case in which the last rule different from (Exchange) and (Contraction) applied in $\Phi$ is ($\multimap$ Elim\textsubscript{t}):\\
    
    Then, before the $n$ applications of the rule (Contraction) and possible applications of (Exchange) what we have is:
    
    \begin{center}
    \begin{mathprooftree}
        \AxiomC{$\Phi_1 \rhd \infertight{ \Gamma_1 }{ b_1 }{ N_1' }{ \rz \multimap \rt }{ \msf{Neutral} }$}
        \AxiomC{$\Phi_2 \rhd \infertight{ \Gamma_2 }{ b_2 }{ M_1' } { \rz }{ \msf{tight} }$}
        \BinaryInfC{$\infertight{ \Gamma_1, \Gamma_2 }{ b_1 + b_2 }{ N_1' M_1' }{ \rt }{ \msf{Neutral} }$}
    \end{mathprooftree}
    \end{center}
    
    and
    
    \begin{itemize}
        \item $\trt = \msf{Neutral}$;
        \item $N_1 = \s(N_1')$;
        \item $M_1 = \s(M_1')$;
        \item $b = b_1 + b_2$.\\
    \end{itemize}
    
    Since $\msf{tight}(\Gamma)$, we have $\msf{tight}(\Gamma_2)$.
    
    Since $M_1 \longrightarrow M_2$ and $M_1 = \s(M_1')$, by applying \autoref{substreduction} $n$ times for the substitutions resulting from the $n$ contractions, we get a term $M_2'$ such that $M_2 = \s(M_2')$ and $M_1' \longrightarrow M_2'$.
    
    Then we can apply the induction hypothesis on $\Phi_2$ and get a derivation ending with
    \[
        \infertight{ \Gamma_2 }{ b_2 - 1 }{ M_2' }{ \rz }{ \msf{tight} }.
    \]

    By rule ($\multimap$ Elim\textsubscript{t}), with $\infertight{ \Gamma_1 }{ b_1 }{ N_1' }{ \rz \multimap \rt }{ \msf{Neutral} }$ from $\Phi_1$, we have:
    
    \begin{center}
    \begin{mathprooftree}
        \AxiomC{$\infertight{ \Gamma_1 }{ b_1 }{ N_1' }{ \rz \multimap \rt }{ \msf{Neutral} }$}
        \AxiomC{$\infertight{ \Gamma_2 }{ b_2 - 1 }{ M_2' }{ \rz }{ \msf{tight} }$}
        \BinaryInfC{$\infertight{ \Gamma_1, \Gamma_2 }{ b_1 + b_2 - 1 }{ N_1' M_2' }{ \rt }{ \msf{Neutral} }$}
    \end{mathprooftree}
    \end{center}
    
    If we now perform the $n$ applications of (Contraction) (and the necessary applications of (Exchange)) over the same variables over which they were performed in $\Phi$ after the application of ($\multimap$ Elim\textsubscript{t}), i.e., over the variables $x_1, x_1', x_2, x_2', \dots, x_n, x_n'$, we can obtain:
    \[
        \infertight{ \Gamma_3 }{ b_1 + b_2 - 1 }{ \s (N_1' M_2') }{ \rt }{ \msf{Neutral} }
    \]
    
    where $\Gamma_3 \equiv \Gamma$.\\
    
    Finally, since $\Gamma_3 \equiv \Gamma$, we can apply (Exchange) as many times as needed to get the final judgment we wanted:
    \[
        \infertight{ \Gamma }{ b_1 + b_2 - 1 }{ \s (N_1' M_2') }{ \rt }{ \msf{Neutral} }
    \]
    
    \hfill
    
    Since $b = b_1+b_2$, then $b_1+b_2-1 = b-1$.
    
    Also, $\trt = \msf{Neutral}$.
    
    And since $N_1 = \s(N_1')$ and $M_2 = \s(M_2')$, we have $\s (N_1' M_2') = N_1 M_2$.\\
    
    So there is indeed $\Phi' \rhd \infertight{ \Gamma }{ b-1 }{ N_1 M_2 } { \rt }{ \trt }$.

\end{enumerate}
\end{proof}


\end{lemma}

\begin{theorem}[Tight correctness]\label{tightcorrect} 
If $\Phi \rhd \infertight{ \Gamma }{ b }{ M } { \rt }{ \trt }$ is a tight derivation, then there exists $N$ such that $M \longrightarrow^b N$ and $\msf{normal}(N)$. Moreover, if $\trt = \msf{Neutral}$ then $\msf{neutral}(N)$.

\begin{proof}
By induction on the evaluation length of $M \longrightarrow^k N$.

If $M$ is a (leftmost-outermost) normal form, then by taking $N = M$ and $k = 0$, the statement follows from the tightness property of tight typings of normal forms (\autoref{tightproperties}(i)). The \emph{moreover} part follows from the neutrality property (\autoref{tightproperties}(ii)).

Otherwise, $M \longrightarrow M'$ and by quantitative subject reduction (\autoref{subjectreduction}) there exists a derivation $\Phi' \rhd \infertight{ \Gamma }{ b-1 }{ M' } { \rt }{ \trt }$. By induction, there exists $N$ such that $\msf{normal}(N)$ and $M' \longrightarrow^{b-1} N$. Note that $M \longrightarrow M' \longrightarrow^{b-1} N$, that is, $M \longrightarrow^b N$.
\end{proof}
\end{theorem}

\subsection{Type Inference Algorithm}\label{sec:tightinference}

We now extend the type inference algorithm defined in \autoref{chap:linearrank} (\autoref{ti2}) to also infer the number of reduction steps of the typed term to its normal form, when using the leftmost-outermost evaluation strategy.

This is done by slightly modifying the unification algorithm in \autoref{unify} and the algorithm in \autoref{ti2}, which will now carry and update a measure $b$ that relates to the number of reduction steps. First, recall \autoref{transformuni}, presented in \autoref{chap:linearrank}.

\begin{definition}[Quantitative Unification Algorithm]\label{unifyq}
Let $P$ be a unification problem (with types in $\tl{0}$). The new unification function $\msf{UNIFY_Q}(P)$, which decides whether $P$ has a solution and, if so, returns the MGU of $P$ and an integer $b$ used for counting purposes in the inference algorithm, is defined as:
\begin{algorithmic}
    \Function{$\msf{UNIFY_Q}$}{$P$}
        \State $b \coloneqq 0$;
        \While{$P \Rightarrow P'$}
                \If{$P = \{\rz_1 \multimap \rz_2 = \rz_3 \multimap \rz_4\} \cup P_1$ \textbf{and} $P' = \{\rz_1 = \rz_3, \rz_2 = \rz_4\} \cup P_1$}
                    \State $b \coloneqq b+1$;
                \EndIf
                \State $P \coloneqq P'$;
        \EndWhile
        \If{$P$ is in solved form}
            \State \Return $(\ts_P, b)$;
        \Else
            \State $\msf{FAIL}$;
        \EndIf
    \EndFunction
\end{algorithmic}
\end{definition}

Let $\tl{1}$-environment be an environment as defined in \autoref{chap:linearrank}, i.e., just like the definition we use in the current chapter, but the predicates are only the first element of the pair (i.e., a sequence from $\tl{1}$).

\begin{definition}[Quantitative Type Inference Algorithm]\label{qti2}
Let $\Gamma$ be a $\tl{1}$-environment, $M$ a $\lambda$-term, $\rt$ a linear rank~2 intersection type, $b$ a quantitative measure and $\msf{UNIFY_Q}$ the function in \autoref{unifyq}. The function $\msf{T_Q}(M) = (\Gamma, \rt, b)$ defines a new type inference algorithm that gives a quantitative measure for the $\lambda$-calculus in the Linear Rank~2 Quantitative Type System, in the following way:
\begin{enumerate}
    \item \tul{If} $M=x$, \tul{then} $\Gamma = [x:\alpha]$, $\rt = \alpha$ and ${\color{azulciencias} b = 0}$, where $\alpha$ is a new variable;
    
    \item \tul{If} $M=\lambda x.M_1$ and $\msf{T_Q}(M_1) = (\Gamma_1, \rt_1, b_1)$ \tul{then}:
    \begin{enumerate}
        \item \tul{if} $x \notin \dom(\Gamma_1)$, \tul{then} $\msf{FAIL}$;
        
        \item \tul{if} $(x:\rz) \in \Gamma_1$, \tul{then} $\msf{T_Q}(M) = ({\Gamma_1}_x, \rz \multimap \rt_1, {\color{azulciencias} b_1})$;
            
        \item \tul{if} $(x:\rz_1 \cap \cdots \cap \rz_n) \in \Gamma_1$ (with $n \geq 2$), \tul{then} $\msf{T_Q}(M) = ({\Gamma_1}_x, \rz_1 \cap \cdots \cap \rz_n \rightarrow \rt_1, {\color{azulciencias} b_1})$.
        
    \end{enumerate}
    
    \item \tul{If} $M=M_1 M_2$, \tul{then}:
    \begin{enumerate}
        \item \tul{if} $\msf{T_Q}(M_1) = (\Gamma_1, \alpha_1, b_1)$ and $\msf{T_Q}(M_2) = (\Gamma_2, \rz_2, b_2)$,\\
        \tul{then} $\msf{T_Q}(M) = (\ts(\Gamma_1 + \Gamma_2), \ts(\alpha_3), {\color{azulciencias} b_1+b_2})$,\\
        where $(\ts, \_) = \msf{UNIFY_Q}(\{\alpha_1 = \alpha_2 \multimap \alpha_3, \rz_2 = \alpha_2\})$ and $\alpha_2,\alpha_3$ are new variables;
        
        \item \tul{if} $\msf{T_Q}(M_1) = (\Gamma_1', \rz_1' \cap \cdots \cap \rz_n' \rightarrow \rt_1', b_1)$ (with $n \geq 2$) and, for each $1 \leq i \leq n$, $\msf{T_Q}(M_2) = (\Gamma_i, \rz_i, b_i)$,\\
        \tul{then} $\msf{T_Q}(M) = (\ts(\Gamma_1' + \sum_{i=1}^n \Gamma_i), \ts(\rt_1'), {\color{azulciencias} b_1 + \sum_{i=1}^n b_i + b_3 + 1})$,\\
        where $(\ts, b_3) = \msf{UNIFY_Q}(\{\rz_i = \rz_i' \mid 1 \leq i \leq n\})$;
        
        \item \tul{if} $\msf{T_Q}(M_1) = (\Gamma_1, \rz \multimap \rt_1, b_1)$ and $\msf{T_Q}(M_2) = (\Gamma_2, \rz_2, b_2)$,\\
        \tul{then} $\msf{T_Q}(M) = (\ts(\Gamma_1 + \Gamma_2), \ts(\rt_1), {\color{azulciencias} b_1+b_2+b_3+1})$,\\
        where $(\ts, b_3) = \msf{UNIFY_Q}(\{\rz_2 = \rz\})$;
        
        \item \tul{otherwise} $\msf{FAIL}$.
    \end{enumerate}
\end{enumerate}
\end{definition}

Note that $b$ is only increased by $1$ and added the quantity given by $\msf{UNIFY_Q}$ in rules 3.(b) and 3.(c), since these are the only cases in which the term $M$ is a redex.

\begin{example}
Let us show the type inference process for the $\lambda$-term $\lambda x.x x$.

\begin{itemize}
    \item By rule 1., $\msf{T_Q}(x) = ([x:\alpha_1], \alpha_1, 0)$.
    
    \item By rule 1., again, $\msf{T_Q}(x) = ([x:\alpha_2], \alpha_2, 0)$.
    
    \item Then by rule 3.(a), $\msf{T_Q}(x x) = (\ts([x:\alpha_1] + [x:\alpha_2]), \ts(\alpha_4), 0+0) = (\ts([x:\alpha_1 \cap \alpha_2]), \ts(\alpha_4), 0)$,
    
    where $(\ts, \_) = \msf{UNIFY_Q}(\{\alpha_1 = \alpha_3 \multimap \alpha_4, \alpha_2 = \alpha_3\}) = (\tsubst{\alpha_3 \multimap \alpha_4/\alpha_1, \alpha_3/\alpha_2}, 0)$.
    
    So $\msf{T_Q}(x x) = ([x:(\alpha_3 \multimap \alpha_4) \cap \alpha_3], \alpha_4, 0)$.
    
    \item Finally, by rule 2.(c), $\msf{T_Q}(\lambda x.x x) = ([\;], (\alpha_3 \multimap \alpha_4) \cap \alpha_3 \rightarrow \alpha_4, 0)$.
\end{itemize}
\end{example}

\begin{example}
Let us now show the type inference process for the $\lambda$-term $(\lambda x.x x) (\lambda y.y)$.

\begin{itemize}
    \item From the previous example, we have $\msf{T_Q}(\lambda x.x x) = ([\;], (\alpha_3 \multimap \alpha_4) \cap \alpha_3 \rightarrow \alpha_4, 0)$.
    
    \item By rules 1. and 2.(b), for the identity, the algorithm gives $\msf{T_Q}(\lambda y.y) = ([\;], \alpha_1 \multimap \alpha_1, 0)$.
    
    \item By rules 1. and 2.(b), again, for the identity, $\msf{T_Q}(\lambda y.y) = ([\;], \alpha_2 \multimap \alpha_2, 0)$.
    
    \item Then by rule 3.(b), $\msf{T_Q}((\lambda x.x x) (\lambda y.y)) = (\ts([\;] + [\;] + [\;]), \ts(\alpha_4), 0+0+0+b_3+1) = ([\;], \ts(\alpha_4), b_3+1)$, where $(\ts, b_3) = \msf{UNIFY_Q}(\{\alpha_1 \multimap \alpha_1 = \alpha_3 \multimap \alpha_4, \alpha_2 \multimap \alpha_2 = \alpha_3\})$, calculated by performing the following transformations:
    \begin{align*}
    \{ \alpha_1 \multimap \alpha_1 = \alpha_3 \multimap \alpha_4, \alpha_2 \multimap \alpha_2 = \alpha_3 \}
    &\Rightarrow \{ \alpha_1 = \alpha_3, \alpha_1 = \alpha_4, \alpha_2 \multimap \alpha_2 = \alpha_3 \}\\
    &\Rightarrow \{ \alpha_1 = \alpha_3, \alpha_3 = \alpha_4, \alpha_2 \multimap \alpha_2 = \alpha_3 \}\\
    &\Rightarrow \{ \alpha_1 = \alpha_4, \alpha_3 = \alpha_4, \alpha_2 \multimap \alpha_2 = \alpha_4 \}\\
    &\Rightarrow \{ \alpha_1 = \alpha_4, \alpha_3 = \alpha_4, \alpha_4 = \alpha_2 \multimap \alpha_2 \}\\
    &\Rightarrow \{ \alpha_1 = \alpha_2 \multimap \alpha_2, \alpha_3 = \alpha_2 \multimap \alpha_2, \alpha_4 = \alpha_2 \multimap \alpha_2 \}
    \end{align*}
    
    So $\ts = \tsubst{(\alpha_2 \multimap \alpha_2) / \alpha_1, (\alpha_2 \multimap \alpha_2) / \alpha_3, (\alpha_2 \multimap \alpha_2) / \alpha_4}$
    
    and $b_3 = 1$ because there was performed one transformation (the first) of the form $\{\rz_1 \multimap \rz_2 = \rz_3 \multimap \rz_4\} \cup P \Rightarrow \{\rz_1 = \rz_3, \rz_2 = \rz_4\} \cup P$.
    
    And then, $\msf{T_Q}((\lambda x.x x) (\lambda y.y)) = ([\;], \alpha_2 \multimap \alpha_2, 1+1) = ([\;], \alpha_2 \multimap \alpha_2, 2)$.
\end{itemize}
\end{example}

Since the Quantitative Type Inference Algorithm only differs from the algorithm in \autoref{chap:linearrank} on the addition of the quantitative measure, and only infers a linear rank 2 intersection type and not a multi-type, the typing soundness (\autoref{soundness2}) and completeness (\autoref{completeness2}) are formalized in a similar way.

\begin{theorem}[Typing soundness]\label{soundness2}
If $\msf{T_Q}(M) = ([x_1:\ro_1, \dots, x_n:\ro_n], \rt, b)$, then $\infertight{ [x_1:(\ro_1, \tro_1), \dots, x_n:(\ro_n, \tro_n)] }{ b' }{ M }{ \rt }{ \trt }$ (for some measure $b'$ and multi-types $\trt, \tro_1, \dots, \tro_n$).

\begin{proof}
The proof follows as in \autoref{soundness} (only the non-t-indexed rules are necessary).
\end{proof}
\end{theorem}

\begin{theorem}[Typing completeness]\label{completeness2}
If $\Phi \rhd \infertight{ [x_1:(\ro_1, \tro_1), \dots, x_n:(\ro_n, \tro_n)] }{ b }{ M }{ \rt }{ \trt }$, then $\msf{T_Q}(M) = (\Gamma', \rt', b')$ (for some $\tl{1}$-environment $\Gamma'$, type $\rt'$ and measure $b'$) and there is a substitution $\ts$ such that $\ts(\rt') = \rt$ and $\ts(\Gamma') \equiv [x_1:\ro_1, \dots, x_n:\ro_n]$.

\begin{proof}
The proof follows similarly to the proof of \autoref{completeness} (note that even when t-indexed rules are used in the derivation, the resulting linear rank 2 intersection type is the same as when the correspondent non-t-indexed rules are used).
\end{proof}
\end{theorem}

As for the quantitative measure given by the algorithm, we conjecture that it corresponds to the number of evaluation steps of the typed term to normal form, when using the leftmost-outermost evaluation strategy. We strongly believe the conjecture holds, based on the attempted proofs so far and because it holds for every experimental results obtained by our implementation. We have not yet proven this property, which we formalize, in part, in the second point of the strong soundness:\\

\begin{conjecture}[Strong soundness]\label{strongsoundness}
If $\msf{T_Q}(M) = ([x_1:\ro_1, \dots, x_n:\ro_n], \rt, b)$, then:
\begin{enumerate}
    \item There is a derivation $\Phi \rhd \infertight{ [x_1:(\ro_1, \tro_1), \dots, x_n:(\ro_n, \tro_n)] }{ b' }{ M }{ \rt }{ \trt }$ (for some measure~$b'$ and multi-types $\trt, \tro_1, \dots, \tro_n$);
    
    \item If $\Phi$ is a tight derivation, then $b = b'$.
\end{enumerate}
\end{conjecture}

Note that the second point implies, by \autoref{tightcorrect}, that there exists $N$ such that $M \longrightarrow^b N$ and $\msf{normal}(N)$, which is what we conjecture.\\

We believe that proving this conjecture is not a trivial task. A first approach could be to try to use induction on the definition of $\msf{T_Q}(M)$. However, this does not work because the subderivations within a tight derivation are not necessarily tight. For that same reason, it is also not trivial to construct a tight derivation from the result given by the algorithm or from a non-tight derivation.
Thus, in order to prove this conjecture, we believe that it will be necessary to establish a stronger relation between the algorithm and tight derivations.

\section{Conclusions and Future Work}\label{chap:conclusion}

 When developing a non-idempotent intersection type system capable of obtaining quantitative information about a $\lambda$-term while inferring its type, we realized that the classical notion of rank was not a proper fit for non-idempotent intersection types, and that the ranks could be quantitatively more useful if the base case was changed to types that give more quantitative information in comparison to simple types, which is the case for linear types. We then came up with a new definition of rank for intersection types based on linear types, which we call \emph{linear rank} \cite{typesabstract}. Based on the notion of linear rank, we defined a new intersection type system for the $\lambda$-calculus, restricted to linear rank~2 non-idempotent intersection types, and a new type inference algorithm  which we proved to be sound and complete with respect to the type system.

We then merged that intersection type system with the system for the leftmost-outermost evaluation strategy presented in \cite{accattoli2018tight} in order to use the linear rank~2 non-idempotent intersection types to obtain quantitative information about the typed terms, and we proved that the resulting type system gives the correct number of evaluation steps for a kind of derivations. We also extended the type inference algorithm we had defined, in order to also give that measure, and showed that it is sound and complete with respect to the type system for the inferred types, and conjectured that the inferred measures correspond to the ones given by the type system.

In the future, we would like to:
\begin{itemize}
    \item prove \autoref{strongsoundness};
    \item further explore the relation between our definition of linear rank and the classical definition of rank;
    \item extend the type systems and the type inference algorithms for the affine terms;
    \item adapt the Linear Rank~2 Quantitative Type System and the Quantitative Type Inference Algorithm for other evaluation strategies.
\end{itemize}



\bibliography{refs}

\end{document}